\documentclass[%
 reprint,
 preprintnumber,
 amsmath,amssymb,
 aps,prd,
 nofootinbib,
]{revtex4-2}

\pdfoutput=1

\usepackage{graphicx}
\usepackage{dcolumn}
\usepackage{bm}
\usepackage{epsfig}
\usepackage{multirow} 
\usepackage{epstopdf}
\usepackage[titles]{tocloft}
\usepackage{amsmath,amssymb,mathtools}
\usepackage{subfig}
\usepackage{booktabs}
\allowdisplaybreaks
\title{}
\date{}
\usepackage{graphicx,color}
\usepackage{amsmath}
\usepackage{slashed}
\usepackage{mathtools}
\usepackage{amsfonts}
\usepackage{hyperref}
\usepackage{ragged2e}
\usepackage{xr}

\definecolor{prd_blue}{RGB}{41, 41, 133}

\hypersetup{
    colorlinks=true,
    linkcolor=prd_blue,
    filecolor=prd_blue,      
    urlcolor=prd_blue,
    citecolor=prd_blue
}

\newcommand{\alspi}{\ \frac{2\alpha_s}{3\pi}}
\newcommand{\bt}{\begin{widetext}}
\newcommand{\et}{\end{widetext}}
\newcommand{\nn}{\nonumber}

\begin{document}

\title{Next-to-leading power corrections to the event shape variables}

\author{Neelima Agarwal$^{1}$}
\email{dragarwalphysics@gmail.com}
\author{Melissa van Beekveld$^{2}$}
\email{melissa.vanbeekveld@physics.ox.ac.uk}
\author{Eric Laenen$^{3,4,5}$}
\email{eric.laenen@nikhef.nl}
\author{Shubham Mishra$^{6}$}
\email{shubhamhere82@gmail.com}
\author{Ayan Mukhopadhyay$^{6}$}
\email{ayanmukhopadhyay5@gmail.com}
\author{Anurag Tripathi$^{6}$\vspace{0.4cm}}
\email{tripathi@phy.iith.ac.in}

\affiliation{$^{1}$Department of Physics, Chaitanya Bharathi Institute of Technology, Gandipet, Hyderabad, Telangana State 500075, India}
\affiliation{$^{2}$Rudolf Peierls Centre for Theoretical Physics, Clarendon Laboratory, Parks Road, University of Oxford,  Oxford OX1 3PU, UK}
\affiliation{$^{3}$Institute of Physics, University of Amsterdam, Science Park 904,\\ 1098 XH Amsterdam, The Netherlands}
\affiliation{$^{4}$Nikhef, Theory Group, Science Park 105, 1098 XG,\\ Amsterdam, The Netherlands}
\affiliation{$^{5}$Institute for Theoretical Physics, Utrecht University, Leuvenlaan 4, 3584 CE Utrecht, The Netherlands}
\affiliation{$^{6}$Department of Physics, Indian Institute of Technology Hyderabad, Kandi, Sangareddy, Telangana State 502285, India}

\date{\today}

\begin{abstract}
 We investigate the origin of next-to-leading power corrections to 
the event shapes thrust and $c$-parameter, at next-to-leading order.
 For both event shapes we trace the origin of such terms in the exact
 calculation, and compare with a recent approach involving
 the eikonal approximation and momentum shifts that follow from the Low-Burnett-Kroll-Del Duca theorem. 
 We assess the differences both analytically and numerically.
 For the $c$-parameter both exact and approximate results are expressed in terms of elliptic integrals, but 
near the elastic limit it exhibits patterns similar to the thrust results.
\end{abstract}

\maketitle
\tableofcontents

\section{Introduction}

Providing precise estimates for cross-sections in
perturbative QCD is needed to match the ever increasing precision of
collider physics measurements. Two complementary directions are pursued in this endeavour. In one, 
exact higher order calculations in the coupling $\alpha_s$ are
performed, and methods thereto developed. In the other, all-order
results are derived in certain kinematic limits, where associated classes
of logarithmic terms are enhanced; the latter can often be resummed to all
orders, using a varied and ever expanding set of methods. 
A highly relevant region in this regard is the near-elastic region (which
we loosely refer to here also as the threshold region), where
the phase space for emitted particles is limited. In such a situation,
the cancellation of
infrared singularities, guaranteed by the KLN theorem~\cite{Kinoshita:1975bt,
  PhysRev.133.B1549}, is incomplete in the sense that it leaves large logarithmic remainders at any
order in perturbation theory. Our analysis in this paper is relevant
for the second direction.

To be more specific, if $\xi$ is a dimensionless kinematic variable, such
that $\xi\rightarrow 0$ towards the elastic region, the corresponding differential cross-section has the generic form
\bt
\begin{equation}
	\frac{d \sigma}{d \xi} \, = \, \sum_{n = 0}^{\infty} \left( \frac{\alpha_s}{\pi} \right)^n \, 
	\left[   \sum_{m = 0}^{2 n - 1}  c_{n m}^{\text{LP}}
	\left( \frac{\log^m \xi}{\xi} \right)_+  + \,
	c_n^{(\delta)} \delta(\xi) + \,\sum_{m = 0}^{2 n - 1} c_{nm}^{\text{NLP}} \, \log^m \xi \, + \, \ldots   \right] \, .
	\label{eq:NLPintro}
\end{equation}
 \et
The first term on the right in the above equation is well-known to originate
from soft and/or collinear radiation and, together with the second term, makes up the \textit{leading
  power} (LP) terms. Much is known about LP terms to arbitrary order,
and there have been numerous approaches towards their resummation~\cite{Parisi:1979xd, Curci:1979am, STERMAN1987310, Catani:1989ne,
  Catani:1990rp, Gatheral:1983cz, Frenkel:1984pz, Sterman:1981jc,
  Korchemsky:1992xv, Korchemsky:1993uz,Forte:2002ni,
  Contopanagos:1996nh, Becher:2006nr, Schwartz:2007ib,
  PhysRevD.78.034027, PhysRevD.80.094013} and  power correction studies~\cite{Gardi:2001di, Agarwal:2020uxi}. Reviews on some of
these different approaches can be found in~\cite{Laenen:2004pm,Luisoni_2015,Becher:2014oda,Campbell:2017hsr,Agarwal:2021ais}.

The last term on the right represents \textit{next-to-leading power} (NLP)
terms. These originate from both soft gluon
and soft quark emissions. Although suppressed by a power of $\xi$, they can be
relevant, since they grow logarithmically towards threshold. In contrast to the LP terms, the precise organization of these
NLP terms to all orders and arbitrary logarithmic accuracy are not yet clear.
In this paper we focus on such NLP terms for two event shapes in $e^+e^-$ collisions: thrust and the $C$-parameter.
These observables are interesting in this regard because, in contrast to most earlier
studies, all QCD effects reside in the final state, and because their definition
involves special phase space constraints that were not considered so
far~\cite{Bauer:2000ew, Bauer:2001ct, Bauer:2000yr, Bauer:2001yt}. For these observables our aim is to trace the origin
of the NLP terms near the elastic limit, to
examine to what extent there is a common pattern of NLP terms, and
assess their size.

Patterns among NLP terms have been studied for various processes~\cite{Bonocore2015, Bonocore2016, Moult:2018jjd, Beneke:2019oqx,
  Moult:2019mog, Bahjat-Abbas:2019fqa, Beneke:2019mua, Moult:2019vou,
  Ajjath:2020sjk, Beneke:2020ibj, Ajjath:2020ulr, vanBeekveld:2021mxn,
  Liu:2019oav}. For a number of observables
their numerical contribution can be significant~\cite{KRAMER1998523, BALL2013746,PhysRevLett.114.212001,
  vanBeekveld:2019prq, vanBeekveld:2021hhv, Ajjath2022}.
The all-order resummation of NLP terms has been pursued through
different approaches, such as a diagrammatic and a path integral approach given
in~\cite{Laenen_2009, Laenen2011}, while a physical kernel
approach is pursued in~\cite{Soar2010OnHD, Florian2014ApproximateNH,
  Presti2014LeadingLL}.
From direct QCD formalism, the development of factorization
theorems for NLP terms extending from LP terms are studied in~\cite{Bonocore2015,Bonocore2016,Bonocore2020AsymptoticDO,PhysRevD.95.125009,PhysRevD.96.065007,Gervais2017SoftRT,PhysRevD.103.034022} (for NLP factorization see,~\cite{Beneke:2022obx}).
The study of NLP effects in the framework of SCET has many aspects, and the active studies are done in the operators contributing at NLP level~\cite{Kolodrubetz:2016uim, Moult:2016fqy, Feige:2017zci, Beneke:2017ztn, Beneke:2018rbh, Bhattacharya:2018vph, Beneke:2019kgv, Bodwin:2021epw}, the development of factorization~\cite{Moult:2019mog, Beneke:2019oqx, Liu:2019oav, Liu:2020tzd}, and the explicit studies of physical observables~\cite{Boughezal:2016zws, Moult:2017rpl, Chang:2017atu, Moult:2018jjd, Beneke:2018gvs, Ebert:2018gsn, Beneke:2019mua, Moult:2019uhz, Liu:2020ydl, Liu:2020eqe, Wang:2019mym, Beneke:2020ibj}.

Event shapes, which probe final state dynamics through geometrical
constraints, have long been used to develop QCD
ideas~\cite{Banfi:2001bz,Catani:1992jc,Catani:1992ua,Ellis:1991qj},
e.g. the development of resummation and fixed-order
computations~\cite{Dokshitzer_1998,Dasgupta_2002,Banfi:2001bz,Antonelli_2000,Ridder_2007,Becher_2008,PhysRevLett.101.162001}.
They have also played a vital role in extracting the strong
coupling constant~\cite{GehrmannDeRidder:2007bj,Becher:2008cf,Dasgupta:2003iq,
  Catani:1991kz}. 
Here we examine to what extent NLP terms for two event shapes can be
predicted using the kinematical shift method~\cite{DelDuca:2017twk,vanBeekveld:2019prq}, as well as the soft quark
emission approximation. 

Our paper is organized as follows. We compute the thrust
distribution in shifted approximation in section~\ref{THRUSTNLP}
to assess the relevance of NLP terms. We perform a similar assessment
in section~\ref{CPARANLP}, which is considerably more complicated.
We conclude in section~\ref{sec:conclusions},
while appendices contain certain technical
aspects of elliptic functions, and a summary table of our results.

\section{Next-to-leading power terms  and  kinematical shifts}
\label{sec:next-leading-power}

It was recently shown \cite{DelDuca:2017twk} that NLP terms,  at next-to-leading order (NLO) and to the
extent that they are due to soft gluon radiation, may be derived efficiently
using a combination of the eikonal approximation and kinematical
shifts, in first instance for colour singlet final states.
The method holds for matrix elements and thus for differential
distributions, and was subsequently extended to final state radiation in~\cite{vanBeekveld:2019prq}.
It rests in essence on the Low-Burnett-Kroll-Del Duca (LBKD) theorem~\cite{Low:1958sn, Burnett:1967km, DelDuca:1990gz}, which enables the expression
of the one-gluon emission amplitude in terms of the elastic amplitude
(even if the latter contains loops, such as in the case of Higgs, or
multi-Higgs production).
The NLO matrix element up to NLP accuracy can be written as a combination of
scalar, spin, and orbital terms:
\begin{align}
  \mathcal{M}_{\text{NLP}} \, =\,  \mathcal{M}_{\text{scal}}+\mathcal{M}_{\text{spin}}+\mathcal{M}_{\text{orb}}\,.
\end{align}
The scalar term is essentially a multiplicative term containing
the LP eikonal approximation. The spin-dependent
term is of NLP accuracy, as the eikonal approximation cannot resolve
emitter spin. The orbital term, also of NLP accuracy, involves derivative operators. These
can be represented as a first-order Taylor expansion of kinematically
shifted momenta.  Combining these terms for the production of colour singlet particles yields the expression 
\begin{align}
  \overline{\sum} |\mathcal{M}_{\text{shift}}|^{2} \, =\,  & g_s^2 N_c (N_c^2-1)  \frac{2p_1\cdot p_2}{(p_1\cdot p_3)(p_2\cdot p_3)} \nn \\ & \times
  |\mathcal{M}_{0}(p_1-\delta p_1,p_2- \delta p_2)|^2,
  \label{eq:MainFormula}
\end{align}
 where $|\mathcal{M}_0(p_1,p_2)|^2 $ is the matrix element squared at the leading order (LO), and $\overline{\sum}$ denotes the sum (average) over the final (initial)
state spins and colours, $p_3$ is the momentum of the emitted radiation, and $p_1$, $p_2$
are the momenta of the particles already present at the born
level. The shifts in the momenta are given by

\begin{align}
\delta p_1^{\mu} & \,  \,=\, -\frac{1}{2}\left(\frac{p_2 \cdot p_3}{p_1 \cdot p_2}p_1^{\mu}- \frac{p_1 \cdot p_3}{p_1 \cdot p_2}p_2^{\mu}+p_3^{\mu}\right) ,  \nn \\ 
\delta p_2^{\mu} & \,  \,=\, -\frac{1}{2}\left(\frac{p_1 \cdot p_3}{p_1 \cdot p_2}p_2^{\mu}- \frac{p_2 \cdot p_3}{p_1 \cdot p_2}p_1^{\mu}+p_3^{\mu}\right). 
\label{shifts}
\end{align}
Expressions \eqref{eq:MainFormula} and \eqref{shifts} yield  the
dominant NLP contributions to the NLO matrix element. We now turn to examine
the implications of this for the two event shapes.

\section{Thrust} 
\label{THRUSTNLP}

\begin{figure*}
  \centering \includegraphics[scale=0.64]{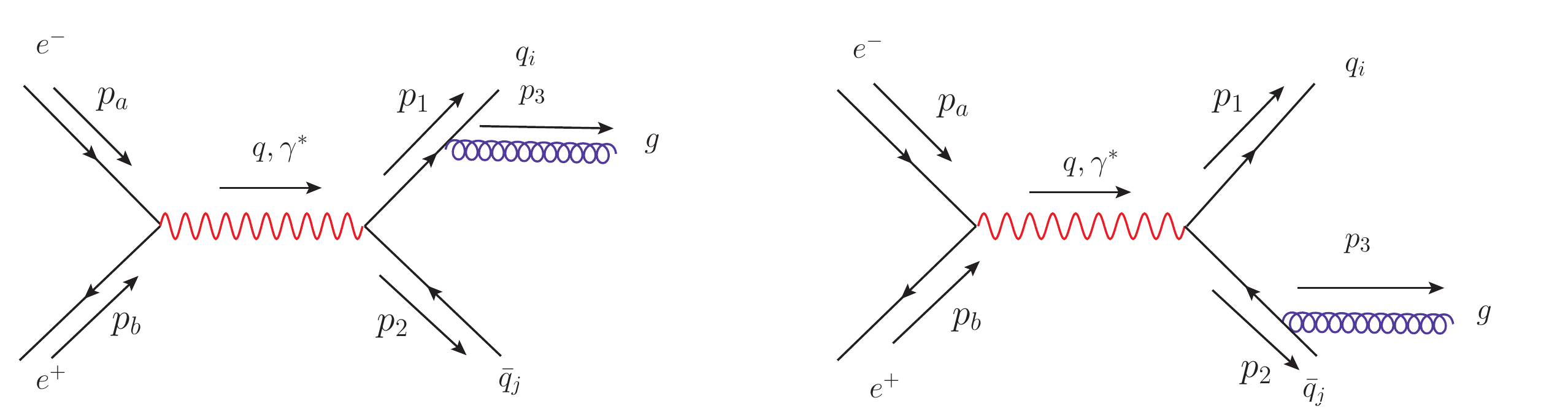}
  \caption{Feynman diagrams for the real emission of a gluon from the
    final state quark or anti-quark.}
  \label{processdiag}
\end{figure*}

\noindent
Thrust~\cite{Farhi:1977sg} in $e^+e^-$ collisions is defined as
\begin{align}
  T  \,=\,   \underset{\bf{\hat{n}}}{\rm{max}} \ \frac{ \sum_{i}|{\bf{p}_{i}} \cdot {\bf{\hat{n}}}|}{\sum_{i}E_{i}}\,,
  \label{tdf}
\end{align}
where $\bf{p}_{i}$ and $E_i$ are the three-momentum and energy of the
$i^{\text{th}}$ particle present in the final state.  The unit
vector $\bf{\hat n}$ that maximizes the sum in the numerator is 
the {\it thrust axis}.  The range of thrust is $[1/2, 1]$ where
$T = 1/2$ for a spherically symmetric event and 
$T = 1$ for a pencil-like (dijet) event. It is an infrared-safe
event shape variable, i.e.  it is insensitive to the soft emissions or
collinear splittings.
The LO reaction at the parton level is
\begin{align}
  e^{+}(p_b)+e^{-}(p_a)\rightarrow \gamma^*(q) \rightarrow q(p_1)  + {\overline q}(p_2)\,,
  \label{processeqnborn}
\end{align}
where we assume all particles to be massless.  For this case, $T = 1$.
At NLO the real emission process is
\begin{align}
  e^{+}(p_b)+e^{-}(p_a)\rightarrow \gamma^*(q) \rightarrow q(p_1)  + {\overline q}(p_2) + g(p_3)\,.
  \label{processeqn}
\end{align}
The diagrams for this process are shown in fig.~(\ref{processdiag}).
For a three-body final state $ T $ takes values in the range $ [2/3,1] $.
The limit $T=1$ is approached when either $(i)$ the emitted gluon is soft
($p_3 \to 0$),  $(ii)$ the quark or the anti-quark is soft
($p_1\to 0 \ \text{or} \ p_2 \to 0$), or $(iii)$ any two final state
partons are collinear.
It is standard practice to define the dimensionless energy fractions for the
final state particles,
\begin{align}
  x_i \, = \, \frac{2E_i}{Q} \quad \quad (i=1,2,3)\,,
\end{align}
where $Q$ is the total center of mass energy, with   $x_1+x_2+x_3 \,=\,2$.
One can readily derive the relations
\begin{align}
  (p_2+p_3)^2\,=\,&2p_2\cdotp p_3 \,=\, Q^2(1-x_1),\nonumber  \\
  (p_1+p_3)^2 \,=\, &2p_1\cdotp p_3 \,=\, Q^2(1-x_2), \label{xi's} \\
  (p_1+p_2)^2 \,=\, &2p_1\cdotp p_2 \,=\, Q^2(1-x_3)\nonumber,
\end{align}
using momentum conservation.
For a final state with three massless particles, eq.~(\ref{tdf}) takes
the simple form
\begin{align}
  T \, = \,\text{max}(x_1,x_2,x_3)\,.
  \label{thrustDef}
\end{align}
\subsection{Thrust distribution at NLO} \label{thrustNLO}

\begin{figure*}[hbtp!]
	\centering
	\includegraphics[width=0.45\linewidth]{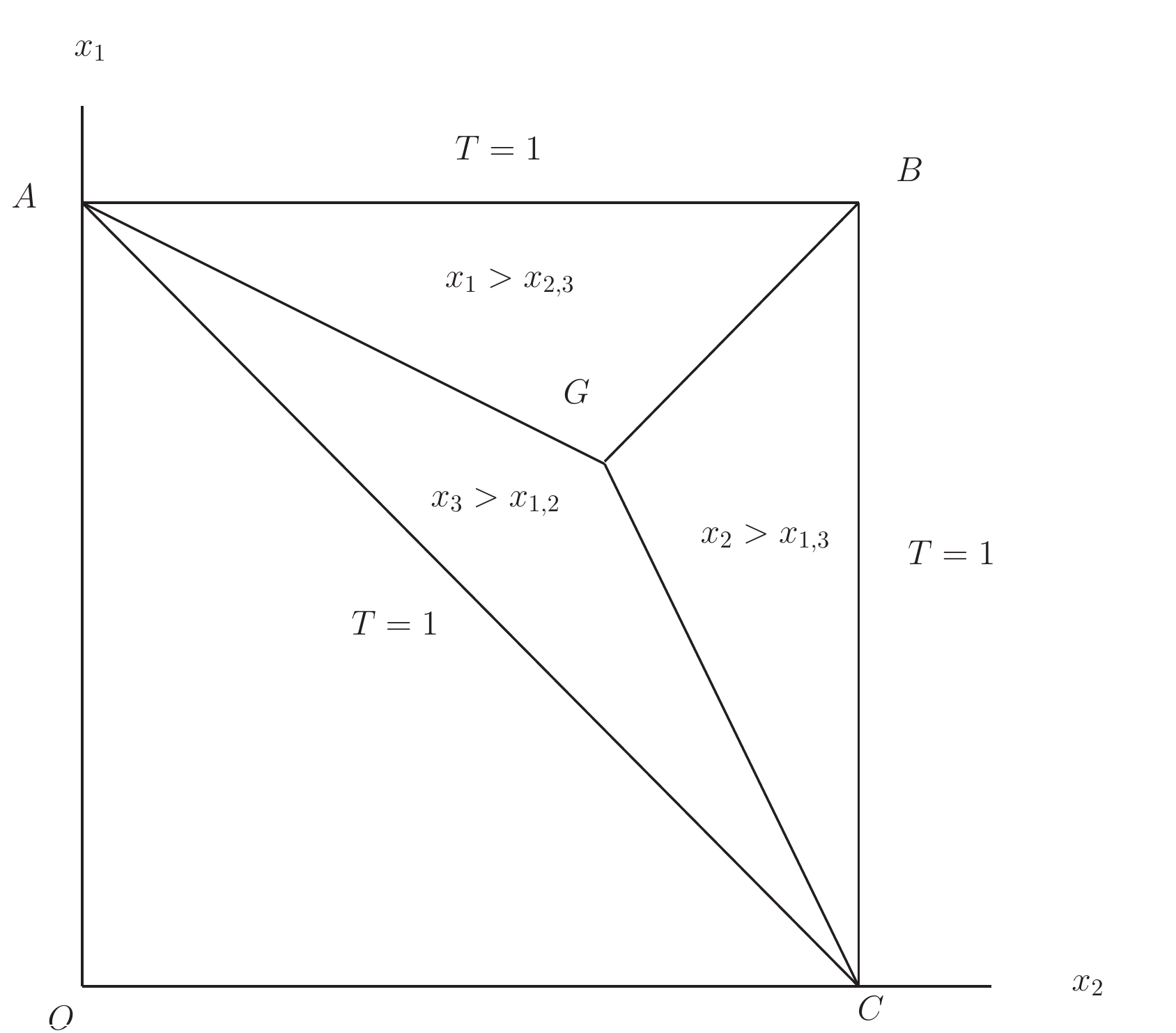}
	\caption{The Dalitz plot for the production of three massless particles.
		The phase space can be divided into three different
		regions depending upon which energy fraction is the largest. In
		region ABG, $x_1$ is the largest of the three $x_i$, while in region BGC, (AGC) $x_2$ ($x_3$) is the largest. 
		In this figure, we set $BA=BC=1$. Along the lines AB, BC, and AC
		one has $T=1$.}
	\label{fig:dalitz}
\end{figure*}

The thrust distribution at NLO is given by
\begin{align}
  \frac{d\sigma}{dT} \,=\,\frac{1}{2s}\int d\Phi_{3} \overline{\sum} |\mathcal{M}(x_1,x_2)|^2  \delta(T-\text{max} (x_1,x_2,x_3)),
  \label{tdstr}
\end{align}
where $s$ is the center of mass energy squared. 
The matrix element squared for the process in
eq.~(\ref{processeqn}) is
\begin{align}
  \overline{\sum} & |\mathcal{M}(x_1,x_2)|^2
  \nn  \\  \,=\, & 8(e^2e_q)^2 g_s^2 C_F N_c \frac{1}{3Q^2} \, \frac{x_1^2+x_2^2}{(1-x_1)(1-x_2)}\,,
  \label{mful}
\end{align}
 where $\alpha=\left( e^2/4 \pi\right)$, $e_q$ is the charge of quarks in the unit of fundamental electric charge $e$,
 $\alpha_s= g_s^2/4\pi$,
$C_F =(N_c^2-1)/2N_c $, and $N_c$ is the number of quark colours.
The three-particle phase space measure, expressed in terms of $x_i$ reads
\begin{align}
  d \Phi_3 \,=\, &\frac{Q^2}{16(2\pi)^3} \, dx_1 \, dx_2 \,  dx_3 \, \delta(x_1+x_2+x_3-2).
    \label{3bx1x2}
\end{align}
The phase space in eq.~(\ref{3bx1x2}) is depicted in fig.~(\ref{fig:dalitz}),
with every point in the plane fulfilling the
constraint $x_1+x_2+x_3=2$.  The region surrounding point $A$ ($C$)
corresponds to the soft anti-quark (quark) region, and the region around
$B$ to the soft gluon region.  On line $BA$, where $x_1=1$, the anti-quark and gluon move
collinearly in opposite directions, while
on line $BC$, where $x_2=1$, the quark and the
gluon are collinear. On line $AC$, where $x_3=1$, the quark and
the anti-quark are collinear, with the gluon moving in the opposite direction.  Along these three lines, $T=1$.

To obtain the thrust distribution one integrates over the
$x_1, x_2$ variables in eq.~(\ref{tdstr}) with the appropriate limits:
\begin{align} \label{fint}
  \frac{1}{\sigma_0(s)}\frac{d\sigma}{dT} \,=\,\frac{2\alpha_s}{3\pi}  \int
  _{0}^{1}  dx_1  \int_{0}^{1-x_1} & dx_2 \,\,
  \frac{x_1^2+x_2^2}{(1-x_1)(1-x_2)}
  & \,\, \nn  \\
  &  \hspace{-0.3cm} \times \delta\left(T-\text{max}(x_1,x_2,x_3)\right)\,,
\end{align}
where $x_{3}= 2 -x_{1} -x_{2}$, and $\sigma_0(s)$ is the LO
cross-section, given by
\begin{align}
\sigma_0(s) \,=\,\frac{4\pi\alpha^2  e_q^2\, N_c}{3s}\,,
\label{born cross section}
\end{align}
For our purposes, we label the contributions from three distinct contributions
in the phase space integration in eq.~(\ref{fint}), defined
by  either $x_1$, $x_2$, and $x_3$ being the largest, as follows:
\begin{enumerate}
\item[I\,:] Quark has the largest energy ($x_1>x_2,x_3$) 
\item[II\,:] Anti-quark has the largest energy ($x_2>x_1,x_3$)
\item[III\,:]  Gluon has the largest energy ($x_3>x_1,x_2$)
\end{enumerate}
The contribution from region I is then
\begin{align}
 \frac{1}{\sigma_0(s)} \frac{d \sigma}{d T} {\Bigg \vert_{\text{I}}} \,=\,\frac{2\alpha_s}{3\pi}  \int _{0}^{1} dx_2 \int_{0}^{1-x_2}  dx_1 \, &  \frac{x_1^2+x_2^2}{(1-x_1)(1-x_2)} \,\, \nn \\  \times & \delta(T-x_1)\,.
\end{align}
With the appropriate limits of integration we have
\begin{align}
\frac{1}{\sigma_0(s)} \frac{d \sigma}{d T} {\Bigg \vert_{\text{I}}} \,=\,\frac{2\alpha_s}{3\pi}  \int _{2(1-T)}^{T} dx_2 \,  \frac{T^2+x_2^2}{(1-T)(1-x_2)}.
\label{x1largeIntegrand}
\end{align}
Instead of $T$, we shall mostly use the variable $\tau \,=\, 1-T$, which vanishes
in the zero-radius dijet limit. The integration in
eq.~(\ref{x1largeIntegrand}) leads to
\begin{align}
& \frac{1}{\sigma_0(s)}  \frac{d\sigma}{d \tau}{\Bigg \vert_{\text{I}}} \nn \\   \,=\, &  \alspi \biggl[ \frac{3\tau^2+8\tau-3}{2\tau}+  \left(\frac{ \tau^2-2\tau+2}{\tau}  \right) \log\left( \frac{1-2\tau}{\tau} \right) \biggr] .
\label{x1maxreg}
\end{align}
Expanding the above expression around $\tau = 0$ gives
\begin{align}
& \frac{1}{\sigma_0(s)}\frac{d\sigma}{d \tau}{\Bigg \vert_{\text{I}}}  \nn \\ & =\alspi \biggl( \frac{-3-4 \log \tau }{2 \tau }+ 2 \log \tau  + \frac{3\tau}{2}  -\tau \log\tau  +\mathcal{O}(\tau^2)  \biggr). 
\label{x1maxcontri}
\end{align}
The first term in the brackets forms the LP term (in $\tau$), with both leading-logarithmic (LL) and next-to-leading
logarithmic (NLL) terms. The former derives from soft and collinear gluon emission, the latter
from hard collinear gluon emission.
The next term constitutes the (NLP)
term, which is our main focus here.
Subsequent terms are of NNLP and beyond accuracy.
The result for thrust distribution from region I is provided in
eq.~\eqref{x1maxcontri}; however, this expression does not show the
correspondence between the various kinematical configurations and the
contributions they produce at the LP and NLP. In order to extract
such information, we shall split the integration result
of eq.~\eqref{x1largeIntegrand} into its upper and lower limit
contributions separately, as these limits reflect the different
regions of the phase space.
 The contributions from the upper (I,u) and lower (I,l) limits of $x_2$ from the region
 I are, respectively, after expansion around $\tau \,=\,0$
\begin{align}
 & \frac{1}{\sigma_0(s)}  \frac{d\sigma}{d \tau}{\Bigg \vert_{\text{I,u}}}   \nn \\   = & \alspi  \biggl( \frac{-2 \log \tau }{\tau } +   2 +   2  \log \tau    - \frac{\tau}{2} +  \tau \log \tau    + \mathcal{O}\left(\tau ^2  \right)\biggr),
\label{x1upperExact}
\end{align}
 \begin{align}
\frac{1}{\sigma_0(s)}\frac{d\sigma}{d \tau}{\Bigg \vert_{\text{I,l}}} \,=\,&\alspi\left( \frac{3}{2\tau}+2 -2  \tau +\mathcal{O}(\tau^2)\right).
\label{x1lowerExact}
\end{align}
Note that here and henceforth the lower limit term is to be \emph{subtracted}
from the upper limit result.
In region I, the upper limit of $x_2$ for small $\tau$ corresponds to a back-to-back
quark and anti-quark configuration allowing the gluon to go soft while the lower limit
corresponds to the quark and gluon being back-to-back, allowing the gluon to be hard.
The LL terms at LP and NLP follow only from the upper limit ($x_
2 \,=\,1-\tau$),  whereas the lower limit ($x_2 \,=\,2\tau$) yields NLL terms
both at LP and NLP. In fact, identical NLL contributions at NLP follow
from both limits; they therefore cancel in the final result for region I.

Thus the  upper and lower limit contributions 
unveil the relation between the particular kinematical
configurations and their respective contributions at LP and NLP. 
To better comprehend the origin of LP and
NLP terms, we will from here on first provide our full results and
then their breakdown into upper and lower limit contributions.

In region II, due to the symmetry under the interchange of $x_{1} \leftrightarrow
x_{2}$, the contribution from this region is identical to eq.~(\ref{x1maxcontri}). Thus
\begin{align}
\frac{1}{\sigma_0(s)}\frac{d\sigma}{d \tau}{\Bigg
  \vert_{\text{II}}} \,=\, \frac{1}{\sigma_0(s)}\frac{d\sigma}{d \tau}{\Bigg
  \vert_{\text{I}}}\,.\label{x2maxcontri}
\end{align}
In region III, the thrust axis is aligned with the momentum of the
gluon. The contribution is
\begin{align}
\frac{1}{\sigma_0(s)}\frac{d\sigma}{dT}{\Bigg \vert_{\text{III}}}  \,=\,  \frac{2\alpha_s}{3\pi}  \int _{0}^{1} dx_2 & \int_{0}^{1-x_2}  dx_1 \, \frac{x_1^2+x_2^2}{(1-x_1)(1-x_2)} \,\, \nn \\ &  \times \delta(T+x_1+x_2-2)\,,
\label{x3largeI}
\end{align}
which gives
\begin{align}
\frac{1}{\sigma_0(s)}\frac{d\sigma}{dT}{\Bigg \vert_{\text{III}}} \,=\,\frac{2\alpha_s}{3\pi}  \int _{2(1-T)}^{T} dx_2 \, \frac{(2-x_2-T)^2+x_2^2}{(x_2+T-1)(1-x_2)}\,.
\label{x3largeFE}
\end{align}
Expanding in $\tau$ gives
\begin{align}
 &\frac{1}{\sigma_0(s)}  \frac{d\sigma}{d \tau}{\Bigg \vert_{\text{III}}}\nn \\  & \,=\,   \alspi \bigg(-2-2\log{\tau} + (2-2\log\tau)\tau+\mathcal{O}(\tau^2)\bigg) .
\label{x3maxcontri}
\end{align}
Thus no LP contributions arise from region III, as expected. The $\tau=0$ configurations in this region
correspond to either the quark or anti-quark being soft, 
 or the quark and anti-quark pair moving collinearly opposite to the gluon.
The latter configuration does not lead to a large contribution because 
the propagator is then far off the mass-shell. 
The contribution from upper and lower limits from this region
respectively are, for small $\tau$,
\begin{align}
 \frac{1}{\sigma_0(s)} &   \frac{d\sigma}{d \tau}{\Bigg \vert_{\text{III,u}}} \nn \\   \,=\, &  \alspi \biggl(-2 - \log \tau +( 1- \log \tau ) \tau   +\mathcal{O}\left(\tau ^2\right) \biggr),
\label{x3upperExact}
\end{align}
\begin{align}
\frac{1}{\sigma_0(s)}   \frac{d\sigma}{d \tau}{\Bigg \vert_{\text{III,l}}}  \,=\,   \alspi \bigg(\log \tau +(-1+\log \tau) \tau+\mathcal{O}(\tau^2) \bigg).
\label{x3lowerExact}
\end{align}
The upper limit of $x_2$ in region III corresponds to the gluon and
anti-quark being back-to-back, and the lower limit has the gluon and quark back-to-back.
LL terms at NLP arise from both limits.

Combining contributions from all three regions, the thrust
distribution at NLO yields the
well-known result
\begin{align}
\frac{1}{\sigma_0(s)}\frac{d\sigma}{d \tau}  \Bigg|_{\text{NLO}} \,=\,\frac{2\alpha_s}{3\pi}&\biggl[\frac{2(3\tau^2-3\tau+2)}{\tau(1-\tau)}\,  \log\left(\frac{1-2\tau}{\tau}\right) \nn \\
& \hspace{1.5cm} -\frac{3(1-3\tau)(1+\tau)}{\tau}\biggr]\,,
\label{taudstr}
\end{align}
with expansion
\begin{align}
\frac{1}{\sigma_0(s)}\frac{d\sigma}{d \tau}  \Bigg|_{\text{NLO}} \,=\, \alspi & \biggl(\frac{-3-4\log{\tau}}{\tau}-2+2\log{\tau}+(5\tau \nn \\ 
& \hspace{1.5cm}-4\log\tau)\tau +\mathcal{O}(\tau^2) \biggr).
 \label{Fthrustnlp}
\end{align}
Note that in eq.~(\ref{Fthrustnlp}) at NLP all
three regions produce LL terms, but a partial cancellation
takes place when combined.
%
\subsection{Next-to-leading power corrections for thrust from kinematical shifts}
\label{ThrustShift} 
Next we use the method of kinematical shifts~\cite{DelDuca:2017twk, vanBeekveld:2019prq} to compute
dominant NLP terms for the thrust distribution due to gluon radiation.
The expression for the shifted matrix
elements is given in eq.~\eqref{eq:MainFormula}, with the momenta shifts given by eq.~(\ref{shifts}). 
%
%
Note that because the emitted gluon is emitted from final particles the momenta shifts in
eq.~(\ref{eq:MainFormula}) are subtracted~\cite{vanBeekveld:2019prq} from the
radiating particle momenta, rather than added as for the initial state case.
The squared, spin-summed, and averaged LO matrix element 
$| \mathcal{M}_{0}|^2$ for the process in eq.~(\ref{processeqnborn}) is
\begin{align}
  \overline{\sum} |\mathcal{M}_{0}(p_1,p_2)|^2 \,=\, \frac{4(e^2 e_q)^2 N_c}{3s}  (2 p_1 \cdot p_2)\,.
\end{align}
Inserting shifted momenta according to eq.~\eqref{shifts} leads to
\begin{align}
  \overline{\sum} & |\mathcal{M}_{0}(p_1-\delta p_1,p_2-\delta p_2)|^2
   \nn \\  \,=\, & \frac{8(e^2 e_q)^2 N_c}{3Q^2} \biggl[ ( p_1\cdot p_2 ) +     p_3 \cdot
  (p_1+p_2)    +  {\cal O}(p_3^2) \biggr]\,.
  \label{MshiftbornF} 
\end{align}
The last term contains two powers of the momentum $p_3$, and can thus be omitted.
Combining eqs.~(\ref{eq:MainFormula}) and~(\ref{MshiftbornF}) gives the expression for shifted matrix element squared as
\begin{align}
  \overline{\sum} & |{\mathcal{M}}_{\text{shift}}|^2 \nn \\
   \,=\, & 8(e^2 e_q)^2  N_c  C_F g_s^2 \,  \frac{1}{3Q^2} \, \nn \\ & \times \left(\frac{2(p_1 \cdot p_2)^2}{(p_1 \cdot p_3)(p_2 \cdot p_3)}+\frac{2p_1 \cdot  p_2}{p_2 \cdot  p_3}+\frac{2p_1 \cdot  p_2}{p_1  \cdot p_3}\right)\,,
  \label{Mshiftp1p2}
\end{align}
which, expressed in terms of $x_i$'s, reads
\begin{align}
  \overline{\sum} |{\mathcal{M}}_{\text{shift}}(p_1,p_2)|^2  \,=\,& 8 (e^2e_q)^2 N_c  g_s^2 C_F \,  \frac{1}{3Q^2} \, \nn \\ & \hspace{.5cm} \times \left(  \frac{2x_1+2x_2-2}{(1-x_1)(1-x_2)} \right) \,.
    \label{shiftexpr}
\end{align}
Note that when the emitted gluon becomes
soft (i.e. $x_1,x_2\to 1$), the above matrix element diverges, whereas when the
gluon is hard ($x_3\to 1$), the numerator $(2x_1+2x_2-2)$ vanishes. 
The thrust distribution under this approximation is given by
\begin{align}
  \frac{d \sigma}{d T} \Bigg|_{\text{shift}} \,=\,\frac{1}{2s}\int d\Phi_{3}\,\, & 
  \overline{\sum} \ |\mathcal{M}_{\text{shift}}(x_1,x_2)|^2  \,\, 
  \nn \\ & \times \delta(T-\text{max}(x_1,x_2,x_3)).
  \label{shiftintegral}
\end{align}
We repeat the steps to compute the thrust
distribution from the exact computation for each of the three
regions. Region I gives
\begin{align}
  \frac{1}{\sigma_0(s)}\frac{d\sigma}{dT}{\Bigg \vert_{\text{I}}} \,=\,\frac{2\alpha_s}{3\pi} \int _{0}^{1} dx_2 \int_{0}^{1-x_2}  dx_1 \,\, & \frac{2x_1+2x_2-2}{(1-x_1)(1-x_2)} \,\, \nn \\ & \times  \delta(T-x_1),
  \label{x1largeshiftI}
\end{align}
which leads to the expansion about $\tau \,=\,0$
\begin{align}
 & \frac{1}{\sigma_0(s)}  \frac{d\sigma}{d \tau}{\Bigg \vert_{\text{I}}}\nn \\ &  \,=\,\alspi \left( \frac{-2-2\log \tau}{\tau}+2+2\log \tau +\mathcal{O}(\tau^2) \right).
  \label{x1largeshift}
\end{align}
When we compare the above expression to eq.~(\ref{x1maxcontri}), we
see that the LLs at both the LP and NLP are correctly captured, while 
the NLL terms at LP  are only partially reproduced.  
The contributions from upper and lower limits are
\begin{align}
  \frac{1}{\sigma_0(s)}\frac{d\sigma}{d \tau}{\Bigg \vert_{\text{I,u}}}  \,=\,&\alspi \left[-\frac{2 (1-\tau ) (1+\log \tau )}{\tau }\right],
\label{x1largeshiftUL} \\
  \frac{1}{\sigma_0(s)}\frac{d\sigma}{d \tau}{\Bigg \vert_{\text{I,l}}} \,=\,&\alspi\left[ -4 -\frac{2 (1-\tau ) \log (1-2 \tau )}{\tau } \right],
    \label{x1largeshiftLL}
\end{align}
which, expanded around $\tau \,=\,0$, gives 
\begin{align}
  \frac{1}{\sigma_0(s)}\frac{d\sigma}{d \tau}{\Bigg \vert_{\text{I,u}}} \,=\,\alspi \biggl(\frac{-2-2\log \tau}{\tau}+2+&2\log \tau \nn \\ & +\mathcal{O}(\tau^2)  \biggr),
  \end{align}
\begin{align}
  \frac{1}{\sigma_0(s)}\frac{d\sigma}{d \tau}{\Bigg \vert_{\text{I,l}}} \,=\,&\alspi  \bigg( \mathcal{O}(\tau^2)\bigg).
\end{align}

The upper limit contribution in region I contains the LL and NLL terms at 
LP and NLP, while the lower limit acts beyond NLP.
The contribution from region II is identical to I.
The contribution from region III is

\begin{align}
  \frac{1}{\sigma_0(s)}  \frac{d \sigma}{d T}{\Bigg \vert_{\text{III}}} \, \nn = \frac{2\alpha_s}{3\pi}  \int _{0}^{1} dx_2 &\int_{0}^{1-x_2}  dx_1  \, \, \nn \\  \times  \delta&(T+x_1+x_2-2)   \nn \\ \times & \frac{2x_1+2x_2-2}{(1-x_1)(1-x_2)}   \,,
  \label{x3largeshiftI}
\end{align}
which gives
\begin{align}
  \frac{1}{\sigma_0(s)}\frac{d\sigma}{d \tau}{\Bigg \vert_{\text{III}}} \,=\,\alspi \left[\frac{4\tau}{1-\tau}\log\left( \frac{1-2\tau}{\tau}\right)  \right]\,.
  \label{x3largeshiftII}
\end{align}
Upon expansion around $\tau  \,=\, 0$ this yields only terms beyond NLP accuracy
\begin{align}
  \frac{1}{\sigma_0(s)}\frac{d\sigma}{d \tau}{\Bigg \vert_{\text{III}}} \,=\,\alspi \bigg(-4\tau \log \tau +\mathcal{O}(\tau^2) \bigg)\,,
  \label{x3largeshift}
\end{align}
\begin{align}
  \frac{1}{\sigma_0(s)}\frac{d\sigma}{d \tau}{\Bigg \vert_{\text{III,u}}}& \,=\, \alspi \bigg(-2\tau\log \tau+ \mathcal{O}(\tau^2)\bigg), \\
  \frac{1}{\sigma_0(s)}\frac{d\sigma}{d \tau}{\Bigg \vert_{\text{III,l}}}& \,=\, \alspi \bigg(2\tau \log\tau+\mathcal{O}(\tau^2) \bigg)\,.
\end{align}
Indeed the hard gluon/soft quark region is not part of the shifted
kinematics method.

Combining contributions from all three regions
 the thrust distribution from
shifted kinematics formalism at NLO reads
\begin{align}
 \frac{1}{\sigma_0(s)}\frac{d\sigma}{d \tau}   \Bigg\vert_{\text{shift}} \,=\,\frac{2\alpha_s}{3\pi}\biggl[\frac{8\tau^2-8\tau+4}{\tau(1-\tau)} & \log\left(\frac{1-2\tau}{\tau}\right)\nn \\ & \hspace{-0.5cm} -\frac{3\tau^2-4\tau+1}{\tau(1-\tau)}\biggr]\, ,
  \label{taudstrshift}
\end{align}
or
\begin{align}
  \frac{1}{\sigma_0(s)} \frac{d\sigma}{d \tau}  \Bigg|_{\text{shift}} \,=\,\alspi & \bigg[\frac{-4-4\log{\tau}}{\tau}+4+4\log{\tau}  \nn \\  &  \hspace{0.9cm}-(4\log\tau)\tau +\mathcal{O}(\tau^2) \bigg]. 
\label{nlpfull}
\end{align}
When compared to the exact computation in eq.~(\ref{Fthrustnlp}), we
see that the LL terms at LP have been
reproduced, in addition to some NLL terms.
At the NLP level, we have captured some, but not all, of the
LL terms, in addition to some NLL contributions.
The missing LL terms at NLP should arise from soft quark 
(anti-quark) contributions.


The exact squared matrix element in eq.~(\ref{mful}) has the following $x_i$ dependence

\bt
\begin{align}
 \overline{\sum}  |{\mathcal{M}}(x_1,x_2)|^2  \, = \, 8(e^2 e_q)^2 N_c C_F g_s^2 \frac{1}{3Q^2} \, \biggl( \frac{2}{(1-x_1)(1-x_2)}  -  \frac{2}{1-x_1} -  \frac{2}{1-x_2}  + 
  \frac{1-x_2}{1-x_1}    + \frac{1-x_1}{1-x_2} \biggr).
\label{Mexpanded}
\end{align}
\et
The first three terms constitute the shifted matrix element
squared, as shown in eq.~\eqref{shiftexpr}. The eikonal
approximation in fact consists of only the first term of eq.~(\ref{Mexpanded}).
The remaining last two terms in eq.~(\ref{Mexpanded}) we refer to  as remainder matrix element squared and it has following form
\begin{align}
\overline{\sum} & |{\mathcal{M}}_{\text{rem}}(x_1,x_2)|^2 \nn \\ &  \,=\, 8(e^2 e_q)^2 N_c C_F g_s^2 \frac{1}{3Q^2}  \biggl(  \frac{1-x_2}{1-x_1}  +\frac{1-x_1}{1-x_2}\biggr)\,.
 \label{Mrem}
\end{align}
Let us take the first term on the right, which
diverges for $x_1 \to 1$.  It is 
most relevant when $x_2 \to 0$
and $x_1 \to 1$, i.e.  $x_3 \to 1$, which
corresponds to the emission of a soft anti-quark.  Similarly the
second term is dominant when
$x_1 \to 0$ and $x_2 \to 1$ i.e. the emission of soft quark.

The contributions from soft quark and anti-quark emissions can
alternatively be computed using the quark emission operator defined
in~\cite{vanBeekveld:2019prq}. The matrix element for the emission of a
soft quark ($p_1 \to 0$) in fig.~(\ref{processdiag}) is
\begin{align}
i \mathcal{M}_{1,q} \,=\,&\frac{i g_s t_a}{2 p_1\cdot p_3} \left[ \mathcal{M}_{H_{1}}(p_1, p_2, p_3) \slashed{p}_3 \gamma^{\mu} \bar\epsilon_\mu  \bar{u}(p_1)   \right]\,,
\end{align}
and similarly for the emission of soft anti-quark ($p_2 \to 0$)
\begin{align}
i\mathcal{M}_{2,q} \,=\, &\frac{i g_s t_a}{2 p_2\cdot p_3} \left[ \mathcal{M}_{H_{2}}(p_1, p_2, p_3) \slashed{p}_3 \gamma^{\mu} \bar\epsilon_\mu  v(p_2)   \right]\,.
\end{align}
Here the hard scattering matrix elements $\mathcal{M}_{H_{1}}$ and
$\mathcal{M}_{H_{2}}$ are defined to contain all the external states
except for the polarization vector $\epsilon_{\mu}$ and spinors for
soft fermion emission. The matrix element squared after the sum (average) over
the final (initial) state spins and colours
\begin{align}
\overline{\sum} &	|\mathcal{M}_q|^2 \,=\, \overline{\sum}	|{\mathcal{M}}_{\text{rem}}(x_1,x_2)|^2 
  \, \nn \\ &    =\,  8(e^2 e_q)^2 N_c C_F g_s^2 \frac{1}{3Q^2}  \left(\frac{1-x_2}{1-x_1} +\frac{1-x_1}{1-x_2}\right)\,,
\end{align}
which indeed reproduces eq.~\eqref{Mrem}.
Thus eq.~(\ref{Mrem}) features a clean separation of the soft quark and anti-quark contributions from the next-to-soft gluon contributions. 
Soft quark emissions contribute to the LL terms at NLP~\cite{vanBeekveld:2019prq}. Their contribution to the thrust distribution from region I is
\begin{align}
\frac{1}{\sigma_0(s)}  \frac{d \sigma}{d T}{\Bigg \vert_{\text{I}}} \,  =  \,\frac{2\alpha_s}{3\pi}  \int _{0}^{1} dx_2 \int_{0}^{1-x_2} & dx_1  \,\, \delta(T-x_1) \nn \\ \times &  \left( \frac{1-x_1}{1-x_2}+\frac{1-x_2}{1-x_1}\right) \,\, ,
\label{x1largeMR1}
\end{align}
which gives
\begin{align}
\frac{1}{\sigma_0(s)}\frac{d\sigma}{d \tau}{\Bigg \vert_{\text{I}}} \,=\,\alspi\left(\frac{1}{2\tau}-2+\frac{3\tau}{2}-\tau \log \tau +\mathcal{O}(\tau^2)   \right).
\label{x1largeMR}
\end{align}
The upper and lower limit components are
\begin{align}
\frac{1}{\sigma_0(s)}\frac{d\sigma}{d \tau}{\Bigg \vert_{\text{I,u}}}& \,=\, \alspi \bigg(\frac{1}{2\tau}+  -\frac{\tau}{2}-\tau \log \tau +\mathcal{O}(\tau^2)   \bigg), \\
\frac{1}{\sigma_0(s)}\frac{d\sigma}{d \tau}{\Bigg \vert_{\text{I,l}}}& \,=\, \alspi \bigg( 2-2\tau +\mathcal{O}(\tau^2) \bigg)\,.
\end{align}
In this region, the hard gluon can become collinear to the anti-quark, and the contribution from this kinematical configuration is captured by the upper limit of $x_2$, which yields the NLL at LP. At the same time, the lower limit captures the NLL at NLP. Region II again gives a contribution identical to eq.~(\ref{x1largeMR}). 
Region III gives

\begin{align}
\hspace{-0.15cm}\frac{1}{\sigma_0(s)} \frac{d \sigma}{d T}{\Bigg \vert_{\text{III}}} \,=\,\frac{2\alpha_s}{3\pi}  \int _{0}^{1} dx_2 & \int_{0}^{1-x_2}  dx_1     \nn \\ & \times \left( \frac{1-x_1}{1-x_2}+\frac{1-x_2}{1-x_1}\right) \,\, \nn \\ & \times  \delta(T+x_1+x_2-2),
\end{align} 
\begin{align}
\frac{1}{\sigma_0(s)}\frac{d\sigma}{d\tau}{\Bigg \vert_{\text{III}}} \,=\,\alspi\bigg(-2-2\log\tau+ & 2(1+\log \tau)\tau \nn \\ & +\mathcal{O}(\tau^2) \bigg).
\label{x3largeMR}
\end{align} 
The contributions from upper and lower limits are
 \begin{align}
\frac{1}{\sigma_0(s)}\frac{d\sigma}{d \tau}{\Bigg \vert_{\text{III,u}}}& \,=\,&\frac{2\alpha_s}{3\pi} \bigg( -2 -\log \tau - \tau\log \tau +\mathcal{O}(\tau^2)  \bigg), 
\end{align}
\begin{align}
\frac{1}{\sigma_0(s)}\frac{d\sigma}{d\tau}{\Bigg \vert_{\text{III,l}}}& \,=\,&\frac{2\alpha_s}{3\pi} \bigg( \log \tau +(-2-\log \tau) \tau +\mathcal{O}(\tau^2) \bigg).
\end{align}
As the upper limit corresponds to soft quark emission and the lower
limit to soft anti-quark contributions, one finds LL contributions at
NLP from both limits.
The remained matrix element squared
 in region III reproduces the missing LL contributions at NLP in
eq.~(\ref{nlpfull}), which the shifted kinematics methods do not capture. 
\begin{align}
 \frac{1}{\sigma_0(s)}\frac{d\sigma}{d\tau}{\Bigg \vert_{\text{rem}}} \,=\,\alspi\bigg ( \frac{1}{\tau}-6-2\log\tau+5\tau   \bigg).
  \label{remain_terms}
 \end{align}
These are, as expected, the terms that were missing from eq.~(\ref{nlpfull}).
It is now interesting to assess the quality of the shifted kinematics
approximation.

\subsection{Numerical assessment of the shifted kinematics approximations for thrust}
\label{sec:numer-assess-eikon-thrust}

\begin{figure*}[hbtp!]
	\centering
	\subfloat[][]{\includegraphics[scale=0.6]{"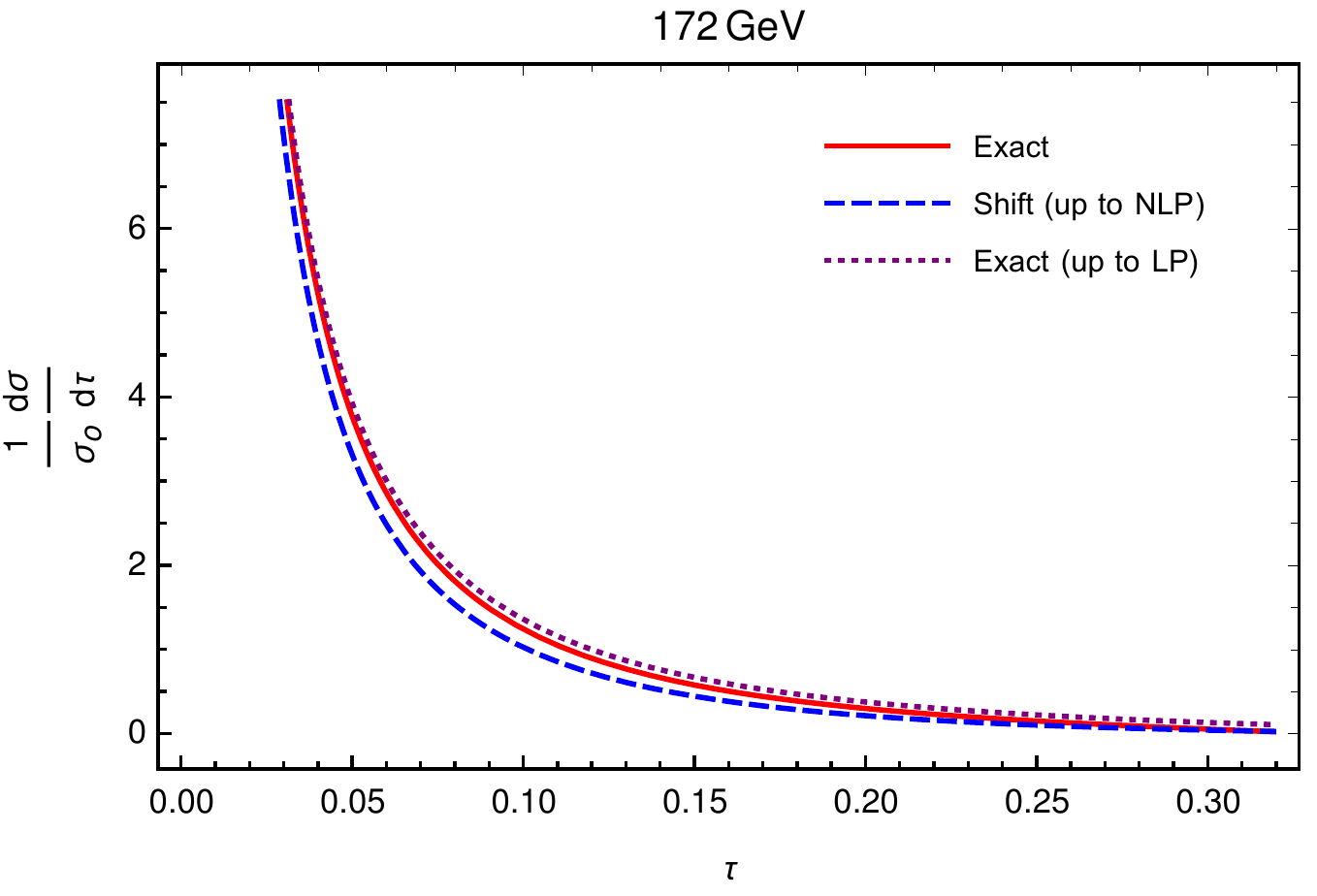"}}
	\qquad 
	\subfloat[][]{\includegraphics[scale=0.62]{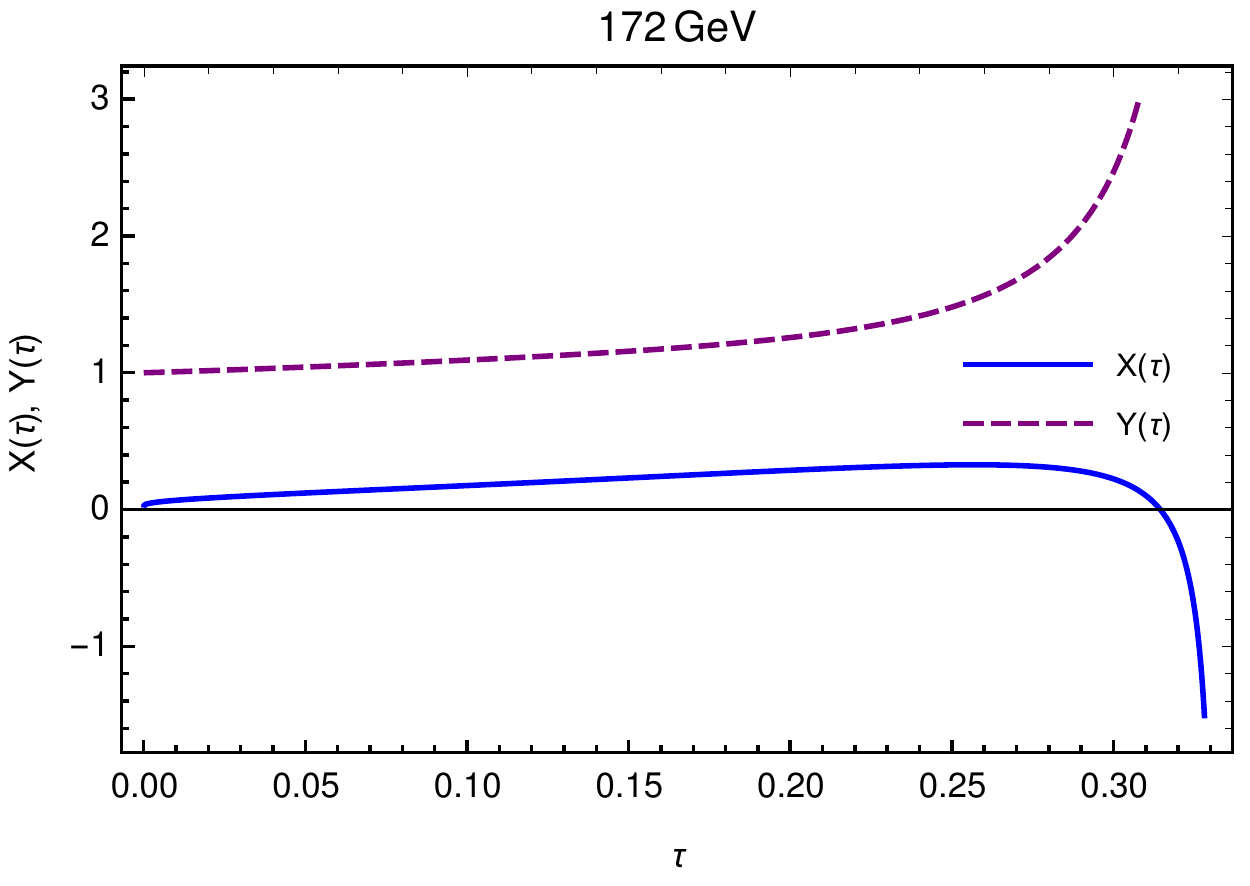}}    
	\caption{ In (a), we plot comparison of the thrust distribution. We show the exact result (solid red curve)  in eq.~\eqref{taudstr}, the shifted approximation (dashed blue curve) up to NLP terms given in eq.~\eqref{nlpfull}. We also plot the exact LP term 
		(dotted purple curve). In (b), we show $X(\tau)$ in eq.~\eqref{eq:Xtau} (solid blue
		curve) and $Y(\tau)$ in eq.~\eqref{eq:Ytau} (purple dashed curve)
		vs.  $\tau$. }
	\label{plot:Thrust_Combined_PLot}
\end{figure*}

The shifted kinematics approximation at the cross-section level
changes the Born cross-section  $\sigma_0(s)$  in eq.~(\ref{born
	cross section}) by replacing $s \to s(1-\tau)^{-1}$ 
\begin{align}
\sigma_0\left(\frac{s}{1-\tau}\right) \,=\, \sigma_0(s) (1-\tau).
\label{eq:Born-shift}
\end{align}
The LP term of thrust distribution computed under the formalism of shifted kinematics in eq.~(\ref{nlpfull}) is 
\begin{align}
\frac{d\sigma}{d\tau}\Bigg|_{\text{LP}} \,=\,\alspi \left( \frac{-4-4\log\tau}{\tau} \right) \sigma_0(s),
\end{align}
when multiplied by the \emph{shifted} born cross-section in eq.~(\ref{eq:Born-shift}) yields
\begin{align} 
\sigma_0 & \left(\frac{s}{1-\tau}\right)   \,\times \frac{d\sigma}{d\tau}\Bigg|_{\text{LP}}  \nn \\&  \,=\, \alspi \biggl(\frac{-4-4\log{\tau}}{\tau}+4+4\log{\tau}\biggr) \sigma_0(s)\,.
\end{align}
which reproduces eq.~\eqref{nlpfull}.
In fig.~(\ref{plot:Thrust_Combined_PLot}\textcolor{prd_blue}{a}) we compare the thrust distribution as computed from the
shifted kinematics approximation in eq.~(\ref{nlpfull}), up to NLP, with the exact computation in eq.~(\ref{taudstr}).\footnote{We have taken $\alpha_{s}= 0.1193$ at $172$ GeV from LEP2 from two loop result in~\cite{ALEPH:2003obs}, the numerical results are also given in~\cite{ALEPHsite}.}\

In the small $\tau$ region, the LP terms alone  provide
an excellent approximation to the exact result.
However, as $\tau$ increases, we notice that the
shifted kinematic curve approximates the
exact results better. The reason LP terms are such an excellent approximation
for thrust is the small coefficient of the NLP LL term; as such the NLL
at LP for thrust dominates the LL term at NLP for small
values of $\tau$. This is, as we shall see, in contrast to the
$C$-parameter, the LL at NLP dominates the NLL term at
LP~\cite{Catani:1998sf, Gardi:2003iv} in the dijet limit.

In fig.~(\ref{plot:Thrust_Combined_PLot}\textcolor{prd_blue}{b}) we provide two different ratio plots that involve expressions of thrust distribution computed from both of the approaches viz.~the exact approach~\eqref{taudstr} and the shifted approach~\eqref{nlpfull}. We define $X(\tau)$  and $Y(\tau)$ as
\begin{align}
	X(\tau) \,=\,& \frac{\frac{1}{\sigma_0(s)}\frac{d\sigma}{d \tau}  \Big|_{\text{NLO}}- \frac{1}{\sigma_0(s)}\frac{d\sigma}{d \tau}  \Big|_{\text{shift-NLP}}}{\frac{1}{\sigma_0(s)}\frac{d\sigma}{d \tau}  \Big|_{\text{NLO}}}, \label{eq:Xtau} 
\end{align}
\begin{align}
	Y(\tau) \,=\, &\frac{\frac{1}{\sigma_0(s)}\frac{d\sigma}{d \tau}  \Big|_{\text{NLO-LP}}}{\frac{1}{\sigma_0(s)}\frac{d\sigma}{d \tau}  \Big|_{\text{NLO}}}. \label{eq:Ytau}
\end{align}
As is evident from their expressions, $X(\tau)$ measures the accuracy of
the shifted approximation, whereas $Y(\tau)$ measures the importance of
NLP terms (and beyond) compared to the LP terms. The plot of
$X(\tau)$ shows that the
shifted kinematics method approximates the exact result well near the dijet limit.
The plot of $Y(\tau)$ shows that the LP contribution is 
dominant near $\tau \to 0$. As $\tau$ increases, the denominator falls faster than the
numerator due to the presence of the NLP terms in addition to the LP
contribution, and thus the ratio rises above one.

\section{$C$-parameter} 
\label{CPARANLP}

In the previous section, we compared the shifted kinematic approximation of the thrust
distribution with the exact NLO result, to understand the origin of
NLP corrections. Here we  perform the
same comparison for the more complicated $C$-parameter distribution. 
The $C$-parameter for massless particles~\cite{Donoghue:1979vi, Fox:1978vu, Parisi:1974sq, Ellis:1980wv} is defined as
\begin{align}
 C  \,=\, 3-\frac{3}{2} \sum_{i, \ j} \frac{\left( p_{i} \cdot p_{j}\right)^{2} }{\left(p_{i} \cdot q \right) \left(p_{j} \cdot q \right) } \,,
\end{align}
where $p_{i}$ is the four-momentum of $i$-th particle, the total
four-momentum is given by $q  = \sum_{i}p_{i}$, and the sums over $i $ and $j$ run over all the particles present in the final state. The minimum value taken
by $C$ equals $0$ (for a zero-radius dijet event), and the maximum value is $1$ (for an
isotropic event).  Here we consider a three-body final state, for which the maximal value of $C$ is 
$3/4$; the value $C =1$ can only be obtained if more than three
particles are present in the final state.
For the three-body final state in fig.~(\ref{processdiag}), it takes the form
\begin{align}
  C  \,=\, \frac{ 6 \left(1-x_{1}\right)\left(1-x_{2}\right)\left(1-x_{3}\right)}{x_{1} \ x_{2} \ x_{3}}\,,
  \label{Cdefn}
\end{align}
where $x_i$ are the energy fractions given in
eq.~(\ref{xi's}). The $C$-parameter has a critical point at $C= 3/4$,
which gives rise to a Sudakov shoulder~\cite{Catani:1997xc}. However,
these first occur at second order in $\alpha_{s}$, where four-particle
final states are possible, hence our study does not encounter them.
Henceforth we work with a rescaled definition of the $C$-parameter
\begin{align}
  c \,=\, \frac{C}{6}  \,=\, \frac{ \left(1-x_{1}\right)\left(1-x_{2}\right)\left(1-x_{3}\right)}{x_{1} \ x_{2} \ x_{3}}\,,
  \label{rana1}
\end{align}
for which the range is $0<c<1/8$. Note that the definition of
the $c$-parameter does not involve the selection of a special axis,
which thereby distinguishes it from a group of other event shapes
variables such as thrust, jet broadening, jet mass, and angularities.
\subsection{The $c$-parameter distribution at NLO} 
\label{cpDistribution}
\begin{figure*}[hbtp!]
	\centering
	\subfloat[][]{\includegraphics[scale=0.6]{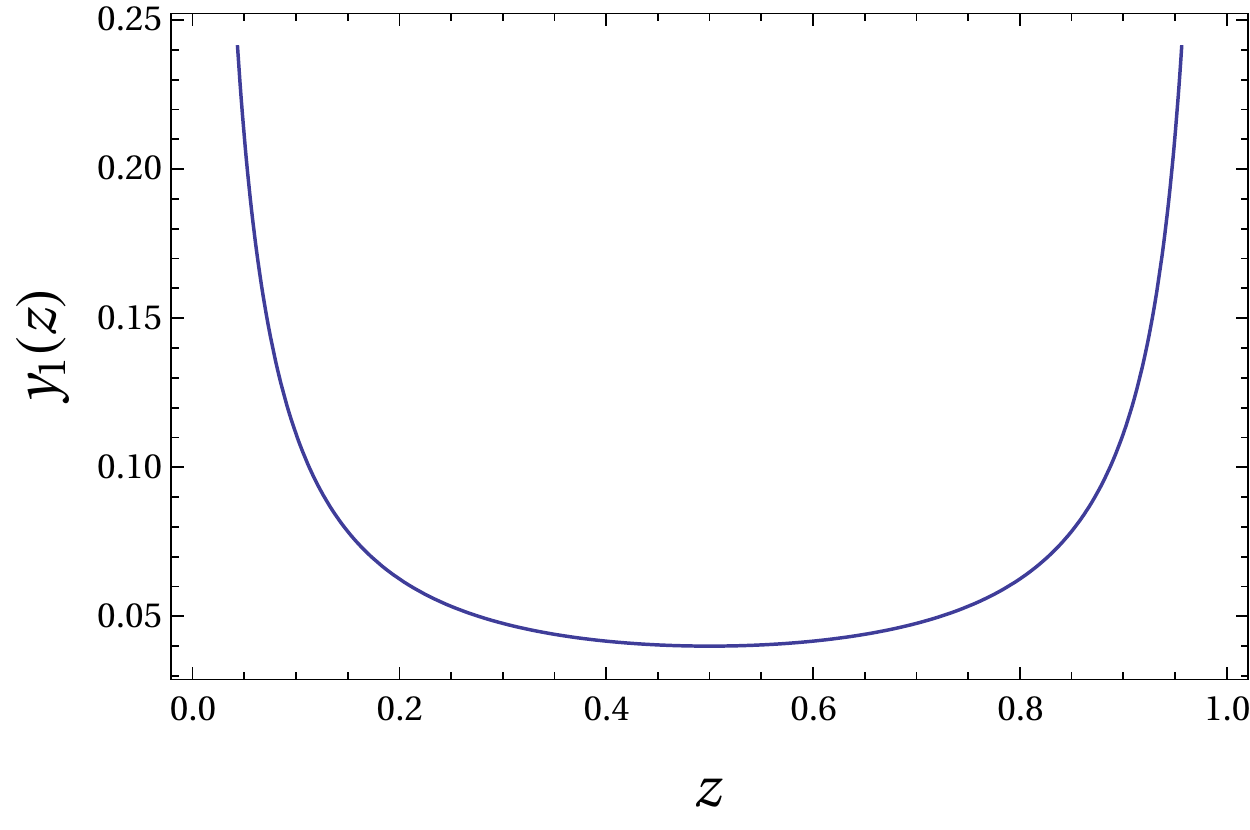}}
	\qquad 
	\subfloat[][]{\includegraphics[scale=0.62]{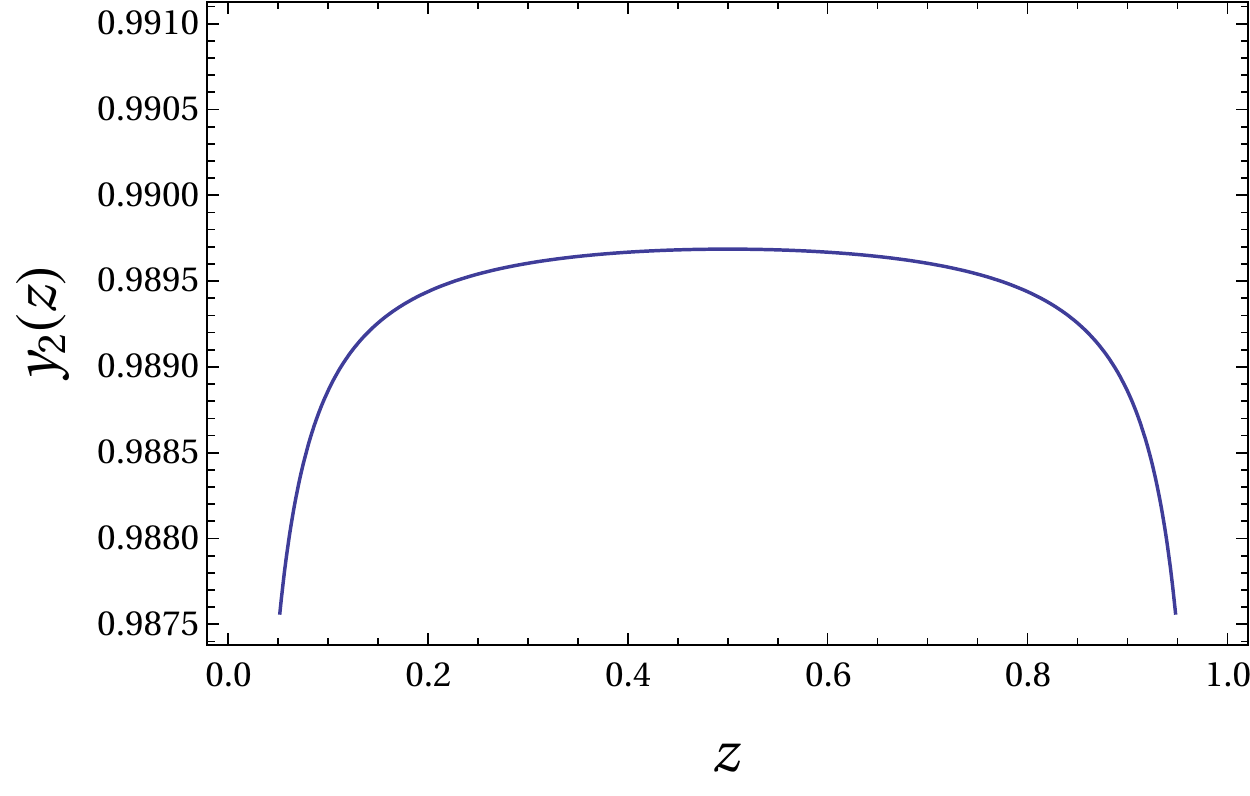}}    
	\caption{Behavior of $y_1(z)$ and $y_2(z)$ as functions of $z$ at $c \,=\,0.01$. }
	\label{plot:y1zy2z}
\end{figure*}

The $c$-parameter distribution is defined as
\begin{align}
 \frac{d \sigma}{d c}  \,=\, \frac{1}{2s} \int  d \Phi_3 \  \overline{\sum} \ |\mathcal{M}(x_1,x_2)|^{2} \ \delta\left( c (x_{1},x_{2}) -c \right) ,
 \label{eq:opk}
\end{align}
where $\overline{\sum} \ |\mathcal{M}(x_1,x_2)|^{2}$ and $d \Phi_3$
are given in eq.~(\ref{mful})  and eq.~(\ref{3bx1x2}) respectively. 
The normalized expression at NLO takes the form 

\begin{align}
\frac{1}{\sigma_{0}(s)} \frac{d \sigma}{d c}  \,=\, \alspi \int_0^1 dx_1  \int_0^{1-x_1} &   dx_2 \  \frac{x_1^2+x_2^2}{(1-x_1)(1-x_2)} \nn \\ & \times \delta\left( c (x_{1},x_{2}) -c \right).
\end{align}
The integrations over energy fractions are considerably more involved
than for thrust due to the rational polynomial form in eq.~\eqref{rana1}. It
is advantageous to 
convert our kinematic variables to $(y,z)$~\cite{Gardi:2003iv} where
\begin{align}
y \,=\, & 2-x_1-x_2,\nonumber \\
z \,=\, &  \frac{1-x_2}{y}\,,
\label{varTransf}
\end{align}
with Jacobian  $J(z,y) =y$. Note that $y$ is in fact the gluon energy fraction. The expressions for the matrix element and $c$-parameter in terms of the new variables read
\begin{align}
& \overline{\sum}  \ |{\mathcal{M}(y,z)}|^2 \nn \\   &  \,=\,   8(e^2 e_q)^2 N_c C_F g_s^2 \frac{1}{3Q^2}   \left(\frac{ 2 + y ( y - 2 y z(1 - z)-2 )}{y^2 z (1-z)}\right),
\label{Myz}
\end{align}
and
\begin{align}
c (y,z)  \,=\,\frac{(1-y)(1-z)yz}{(1-y(1-z))(1-yz)}\,.
 \label{eq:cpdefn}
\end{align}
The $x_1 \leftrightarrow x_2$ symmetry in the matrix element squared in
eq.~(\ref{mful}) now appears as $z \leftrightarrow (1-z)$ symmetry in
eq.~(\ref{Myz}).
Substitution into eq.~(\ref{eq:opk}) yields
\begin{align}
 \frac{1}{\sigma_{0}(s)} &  \frac{d \sigma}{d c} \Bigg\vert_{\text{NLO}} \nn \\ &  \,=\,  \frac{2 \alpha_{s}}{3 \pi}  \int^{1}_{0} dy \ \int^{1}_{0}  dz  \  \frac{ 2 + y ( y - 2 y z(1 - z)-2 )}{y z (1-z)}  \ \nn \\  &  \hspace{3cm}  \times \delta\left(  c (y,z) -c \right). 
 \label{eq:opk1}
\end{align}
To determine the limits of the $z$-integration, we use the phase
space in fig.~(\ref{fig:dalitz}) as follows.  From
eq.~(\ref{varTransf}), it is evident that $z$ will attain its lowest
value when $x_2 =1$, yielding $z =0$ as the lower limit of $z$. Similarly,
$z$ will attain its largest value when $x_2$ and $y$ both are
smallest.  However $x_2$ and $y$
cannot be zero at the same time.
Thus with $x_2 \to 0$ (and thus $y \to 1$), one finds that
$z =1$ is the upper limit.
The limits of the $y$-integration can be found from the $\delta$-function constraint and
the limits of the $z$-integration. We first rewrite
eq.~(\ref{eq:opk1}) in the form
\bt
\begin{align}
\frac{1}{\sigma_{0}(s)} \frac{d \sigma}{d c} \Bigg\vert_{\text{NLO}}  \,=\,  \frac{2 \alpha_{s}}{3 \pi}   \int^{1}_{0} dy  \int^{1}_{0} dz    \frac{2 (y (z-1)+1)^2 (y z-1)^2 (y (2 y (z-1) z+y-2)+2)}{(y-1)^2 y^2 (z-1) z (2 z-1)}   \biggl( \delta(z-z_1)+\delta(z-z_2)\biggr)   \,,
\label{fcpar2}
\end{align}
where we used the argument of the
 $\delta$-function in eq.~\eqref{eq:opk1} and it  has following two roots
\begin{align}
z_{1,2} &  \,=\,\frac{1}{2} \left(  1  \pm \frac{\sqrt{y(y(1+c)-1)(c(y-2)^2+y(y-1))}}{y(y(1+c)-1)} \right)\,. \label{eq:zroots}
\end{align}
\et
Here $z_1 (z_2)$ corresponds to the $+(-)$ solution. Note
that as $c \to 0$, we have  $z_1 \to 1 $ and $z_2 \to 0$ as expected.
The argument of the $\delta$-function in eq.~(\ref{eq:opk1}) is a
complicated function of $y$ and $c$. 
As the integral has a symmetry $z \leftrightarrow (1-z)$, the integral
over $z$ in eq.~(\ref{fcpar2}) equals twice the integral
between $z =1/2$ and the upper limit, where only $\delta(z-z_1)$ is
relevant. After the $z$ integration
the limits of $y$ change to $(y_1, y_2)$, see below in eq.~(\ref{Ylimits}). 
We have now
The above expressions are smooth functions of $z$, as can be seen
from fig.~(\ref{plot:y1zy2z}).
\bt
\begin{align}
\frac{1}{\sigma_0(s)}\frac{\rm{d}\sigma}{\rm{d} c} \Bigg\vert_{\text{NLO}}   \,=\,\frac{2\alpha_s}{3\pi} \int^{y_2}_{y_1} dy  \frac{2 (1-y) \left(y \left(c (y-2)^2+(y-3) y+4\right)-2\right)}{c (c y+y-1) \sqrt{y (c y+y-1) \left(c (y-2)^2+(y-1) y\right)}}\,.
\label{ExactInt}
\end{align}
To find the integration limits of  $y_1$  and  $  y_2$ we proceed as follows.
First, going back to eq.~(\ref{eq:opk1}), the $\delta$-function can be
also used to solve for $y$, for which there are two solutions,
\begin{align}
y_1(z)& \,=\, \frac{2 c}{\left[c^2 (1-2 z)^2+2 c (z-1) z+(z-1)^2 z^2\right]^{1/2}+c-z^2+z}\,,
\label{eq:y1z} \\
y_2(z)& \,=\,-\frac{\left[c^2 (1-2 z)^2+2 c (z-1) z+(z-1)^2 z^2\right]^{1/2}+c-z^2+z}{2 (c+1) (z-1) z}\,.
\label{eq:y2z}
\end{align}
\et
Clearly, there are extrema for both expressions  at $z =1/2$.
The second derivative of $y_1(z)$ ($y_2(z)$) at
$z=1/2$ is positive (negative).
Thus the expression of $y_1(z)$ in
eq.~(\ref{eq:y1z}) is the lower limit of $y$ and 
$y_2(z)$ in eq.~(\ref{eq:y2z}) is the upper limit of $y$.
Substituting $z =1/2$ in eqs.~(\ref{eq:y1z}) and (\ref{eq:y2z}) we find 
\begin{align}
y_1   \,=\, \frac{1+4c-\sqrt{1-8c}}{2(1+c)} \,,\qquad
y_2   \,=\, \frac{1+4c+\sqrt{1-8c}}{2(1+c)}\,.   
\label{Ylimits}  
\end{align}
\begin{figure*}[hbtp!]
	\centering
	\subfloat[][]{\includegraphics[scale=0.6]{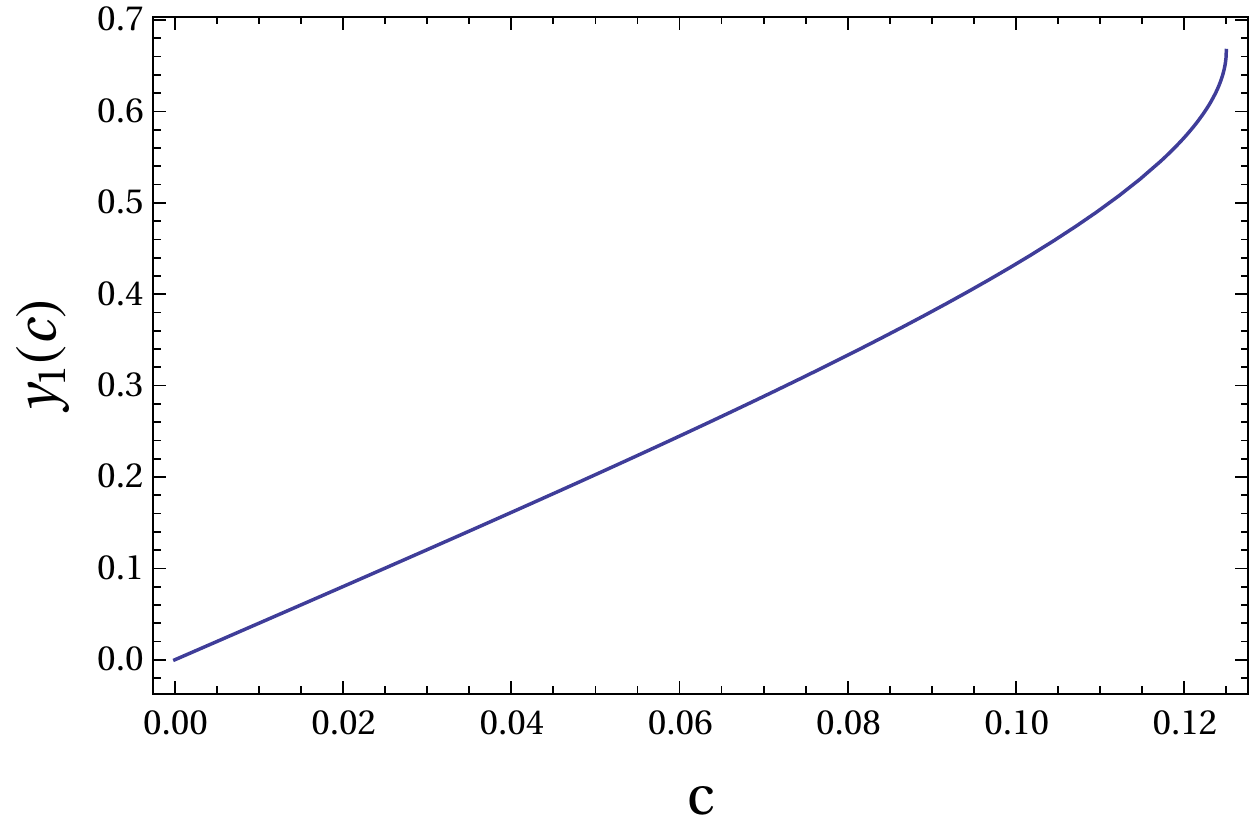}}
	\qquad 
	\subfloat[][]{\includegraphics[scale=0.6]{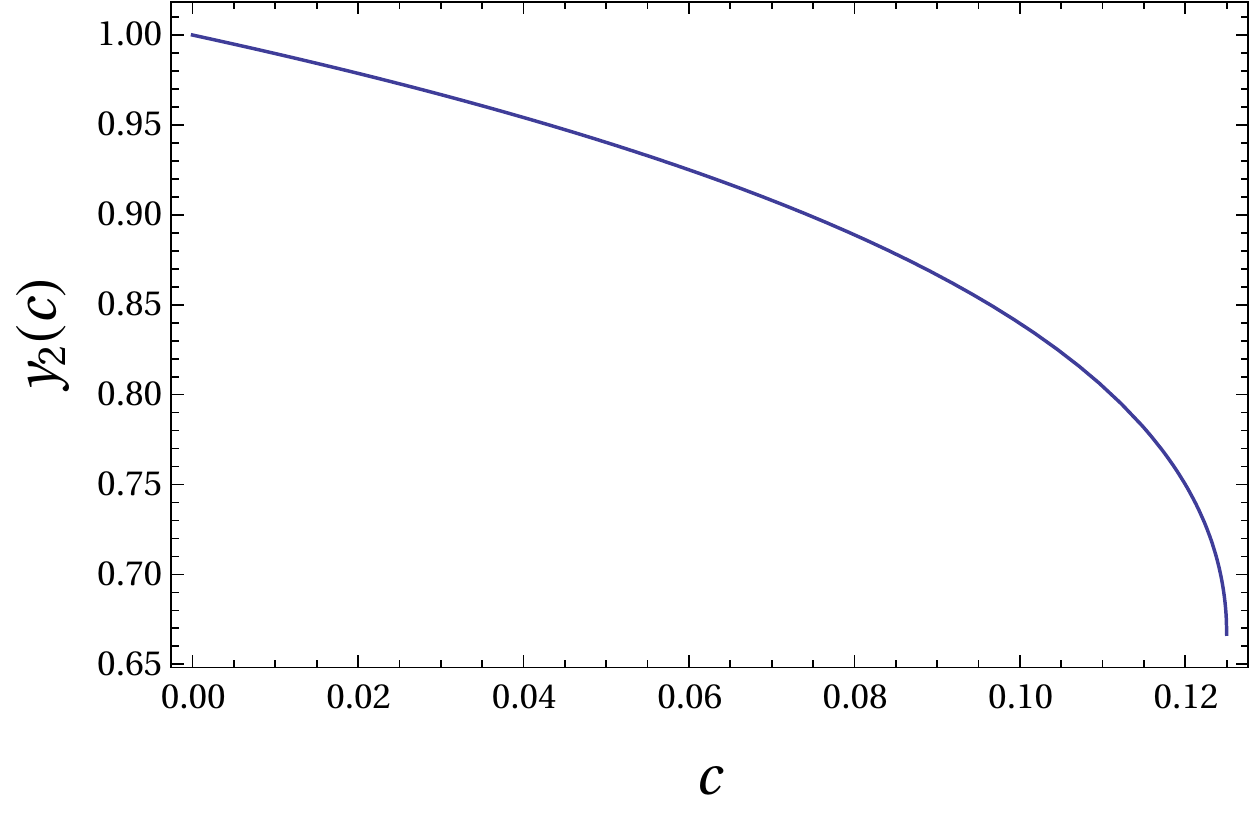}}    
		\caption{Lower and upper limits $y_1(c)$ and $y_2(c)$ respectively as functions of $c$.}
	\label{plot:Ylimits}
\end{figure*}
The behavior of lower and upper limits are shown in figs.~(\ref{plot:Ylimits}\textcolor{prd_blue}{a}) and (\ref{plot:Ylimits}\textcolor{prd_blue}{b}).
For thrust, the integration limits of the final
integration in eq.~(\ref{x1largeIntegrand}) had simple expressions ($2\tau$ and $1-\tau$ ),
and their relation with the respective kinematic configurations were readily visible in fig.~(\ref{fig:dalitz}).
To establish the relation between kinematic configurations and the integration limits $y_1$ and $y_2$ for the case of $c$-parameter, we expand their expressions around $c \,=\,0$ (as $c\to 0 $ is the dijet limit),\footnote{Note that while performing the final integration in eq.~(\ref{ExactInt}) we use the exact expressions of limits as given in eq.~(\ref{Ylimits}).}
\begin{align}
y_1& \,=\,4c+\mathcal{O}(c^3)    \nonumber, \\
  y_2& \,=\,1-c-3c^2+\mathcal{O}(c^3)
\label{YlimitSER}.
\end{align}
In the dijet limit,  the upper limit of $y$  corresponds to either
$(i)$ a back-to-back  hard gluon and 
quark--anti-quark collinear pair $(ii)$ soft quark (or soft anti-quark)
emission, or $(iii)$ the hard gluon collinear to the quark or the
anti-quark.  The lower limit in dijet limit corresponds to a soft gluon
emission and a back-to-back  quark and 
anti-quark. Thus these upper and lower limits correspond to different points in the phase space, as expected. The integration over $y$ in eq.~\eqref{ExactInt} produces incomplete elliptic
integrals of three types, each with somewhat involved arguments and
coefficients. 
After conversion to their so-called complete
counterparts\footnote{More information about these transformations of
  elliptic integrals and their analytical properties can be found in~\cite{ElliptBook}.}, 
and carefully collecting their coefficients, the final expression can
be organized in a compact form~\cite{Gardi:2003iv} as follows
\bt
\begin{align}
\frac{1}{\sigma_0(s)}\frac{d\sigma}{dc} \Bigg|_{\text{NLO}}    \,=\, \frac{2\alpha_s}{3\pi}  \bigg(  e(c) \ {\rm{E}}[m_1(c)]+p(c) \ {\Pi}[n_1(c), m_1(c)]+k(c) \ {\rm{K}}[m_1(c)]\bigg)\,,
\label{eq:elliptForm}
\end{align}
 \et
 where $ {\rm{E}} $, ${\Pi}$, and $ {\rm{K}}$  are the complete elliptic integrals
  of the first, second, and third kind.
 The integral representation of these elliptic integrals are as following
 \begin{align}
     \label{eq:ellipticINTcomplete}  
{\rm{E}}[m]  &  \,=\, \int^{1}_{0} dt \ \frac{\sqrt{1-m^2 t^2}}{\sqrt{1-t^2}}\,, \nonumber \\
{\Pi}[n,m] &   \,=\, \int^{1}_{0} \frac{dt}{(1-n t^2)\sqrt{(1-t^2)(1-m^2t^2)}}\,, \nonumber\\
{\rm{K}}[m] &  \,=\, \int^{1}_{0} \frac{dt}{\sqrt{(1-t^2)(1-m^2t^2)}} \, . 
\end{align} 
Here $ m $, and $ n $ are the \emph{parameter} and \emph{characteristic}  of the elliptic integrals,  respectively.
The arguments of these elliptic integrals in eq.~\eqref{eq:elliptForm} have the following form 
\begin{align}
n_1(c)& \,=\,\frac{2\sqrt{1-8c}}{1+\sqrt{1-8c}-4c}\, , \nn \\    
m_1(c)& \,=\,\frac{2\sqrt{1-8c}}{1+\sqrt{1-8c}-4c-8c^2}\,.
\label{arguEllipt}
\end{align}

\begin{figure*}[hbtp!]
	\centering
	\subfloat[][]{\includegraphics[scale=0.6]{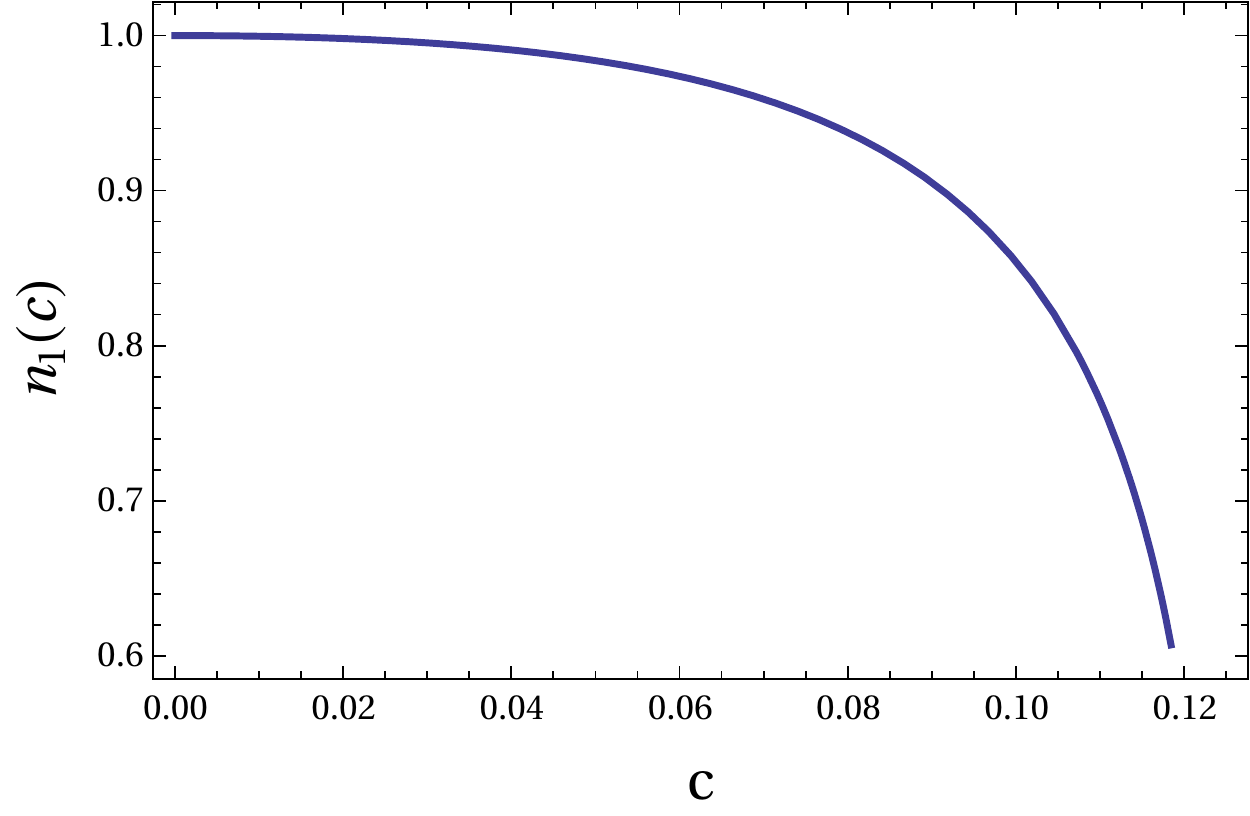}}
	\qquad 
	\subfloat[][]{\includegraphics[scale=0.612]{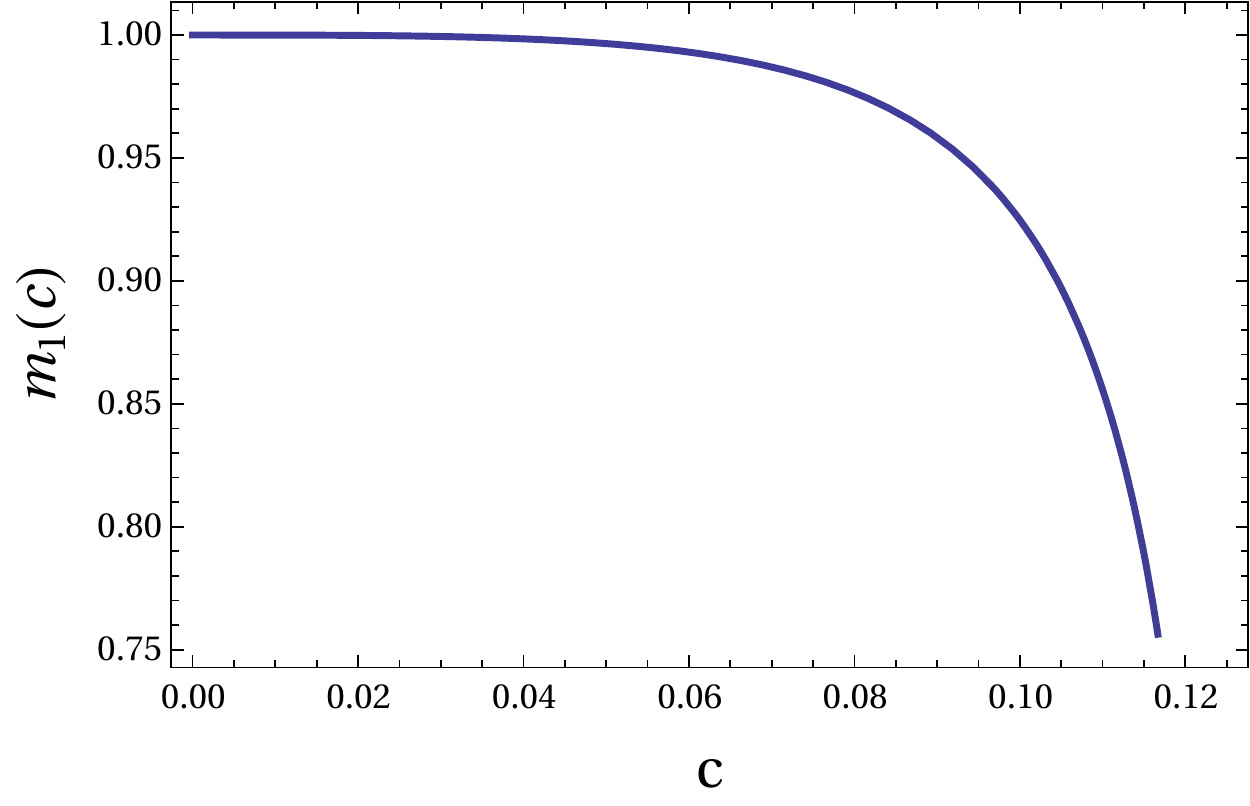}}    
		\caption{Behavior of arguments $n_1(c)$ and $m_1(c)$.}
  \label{fig:ellipArg}
\end{figure*}

The behavior of these arguments is shown in fig.~(\ref{fig:ellipArg}).
We computed  eq.~\eqref{eq:elliptForm} throughout with a
massless on-shell gluon and it agrees with the characteristic function derived
in~\cite{Gardi:2003iv} for finite gluon virtuality $\xi$, in the limit $\xi \to 0$.
Notice that the arguments have monotonic behavior. 
The elliptic integrals in eq.~(\ref{eq:elliptForm}) have the
following asymptotic behavior as $c \to 0$
\begin{align}
{\rm{E}}[m_1(c)] \,=\,&1 + \mathcal{O}(c^3\log{c})\,, \nonumber  \\
{\rm{\Pi}}[n_1(c), m_1(c)]  \,=\,& - \frac{\log{c}}{8c^2} - \frac{1+\log{c}}{4c} + \mathcal{O}(c^0\log{c})\,,  \label{eq:EKPexpanded} \\
{\rm{K}}[m_1(c)] \,=\,&-\frac{3\log{c}}{2} + \mathcal{O}(c^3\log{c})\,. \nonumber
\end{align}
The expressions for the
coefficients appearing in eq.~(\ref{eq:elliptForm}) are
\begin{align}
e(c) \,=\,&\frac{-3(1+2c)\sqrt{1-4c(1+2c)+\sqrt{1-8c}}}{\sqrt{2} c (1+c)^{3}}\,, \nonumber   \\ 
p(c) \,=\,&\frac{\sqrt{2}(2+c+2c^2) (1-\sqrt{1-8c})^2}{ c (1+c)^{3}\sqrt{1-4c(1+2c)+\sqrt{1-8c}}}\,, \label{EFcoeffexact} \\
k(c) \,=\,&\frac{4\sqrt{2}(1-2c(2+c))}{(1+c)^{3}\sqrt{1-4c(1+2c)+\sqrt{1-8c}}}\,. \nonumber 
\end{align}
Their expansions around $c =0$ are
\begin{align}
e(c) \,=\,&-\frac{3}{c}+9+12c+36c^2+\mathcal{O}(c^3)\,, \nonumber \\
p(c) \,=\,&32c+112c^2+\mathcal{O}(c^3)\,, \label{coeffExSER} \\
k(c) \,=\,&4-20c+48c^2+\mathcal{O}(c^3)\,. \nonumber 
\end{align}
The coefficient $e(c)$, when multiplied by the complete elliptic integral of  the second kind $E[m_1(c)]$ yields NLL terms
at LP, the coefficient $p(c)$, multiplied with the complete elliptic integral of the third kind
${\Pi}[n_1(c), m_1(c)]$ produces LL and NLL contributions at both LP and
NLP, while the coefficient $k(c)$ together with the complete elliptic integral of  the first kind $K[m_1(c)]$ also produces LL terms at
NLP.
From eqs.~(\ref{arguEllipt}),  (\ref{eq:EKPexpanded}) and (\ref{coeffExSER})
the $c$-parameter distribution at NLO for small-$c$ reads,
\begin{align}
&\frac{1}{\sigma_{0}(s)}  \frac{d\sigma}{dc} \Bigg\vert_{\text{NLO}}  \,\nn \\ &=\,\frac{2 \alpha_{s}}{3 \pi}\biggl(\frac{-3-4\log{c}}{c}+1-28\log{c}+ \mathcal{O}(c)   \biggr). 
\label{cpnlo}
\end{align}
We have computed the $c$-parameter distribution here without
approximating the event shape variable or the matrix element squared,
similar to what we did for thrust in eq.~(\ref{Fthrustnlp}). The above
result, which agrees with~\cite{Catani:1998sf, Gardi:2003iv},
contains the contributions from all 
regions of the phase space. By
comparing eq.~(\ref{cpnlo}) and eq.~(\ref{Fthrustnlp}) we observe that the
LP terms for both event shape variables have an identical
structure in the dijet limit. Further similarities between these
event-shape variables are discussed in
~\cite{PhysRevD.91.094017, Dokshitzer:1995zt, Akhoury:1995sp, Korchemsky:1994is, PhysRevD.75.014022}.
The expression eq.~\eqref{cpnlo} does not fully expose the relation to 
different kinematical configurations. To do so, we list the upper and
lower limit contributions of the $y$ integral separately. When computing the contributions from the upper and lower limits separately we have observed that it is only the
coefficients $e(c), p(c) \, \text{and} \, k(c)$ of the elliptic
integrals that differ, while the general form of the elliptic integrals remains the same as in eq.~(\ref{eq:elliptForm}). 
From the upper limit of
eq.~(\ref{ExactInt}) we have, for small-$c$,
\begin{align}
e_{\text{u}}(c)& \,=\, -\frac{3}{c}+9+12c+36c^2+\mathcal{O} (c^3)\,, \nonumber \\
p_{\text{u}}(c)& \,=\, 0\,,\label{eq:EcoeffUL} \\
k_{\text{u}}(c)& \,=\, -12-24c-108c^2+\mathcal{O} (c^3)\,, \nonumber 
\end{align} 
For the lower limit, the behavior reads
\begin{align}
e_{\text{l}}(c)& \,=\, 0\,, \nonumber \\
p_{\text{l}}(c)& \,=\, -32c-112c^2-912c^3+\mathcal{O}(c^{7/2})\,,\label{coeffLL} \\
  k_{\text{l}}(c)& \,=\, -16-4c-156c^2+\mathcal{O} (c^3) \nonumber\,.
\end{align}
Since $p_u(c) \,=\,e_l(c) \,=\,0$, no LL terms at LP
derive from the upper limit,  nor do we find NLL terms at LP from the lower
limit. Collecting the upper and lower limit contributions for
the $c$-parameter distribution in the dijet limit leads to
\begin{align}
\frac{1}{\sigma_{0}(s)} \frac{d\sigma}{dc} \Bigg\vert_{\text{u}} & \,=\, \alspi \left( -\frac{3}{c}+9(1+2\log c) + \mathcal{O}(c)  \right), 
\label{ULexact} \\
\frac{1}{\sigma_{0}(s)} \frac{d\sigma}{dc} \Bigg\vert_{\text{l}} & \,=\, \alspi \left( \frac{4  \log c}{c}+2(4+23\log c)+ \mathcal{O}(c)  \right).
\label{LLexact}
\end{align}
Thus the upper limit of $y$  in the dijet limit  corresponds to various kinematic
configurations involving a hard gluon such as hard gluon being back-to-back with quark (anti-quark) and collinear to anti-quark (quark), which produces NLL terms at LP. As discussed the upper limit  also corresponds to the soft quark/anti-quark emissions,
LL and NLL terms at NLP also arise from this limit. The lower
limit of $y$ corresponds to kinematic configurations involving
a soft gluon and thus yields LL terms at both LP and NLP. No NLL terms
at LP are generated here, although there is a NLL contribution at NLP.

\subsection{Next-to-leading power corrections to $c$-parameter from shifted kinematics}

We next  compute the $c$-parameter distribution using
the shifted kinematics method, and again assess to what extent LP and
NLP terms in the exact NLO calculation are reproduced. The
approximation is
\begin{align}
 \frac{1}{\sigma_{0}(s)} \frac{d \sigma}{dc} \Bigg\vert_{\text{shift}}   \,=\, \frac{1}{2s} \int d\Phi_3  \ \overline{\sum}& |\mathcal{M}_{\text{shift}}(x_1,x_2)|^2  \ \nn
 \\ & \times  \delta\left(  c (y,z) - c \right)\,.
 \label{rana5}
\end{align}

The shifted matrix element in eq.~(\ref{shiftexpr}),  when
written in terms of the transformed variables ($y, z$), takes the form
\begin{align}
\overline{\sum} & |\mathcal{M}_{\text{shift}}(y,z)|^2 \nn \\ &  \,=\,8(e^2 e_q)^2 N_c C_F g_s^2 \frac{1}{3Q^2}  \left(\frac{2 (y-1)}{y^2 (z-1) z}\right). 
\label{Mshiftcp}
\end{align}
Again, we make no approximation to the
event shape definition itself. We then proceed in the same manner as with the exact matrix element. We have  
\bt
\begin{align}
 \frac{1}{\sigma_{0}(s)} \frac{d \sigma}{dc} \Bigg\vert_{\text{shift}} & \,=\,   \alspi \int^{1}_{0}  dy \int^{1}_{0} dz \frac{2 (1-y (1-z))^2 (1-y z)^2}{(1-y)^2 y^2 z (z-1)  (2 z-1)} \bigg(\delta(z-z_1)+\delta(z-z_2)\bigg)\,,
\end{align}
where $z_1$ and $z_2$ are provided in eq.~(\ref{eq:zroots}). 
After the $z$ integration we have 
\begin{align}
\frac{1}{\sigma_0(s)}\frac{d\sigma}{dc} \Bigg\vert_{\text{shift}}  \,=\,\alspi \int^{y_2}_{y_1} dy \frac{4 (y-1)^2}{c\sqrt{y  \left( y+cy-1 \right) \left( c  (y-2)^2+(y-1) y \right)}}\,.
\label{shiftINT}
\end{align}
The result of $ y $ integration can again be organized
in a manner similar to eq.~(\ref{eq:elliptForm}) as
\begin{align}
\frac{1}{\sigma_0(s)}\frac{d\sigma}{dc} \Bigg\vert_{\text{shift}}  \,=\,\frac{2\alpha_s}{3\pi} \bigg( e_{\text{s}} (c) E[m_1(c)]+p_{\text{s}} (c) \Pi[n_1(c),m_1(c)]+k_{\text{s}} (c) K[m_1(c)]\bigg)\,,
\label{eq:shiftEllipt}
\end{align}
\et
 A comparison of eq.~(\ref{shiftINT}) with eq.~(\ref{ExactInt}) shows
a significant simplification of the integrand, where the subscript $\text{s}$ on the coefficients on the right indicates the shifted kinematics method. The elliptic integrals and their arguments are given in eqs.~(\ref{eq:EKPexpanded}) and (\ref{arguEllipt}) respectively.

The coefficients in eq.~(\ref{eq:shiftEllipt}) do 
change from the exact result, and we find
\begin{align}
k_{\text{s}} (c) \,=\,&-\frac{\sqrt{2-2\sqrt{1-8c}-8c-16c^2}}{c^{3/2}(1+c)^{5/2}}\,, \nonumber  \\
e_{\text{s}} (c) \,=\,&-\frac{(1+\sqrt{1-8c}-4c) \ \sqrt{1-\sqrt{1-8c}-4c-8c^2}}{2\sqrt{2} c^{5/2}(1+c)^{5/2}}\,,  \\
p_{\text{s}} (c) \,=\,&\frac{ (1-\sqrt{1-8c}-4c) \ \sqrt{1-\sqrt{1-8c}-4c-8c^2} }{\sqrt{2}c^{5/2}(1+c)^{5/2}}\,.\nonumber
\end{align}
Their small-$c$ behavior is
\begin{align}
e_{\text{s}} (c) \,=\,&-\frac{4}{c}+16-4c+72c^2+\mathcal{O}(c^3)\,, \nonumber \\
p_{\text{s}} (c) \,=\,&32c+128c^2+\mathcal{O}(c^3)\,, \\
k_{\text{s}} (c) \,=\,&-8c+\mathcal{O}(c^3)\,. \nonumber 
\end{align}

Expanding around $c =0$, the $c$-parameter distribution from shifted kinematics approximation reads
\begin{align}
&\frac{1}{\sigma_{0}(s)}   \frac{d\sigma}{dc} \Bigg\vert_{\text{shift}}  \,  \nn \\ &=\, \alspi  \bigg(\frac{-4-4\log{c}}{c}+8-24\log{c}+ \mathcal{O}(c)   \bigg). 
\label{shiftedCPdstr}
\end{align} 
For better insight, it is again useful to examine separately the contributions of
the upper and lower of the $y$ integral.
The small-$c$ behavior of the coefficients of the elliptic integrals appearing from the upper limit of eq.~(\ref{shiftINT}) is
\begin{align}
e_{\text{su}}(c)& \,=\, -\frac{4}{c}+16-4c+72c^2+252c^3+\mathcal{O} (c^{7/2}) \nonumber, \\
p_{\text{su}}(c)& \,=\, 0, \label{coeffSUL} \\
k_{\text{su}}(c)& \,=\, -16-16c-144c^2-688c^3+\mathcal{O} (c^{7/2}) \nonumber \,.
\end{align}
For the lower limit we have
\begin{align}
e_{\text{sl}}(c)& \,=\, 0 \nonumber, \\
p_{\text{sl}}(c)& \,=\, -32c-128c^2-928c^3+\mathcal{O}(c^{7/2}),\label{coeffSLL} \\
k_{\text{sl}}(c)& \,=\, -16-8c-144c^2-616c^3+\mathcal{O} (c^{7/2}) \nonumber .
\end{align}
Note that the $\mathcal{O}(c^3)$ terms contribute at NNLP accuracy.
We now have
\begin{align}
\frac{1}{\sigma_{0}(s)} \frac{d\sigma}{dc} \Bigg\vert_{\text{su}}&  \,=\, \alspi \left( -\frac{4}{c}+8(2+3\log c) + \mathcal{O}(c)  \right), 
\label{ULshift} \\
\frac{1}{\sigma_{0}(s)} \frac{d\sigma}{dc} \Bigg\vert_{\text{sl}} & \,=\, \alspi \left( \frac{4  \log c}{c}+8(1+6\log c) + \mathcal{O}(c)  \right).
\label{LLshift}
\end{align}
From the above two expressions, we observe that the lower (upper)
limit fully captures the LL (NLL) at LP, while the LL and NLL at
NLP receive contributions from both limits.
The results for the $c$-parameter distributions at NLO up
to NLP, obtained from the exact and shifted kinematics computation,
are given in eqs.~(\ref{cpnlo}) and (\ref{shiftedCPdstr}), respectively.
The integrand 
in the shifted kinematics method was considerably simpler than for
the exact computation. At LP, the method reproduced the LL term. The
NLL term at LP was not fully reproduced because the contribution from
the hard-collinear gluon is absent. Similarly, not all of the LL at
NLP is captured, due to the absence of from soft quark and soft anti-quark contributions.
Recall that the shifted kinematics method does not 
account for soft quark contributions, which we computed separately
using soft quark emission vertices in section \ref{ThrustShift}.

We again examine the remaining contributions to the $c$-parameter
distribution, such as those from the soft quark, soft
anti-quark, and hard-collinear gluon configurations. This allows the mapping of all of the contributions to
the $c$-parameter from various regions of the phase space. 
 When expressed in terms of the transformed $(y,z)$ variables the remainder matrix element as described in eq.~(\ref{Mrem}) reads 
\begin{align}
\overline{\sum}|{\mathcal{M}}_{\text{rem}}(y,z)|^2 \,=\, 8&(e^2 e_q)^2  N_c C_F g_s^2 \frac{1}{3Q^2} \nn \\ &  \times \biggl( \frac{1}{z}+  \frac{1}{1-z}-2 \biggr). 
\label{mremcp}
\end{align}  

The expression for $c$-parameter distribution using the above matrix element squared is

\bt
\begin{align}
 \frac{1}{\sigma_{0}(s)} \frac{d \sigma}{dc} \Bigg\vert_{\text{rem}}=\frac{2 \alpha_{s}}{3 \pi} \int^{1}_{0}  dy \int^{1}_{0} dz \frac{1}{z(1-z)}  \Big(\delta(z-z_1)+\delta(z-z_2)\Big)  \frac{-y(y-1)^2 \left (1-2z(1-z) \right)}{(c y+y-1) \sqrt{y (c y+y-1) \left(c (y-2)^2+(y-1) y\right)}},
\end{align}
with $z_1$ and $z_2$ are given in eq.~(\ref{eq:zroots}). After performing the $z$ integration we are left with 
\begin{align}
 \frac{1}{\sigma_{0}(s)} \frac{d\sigma}{dc} \Bigg\vert_{\text{rem}} \,=\,  \frac{2\alpha_s}{3\pi}  \int^{y_1}_{y_2} dy \ \frac{-2 y(y-1) (c ((y-2) y+2)+(y-1) y)}{c (c y+y-1) \sqrt{y (c y+y-1) \left(c (y-2)^2+(y-1) y\right)}}\,.
 \label{remINT}
\end{align}
with $y_1$ and $y_2$ given in eq.~(\ref{Ylimits}).
The outcome takes again the form as in eq.~(\ref{eq:elliptForm}),
somewhat more complicated than either of the two approximations. The result can be written as
\begin{align}
\frac{1}{\sigma_0(s)}\frac{d\sigma}{dc} \Bigg\vert_{\text{rem}}  \,=\,  \frac{2\alpha_s}{3\pi} \bigg( e_{\text{r}}(c)  \ E [m_1(c)]  +  p_{\text{r}}(c)  \  \Pi[n_1(c),m_1(c)]  +  k_{\text{r}}(c) \  K [m_1(c)]\bigg), 
\label{mremelliptics}
\end{align}
\et
The small-$c$ behavior of the coefficients reads
\begin{align}
e_r(c)& \,=\, \frac{1}{c}-7+16c -36c^2+9c^3+\mathcal{O} (c^{7/2})\,, \nonumber \\
p_r(c)& \,=\, -16c -16c^3+\mathcal{O} (c^{7/2})\,,\\
k_r(c)& \,=\, 4 -12c +48 c^2+16c^3+\mathcal{O} (c^{7/2})\,,\nonumber 
\end{align} 
\begin{align}
\frac{1}{\sigma_{0}(s)} \frac{d\sigma}{dc} \Bigg\vert_{\text{rem}} \,=\,
\frac{2\alpha_s}{3\pi}  \left( \frac{1}{c} -7-4\log c + \mathcal{O}(c)  \right)\,.
  \label{MrcpFinal}
\end{align}
Here we indeed see contributions from the hard-collinear gluon, the soft quark, and
anti-quark configurations, which yield NLL terms at LP (hard-collinear gluon) as well as LL terms at NLP (soft quark/anti-quark) respectively. Combining these contributions with the outcome
of shifted kinematics approximation eq.~(\ref{shiftedCPdstr}) allows us to fully map the
distributions of the $c$-parameter to the different kinematical
configurations.

Let us finally break the contributions from remaining term also down into upper and
lower limit components.
The coefficients from the upper limit of  eq.~(\ref{remINT}) yield
\begin{align}
e_{\text{ru}}(c)& \,=\, \frac{1}{c}-7+16c-36c^2+9c^3+\mathcal{O} (c^{7/2}) \,, \nn \\
p_{\text{ru}}(c)& \,=\, 0\,,  \\
k_{\text{ru}}(c)& \,=\, 4-8c+36c^2+64c^3+\mathcal{O} (c^{7/2}) \,\nonumber .
\label{eq:coeFFRUL}
\end{align} 
while for the lower limit, we have
\begin{align}
e_{\text{rl}}(c)& \,=\, 0 \nonumber, \\
p_{\text{rl}}(c)& \,=\, 16c+16c^3+\mathcal{O}(c^{7/2}), \\
k_{\text{rl}} (c)& \,=\, 4c-12c^2+48c^3+\mathcal{O} (c^{7/2}) \nonumber .
\label{eq:coeFFRLL}
\end{align}
Using eqs.~(\ref{arguEllipt}),~(\ref{eq:EKPexpanded}) and above two expressions, the
contributions to the $c$-parameter distribution from the upper and
lower limit respectively are
\begin{align}
\frac{1}{\sigma_{0}(s)} \frac{d\sigma}{dc} \Bigg\vert_{\text{ru}}&  \,=\, \alspi \left( \frac{1}{c}-7-6\log c + \mathcal{O}(c)   \right),  \\
\frac{1}{\sigma_{0}(s)} \frac{d\sigma}{dc} \Bigg\vert_{\text{rl}}&  \,=\, \alspi \bigg(-2\log c + \mathcal{O}(c)  \bigg). 
\end{align}
The missing LL terms at NLP have been generated here. The upper limit
contributes to the missing NLL term at LP (related to hard-collinear
gluon emission) and NLP. Both limits
contain LL terms at NLP (soft (anti-)quark emission).

\subsection{Numerical assessment of the shifted kinematics  approximations for $c$-parameter}
\label{sec:numer-assess-eikon-cp}
\begin{figure*}[hbtp!]
	\centering
	\subfloat[][]{\includegraphics[scale=0.6]{"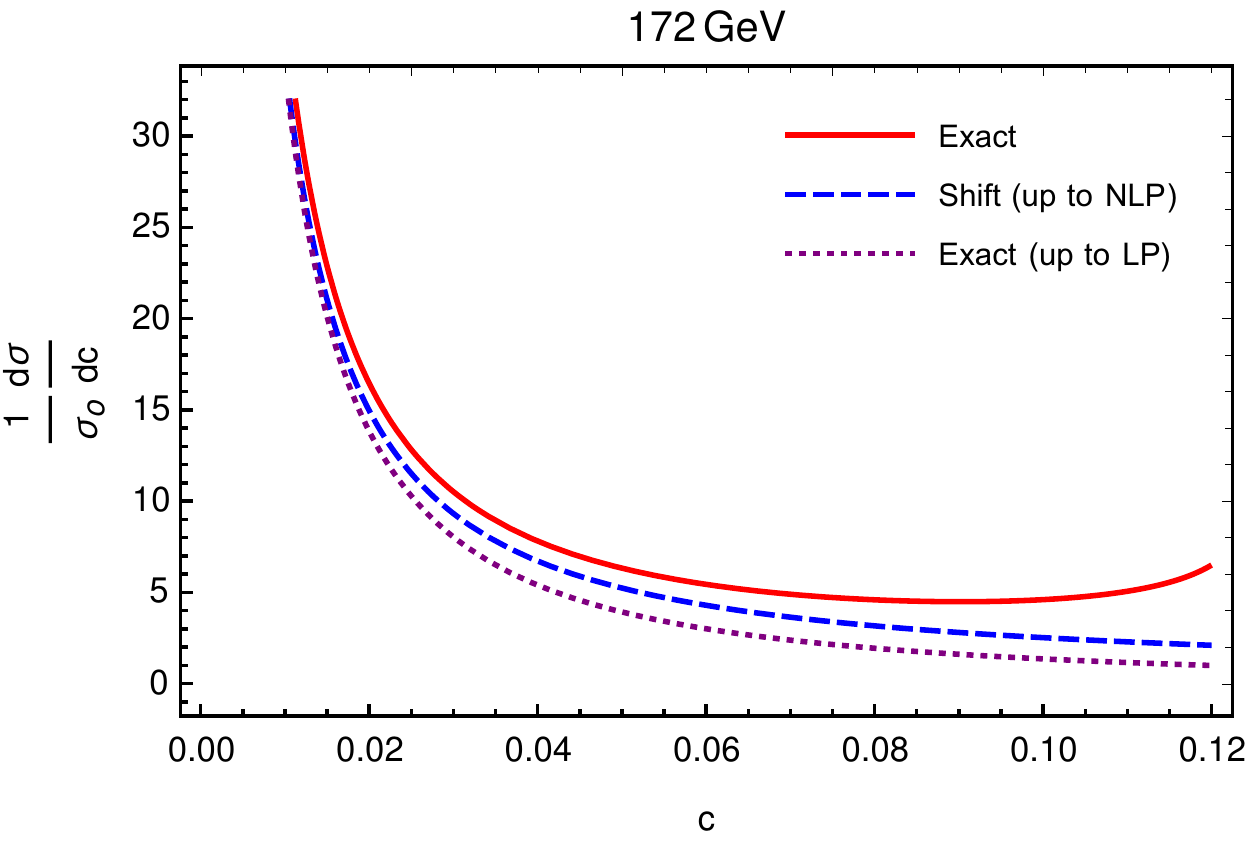"}}
	\qquad 
	\subfloat[][]{\includegraphics[scale=0.6]{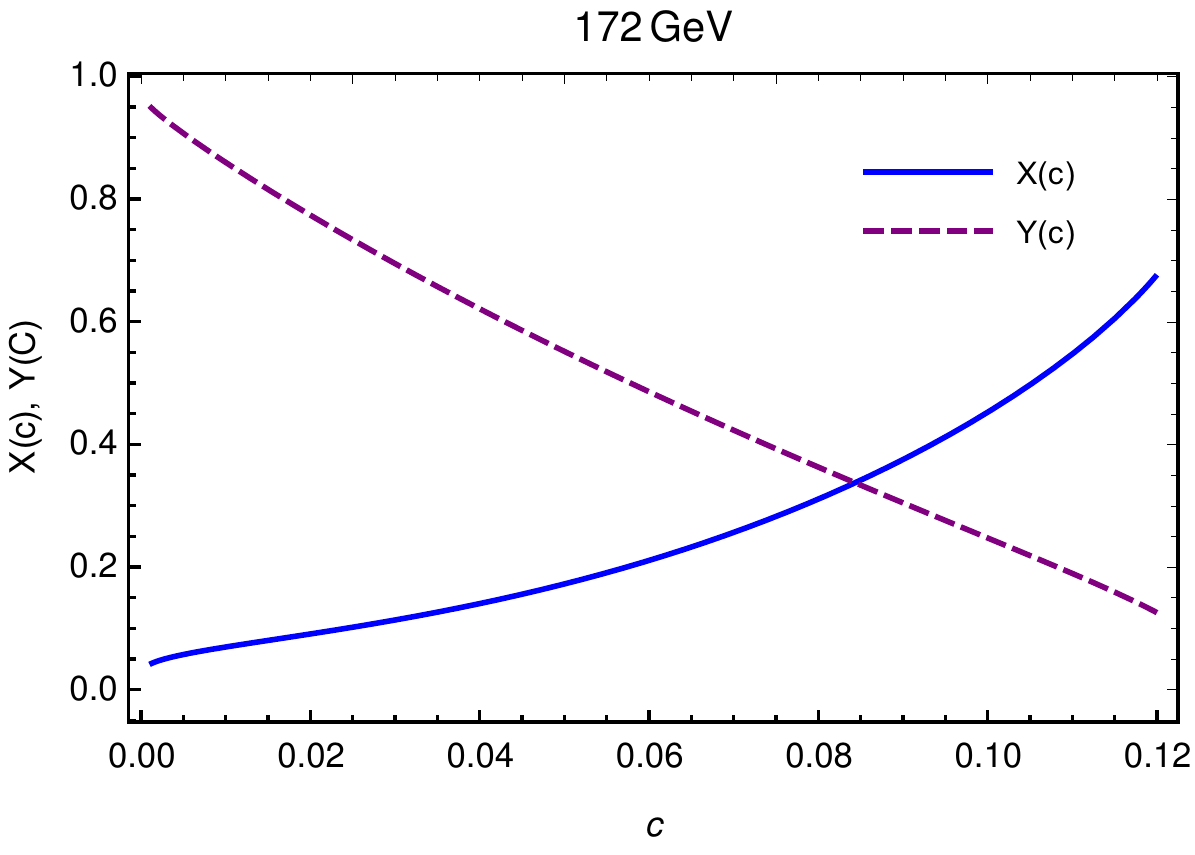}}    
	\caption{ In (a), we plot comparison of the $ c $-parameter distribution. We show the exact result (solid red curve)  in eq.~\eqref{eq:elliptForm}, the shifted approximation (dashed blue curve) up to NLP terms given in eq.~\eqref{shiftedCPdstr}. We also plot the exact LP term given in eq.~\eqref{cpnlo} 
		(dotted purple curve). In (b), we show behavior of  $X(c)$ (solid blue
		curve) and $Y(c)$ (purple dashed curve) from  eqs.~\eqref{eq:Xc} and~\eqref{eq:Yc} respectively 
		vs.  $c$. }
	\label{plot:CPARA_Combined_PLot}
\end{figure*}
We analyze our $c$-parameter results numerically. In the small-$c$
limit, the effectiveness of the shifted kinematics method is
evident from fig.~(\ref{plot:CPARA_Combined_PLot}\textcolor{prd_blue}{a}), the shifted kinematics result
provides a better approximation to the exact result, than the exact result taken
up to LP terms only, even though we did not generate all the dominant terms at LP and NLP from this approximation.
The $c$-parameter distribution in fig.~(\ref{plot:CPARA_Combined_PLot}\textcolor{prd_blue}{a}) has a more pronounced
departure of the LP terms from the exact result
than the thrust distribution in fig.~(\ref{plot:Thrust_Combined_PLot}\textcolor{prd_blue}{a}). This is due to the $c$-parameter distribution having a 
larger LL coefficient at NLP than thrust.
We define $X(c)$ and $Y(c)$ in similar fashion to $X(\tau)$ and $Y(\tau)$ in eqs.~(\ref{eq:Xtau}) and~(\ref{eq:Ytau}) as 

\begin{align}
X(c)& \,=\, \frac{\frac{1}{\sigma_0(s)}\frac{d\sigma}{d c}  \Big|_{\text{NLO}}- \frac{1}{\sigma_0(s)}\frac{d\sigma}{d c}  \Big|_{\text{shift-NLP}}}{\frac{1}{\sigma_0(s)}\frac{d\sigma}{d c}  \Big|_{\text{NLO}}}, \label{eq:Xc} 
\end{align}
\begin{align}
Y(c)& \,=\, \frac{\frac{1}{\sigma_0(s)}\frac{d\sigma}{d c}  \Big|_{\text{NLO-LP}}}{\frac{1}{\sigma_0(s)}\frac{d\sigma}{d c}  \Big|_{\text{NLO}}}\,, \label{eq:Yc}
\end{align}
and exhibit them in fig.~(\ref{plot:CPARA_Combined_PLot}\textcolor{prd_blue}{b}). $X(c)$ 
exhibits similar behavior as the analogous case for thrust in
fig.~(\ref{plot:Thrust_Combined_PLot}\textcolor{prd_blue}{b}), but the deviation of $X(c)$ from zero is
faster with increasing $c$. $Y(c)$ 
decreases sharply from unity as $c$ increases, a behavior opposite to
thrust in fig.~(\ref{plot:CPARA_Combined_PLot}\textcolor{prd_blue}{b}). Indeed, due 
to the large negative LL coefficient at NLP for the $c$-parameter, the denominator here increases sharply. More precisely, for the
$c$-parameter, the LL coefficient at NLP is seven times that at
LP, with the same sign, while for thrust, the LL coefficient at NLP is
half that of LP, with the opposite sign.
%

  \section{Conclusions}
  \label{sec:conclusions}
\noindent  The fixed order results in perturbation theory for massless fields
  contain logarithms of the (small) ratio of energy scales, arising from 
  soft or collinear radiation.  They occur at both leading and next-to-leading
  power of this ratio. This work focuses on 
  NLP terms appearing in the thrust and $C$-parameter event shape distribution.
We identify in exact NLO results the origin
  of large logarithmic terms at LP and NLP.  Subsequently we provide an approximation of
  these results using the recent shifted kinematics
  method, designed to capture large logarithms at NLP
  accuracy due to soft gluon emission. Moreover we compute soft quark emission contributions
  using corresponding effective Feynman rules.

This formalism indeed reproduces the dominant
  contributions at LP and NLP from soft gluon radiation
correctly for both thrust and $c$-parameter.
  The remaining LL terms at NLP are reproduced 
  the soft (anti-)quark 
  emission, as anticipated 
  in~\cite{vanBeekveld:2019prq, vanBeekveld:2019lwy}.
  We were able to map the various sources of
  contributions at LP and NLP, using an integral form, and its limits,
  for the distributions.

  Our detailed diagnosis of NLP terms in event shape variables, and
  demonstration that they can be fully reproduced using a shifted
  kinematics approach together with soft fermion emission vertices,
  should be a useful resource towards further understanding of NLP terms, in
  particular those arising from final state emissions.

\begin{acknowledgements}
\noindent EL and AT would like to thank the MHRD Government of India, for
the SPARC grant SPARC/2018-2019/P578/SL, \textit{Perturbative QCD for Precision Physics at the
LHC}.
SM would like to thank CSIR, Govt. of India, for the SRF fellowship
(09/1001(0052)/2019-EMR-I).
\end{acknowledgements}
 
\appendix
\section{Transformation of incomplete elliptic integrals}
\label{eq:ElliptTransf}
\noindent
In this appendix, we demonstrate how we handle the incomplete elliptic
integrals that appear in the expression of $c$-parameter
distribution. The final expression for the $c$-parameter distribution is
written in a compact manner in eq.~\eqref{eq:elliptForm}, where $ K $,
$ E $ and $ \Pi $ are the complete elliptic integrals of the first,
 second and third kind, respectively. However, when we perform the
integration over final variable $ y $ in
eqs.~(\ref{ExactInt}),~(\ref{shiftINT}) and~(\ref{remINT}), this
integration produces incomplete elliptic integrals  $ F $,
$ E $, and $ \Pi $. These incomplete elliptic integrals are later
converted into complete integrals to arrive at
eq.~\eqref{eq:elliptForm}, as first written in \cite{Gardi:2003iv}.
The three kinds of incomplete elliptic integrals appearing in $c$-parameter  distribution  are
\begin{align} \label{eq:IncompleteElliptics}
	F[\phi, m] \,     \equiv \, &  \int_{0}^{\phi} d \theta \frac{1}{\sqrt{1-m \sin^2\theta}} \nn \\  \,=\, & \int_{0}^{\sin\phi} \frac{dt}{\sqrt{(1-t^2)(1-mt^2)}} \,, 
 \end{align}
 \begin{align}
 \hspace{-1.5cm} E[\phi, m]\,  \equiv  \, & \int_{0}^{\phi} d \theta  \sqrt{1-m \sin^2\theta}\,\nn \\   \,=\, & \, \int_{0}^{\sin\phi}  dt \sqrt{\frac{1-mt^2}{1-t^2}}  \,, 
 \end{align}
 \begin{align}
\hspace{0.7cm}	\Pi[n,\phi, m] \,   \equiv \, &  \int_{0}^{\phi} d \theta \frac{1}{(1-m\sin^2\theta)\sqrt{1-m\sin^2\theta}} \nn \\ \, = \, & \int_{0}^{\sin\phi} \frac{dt}{(1-nt^2)\sqrt{(1-t^2)(1-mt^2)}} \,. 
\end{align}
Here $ \phi, m $, and $ n $ are called the \emph{amplitude},   \emph{parameter}, and \emph{characteristic}  of the elliptic integrals,  respectively. Eq.~\eqref{eq:ellipticINTcomplete} gives their respective complete form. The corresponding transformation into a complete elliptic integral can be performed using the rule
\begin{align} \label{eq:transfTOcomplete}
	F[\phi,m]& \,=\,K[m] , \nonumber\\
	E[\phi,m]& \,=\,E[m] , \\
	\Pi[n,\phi, m]& \,=\, \Pi[n, m]. \nonumber
\end{align} 
 The above transformation is only possible when the amplitude  $ \phi =\pi/2$.
 The indefinite integration of eqs.~(\ref{ExactInt}),~(\ref{shiftINT}) and~(\ref{remINT})  results in  multiple incomplete elliptic integrals with only two unique amplitudes in their arguments, namely
$ \phi_1(c,y) $ and $\phi_2(c,y)$,  given by
\begin{align}\label{eq:Ellipt_amplitudes}
	\phi_1(c,y)& \,=\, \left(\frac{-1+\sqrt{1-8c}-4c+8c/y}{2\sqrt{1-8c}}\right)^{1/2},
	\end{align}

\begin{align}
	\phi_2(c,y) & \,=\,  \left(\frac{1+\sqrt{1-8c}+4c-8c/y}{2\sqrt{1-8c}}\right)^{1/2}.
\end{align}
The amplitudes $ \phi_1(c,y) $ and $\phi_2(c,y) $ are present in the arguments of all three kinds of incomplete elliptic integrals. If we directly substitute the upper limit ($y_2$) after integration, then none of the incomplete elliptic integrals with the amplitude $ \phi_1(c,y) $ can be reduced into complete elliptic integral since 
\begin{align}
	\phi_1(c,y) \Big\vert_{y \,=\,y_2} \,=\,0.
\end{align}
The transformation into complete elliptic integrals holds only when $ \phi_1  =\pi/2 $ as given in eq.~\eqref{eq:transfTOcomplete}. However, the incomplete elliptic integrals with the amplitude $ \phi_2(c,y) $ upon substitution of upper limit can be directly reduced into complete elliptic integral as 
\begin{align}
	\phi_2(c,y) \Big\vert_{y \,=\,y_2} \,=\,\frac{\pi}{2}.
\end{align}
We observe a similar but opposite behavior when we substitute the lower limit ($ y_1 $) into the integration result, this time, the incomplete elliptic integrals with the amplitude $\phi_1(c,y)$ can be reduced into complete elliptic integral as 
\begin{align}
	\phi_1(c,y) \Big\vert_{y \,=\,y_1} \,=\,\frac{\pi}{2},
\end{align}
while the elliptic integrals with amplitude $ \phi_2(c,y) $ cannot be reduced into complete elliptic integral as 
\begin{align}
	\phi_2(c,y) \Big\vert_{y \,=\,y_1} \,=\,0.
\end{align}
In order to resolve the issue of transforming the incomplete elliptic integrals into complete elliptic integrals, we modify the upper and
lower limits of $y$ integration as
\begin{align}
	(y_1, y_2) \to (y_1+e, y_2+e)\, ,
	\label{eq:modified limits}
\end{align}
where $e$ is an infinitesimal real off-set parameter that will be
taken to zero at the end. The relative sign of $ e $ does not affect the final expression for $c$-parameter distribution, which is independent of this parameter $e$.  When we substitute the upper limit  of integration $y_2 $ with the off-set parameter as defined in the above expression, the amplitudes $ \phi_1(c,y) $ and $\phi_2(c,y)  $  modify and have the following form
\bt 
\begin{align}
	\phi_1(c,y,e) \Big \vert_{y \,=\,y_2+e}& \,=\, \left( \frac{\left(-4 c+\sqrt{1-8 c}-1\right) (c+1) e}{\sqrt{1-8 c} \left(2 c (e+2)+\sqrt{1-8 c}+2 e+1\right)}\right)^{1/2} \,,  \\
	\phi_2(c,y,e) \Big \vert_{y \,=\,y_2+e}&  \,=\,\left( \frac{-\frac{16 (c+1) c}{2 c (e+2)+\sqrt{1-8 c}+2 e+1}+4 c+\sqrt{1-8 c}+1}{2 \sqrt{1-8 c}}\right)^{1/2},
\end{align}
similarly, from the lower limit, the amplitudes are
\begin{align}
	\phi_1(c,y,e) \Big \vert_{y \,=\,y_1+e}& \,=\, \left(\frac{\frac{16 (c+1) c}{2 c (e+2)-\sqrt{1-8 c}+2 e+1}-4 c+\sqrt{1-8 c}-1}{2 \sqrt{1-8 c}}\right)^{1/2} \,,  \\
	\phi_2(c,y,e) \Big \vert_{y \,=\,y_1+e}&  \,=\,\left(-\frac{(c+1) \left(4 c+\sqrt{1-8 c}+1\right) e}{\sqrt{1-8 c} \left(-2 c (e+2)+\sqrt{1-8 c}-2 e-1\right)}\right)^{1/2}.
\end{align}
\et
Note that, it is necessary to use this parameter because the
straightforward substitution of limits did not allow the
transformation of every incomplete elliptic integral. Let us
consider a few examples to demonstrate how this off-set parameter
solves the problem with incomplete elliptic integrals that cannot be
reduced into complete elliptic integrals as their amplitude
$ \phi \neq \pi/2 $. We can categorize all the elliptic integrals
appearing after $ y $ integration into two classes according to their
amplitudes $\phi$ as $ (i) $ non-reducible incomplete elliptic
integrals $ (\phi \neq \pi/2) $, and $ (ii) $ reducible incomplete elliptic
integrals $ (\phi  = \pi/2) $.
\subsection{Non-reducible incomplete elliptic integrals}
Here we consider an incomplete elliptic integral that  appears from the upper limit contributions, the expression of the elliptic integral is 
\bt
\begin{align} \label{eq:E1appendix}
	E[\phi_1(c,e),m_1(c)] \,=\,E\left[\sin ^{-1}\left(\frac{\left(-4 c+\sqrt{1-8 c}-1\right) (c+1) e}{\sqrt{1-8 c} \left(2 c (e+2)+\sqrt{1-8 c}+2 e+1\right)}\right)^{1/2}, \   \frac{2\sqrt{1-8c}}{1+\sqrt{1-8c}-4c-8c^2} \right]\, ,
\end{align}
 where $ m_1 $ is given in eq.~(\ref{arguEllipt}), the amplitude of this elliptic function  is
\begin{align}
	\phi_1(c,e) \,=\, \sin ^{-1}\left(\frac{\left(-4 c+\sqrt{1-8 c}-1\right) (c+1) e}{\sqrt{1-8 c} \left(2 c (e+2)+\sqrt{1-8 c}+2 e+1\right)}\right)^{1/2}.
\end{align}
 The elliptic integral in eq.~(\ref{eq:E1appendix}) cannot be reduced into a complete elliptic integral as in the limit $e \to 0$ the amplitude $ \phi_1(c,e) \to 0$.
 However, we can expand this elliptic integral in powers of $ e $  around
$e  = 0$, upon expanding we get
\begin{align}
	E[\phi_1(c,e),m_1(c)]& \,=\,\sqrt{\frac{\left(-4 c+\sqrt{1-8 c}-1\right) (c+1)}{\sqrt{1-8 c} \left(4 c+\sqrt{1-8 c}+1\right)}} \sqrt{e} +\mathcal{O}(e^{3/2})
\end{align}
The expansion's first term, proportional to $e^{1/2}$, is the only
significant term in the limit when $e\to 0$.
The next term in the above expression is proportional to $e^{3/2}$,
and can be dropped. The small-$e$ expression for these specific incomplete elliptic integrals occurring is then stored in the form
\begin{align}\label{eq:E1expanded}
	E[\phi_1(c,e),m_1(c)] \,=\,\sqrt{\frac{\left(-4 c+\sqrt{1-8 c}-1\right) (c+1)}{\sqrt{1-8 c} \left(4 c+\sqrt{1-8 c}+1\right)}} \sqrt{e}.
\end{align}	
\et
Now, we shift our attention toward the elliptic integrals 
 that appears from the lower limit contribution, one such incomplete elliptic integral is
\bt
\begin{align}\label{eq:F2}
	F[\phi_2(c,e),m_1(c)] \,=\,F\left[\sin ^{-1}\left(\frac{(c+1) \left(4 c+\sqrt{1-8 c}+1\right) e}{\sqrt{1-8 c} \left(2 c (e-2)+\sqrt{1-8 c}+2 e-1\right)}\right)^{1/2}, \  \frac{2\sqrt{1-8c}}{1+\sqrt{1-8c}-4c-8c^2} \right]\,.
\end{align}
The above elliptic integral  cannot be reduced into a complete elliptic integral as the amplitude $ \phi_2(c,e) $ is 
\begin{align}
	\phi_2(c,e) \,=\,	\sin ^{-1}\left(\frac{(c+1) \left(4 c+\sqrt{1-8 c}+1\right) e}{\sqrt{1-8 c} \left(2 c (e-2)+\sqrt{1-8 c}+2 e-1\right)}\right)^{1/2}
\end{align}
\et
and in the limit $ e \to 0 $ the amplitude $ \phi_2(c,e) \to 0 $. Following the  similar procedure as in  eq.~\eqref{eq:E1expanded}, the elliptic integral in eq.~(\ref{eq:F2}) is expanded around $ e  = 0 $, we get
\begin{align}\label{eq:F1expanded}
	F[\phi_2(c,e),m_1(c)]=\sqrt{\frac{(c+1) \left(4 c+\sqrt{1-8 c}+1\right)}{\sqrt{1-8 c} \left(-4 c+\sqrt{1-8 c}-1\right)}} \sqrt{e}\, .
\end{align}

The coefficients of these elliptic functions depend on $ e $ when the
integration limits are modified in accordance with
eq.~\eqref{eq:modified limits}. When we expand our final result in
powers of $ e $, the negative powers of $ \sqrt{e} $ from the
coefficients combine with the positive powers of $ \sqrt{e} $ from the
stored expressions of non-reducible incomplete elliptic integrals to
yield few terms independent of $ e $. Subsequently $e$ can be set to zero.
\begin{table*}[hbtp!]
	
	\begin{tabular}{|p{5.3cm}|p{5.3cm}|p{5.3cm}|}
		\hline 
		\[\text{Observable}\] & \[ \text{Matrix element}\] & \[\text{Distribution up to NLP } \] \\ [0.06pt] 
		\hline 
		\[\text{Thrust}\] &  \[\text{Exact}\]  &  \[ \frac{-3-4\log{\tau}}{\tau}-2+2\log{\tau}\]\\ [0.06pt] \hline
		\[\text{Thrust}\] &  \[\text{Shift}\]  & \[\frac{-4-4\log{\tau}}{\tau}+4+4\log{\tau}\]\\ [0.06pt] \hline
		\[\text{Thrust}\] &  \[\text{Remainder}\] & \[\frac{1}{\tau}-6-2\log{\tau} \]   \\[0.06pt] \hline
		\[\text{Thrust}\] &  \[\text{Eikonal}\]  & \[ \frac{-4\log{\tau}}{\tau}-8-4\log{\tau} \]  \\ [0.06pt]
		\hline
		\[\text{$c$-parameter}\] &  \[\text{Exact}\] & \[ \frac{-3-4\log{c}}{c}+1-28\log{c}  \]  \\ [0.06pt] \hline
		\[\text{$c$ -parameter}\] &  \[\text{Shift}\] & \[ \frac{-4-4\log{c}}{c}+8-24\log{c} \]   \\[0.06pt] \hline
		\[\text{$c$ -parameter}\] &  \[\text{Remainder}\] & \[ \frac{1}{c}-7-4\log{c} \]  \\[0.06pt] \hline
		\[\text{$c$ -parameter}\] &  \[\text{Eikonal}\] & \[ \frac{-4\log{c}}{c}-8-40\log c \]   \\[0.06pt] \hline
	\end{tabular}
	\caption{Table for the result of thrust and $c$-parameter distributions calculated using the exact definition of event shape variables and four different definitions of matrix elements. } 
	\label{distributionTable}
\end{table*}


\subsection{Reducible incomplete elliptic integrals}
The category of incomplete elliptic integrals, which can be reduced
into complete elliptic integrals is easier to handle as compared with non-reducible ones. An elliptic
integral appearing from the upper limit contribution is
\bt
\begin{align}
E[\phi_2(c,e),m_1(c)] \,=\,E\left[\sin ^{-1}\left(\frac{-\frac{16 (c+1) c}{2 c (e+2)+\sqrt{1-8 c}+2 e+1}+4 c+\sqrt{1-8 c}+1}{2 \sqrt{1-8 c}}\right)^{1/2}, \  \frac{2\sqrt{1-8c}}{1+\sqrt{1-8c}-4c-8c^2}    \right]\, ,
\end{align}  
where the amplitude $  \phi_2(c,e)$ reads
\begin{align}
\phi_2(c,e) \,=\, \sin ^{-1}\left(\frac{-\frac{16 (c+1) c}{2 c (e+2)+\sqrt{1-8 c}+2 e+1}+4 c+\sqrt{1-8 c}+1}{2 \sqrt{1-8 c}}\right)^{1/2}\,.
\end{align}
\et

In the limit $ e \to 0 $ the amplitude $ \phi_2(c,e) \to \pi /2 $ and
 this  incomplete elliptic   integral can be reduced directly to a complete elliptic integral using reduction formula in eq.~\eqref{eq:transfTOcomplete} 
 
\begin{align}
	E[\phi_2(c,e),m_1(c)] \,=\,E[m_1(c)] \, .
\end{align} 

In this way all reducible incomplete elliptic integrals can be substituted with their corresponding complete elliptic integrals.
Upon substituting all the complete elliptic integrals and the stored
expressions of incomplete elliptic integrals, our final result is
expanded in powers of $ e $. No negative powers occur, and $ e $ can be
taken to zero, yielding eq.~\eqref{eq:elliptForm}.

\section{Eikonal approximation and result summary}
\label{sec:table-event-shape}
\begin{figure*}[hbtp!]
	\centering
	\subfloat[][]{\includegraphics[scale=0.648]{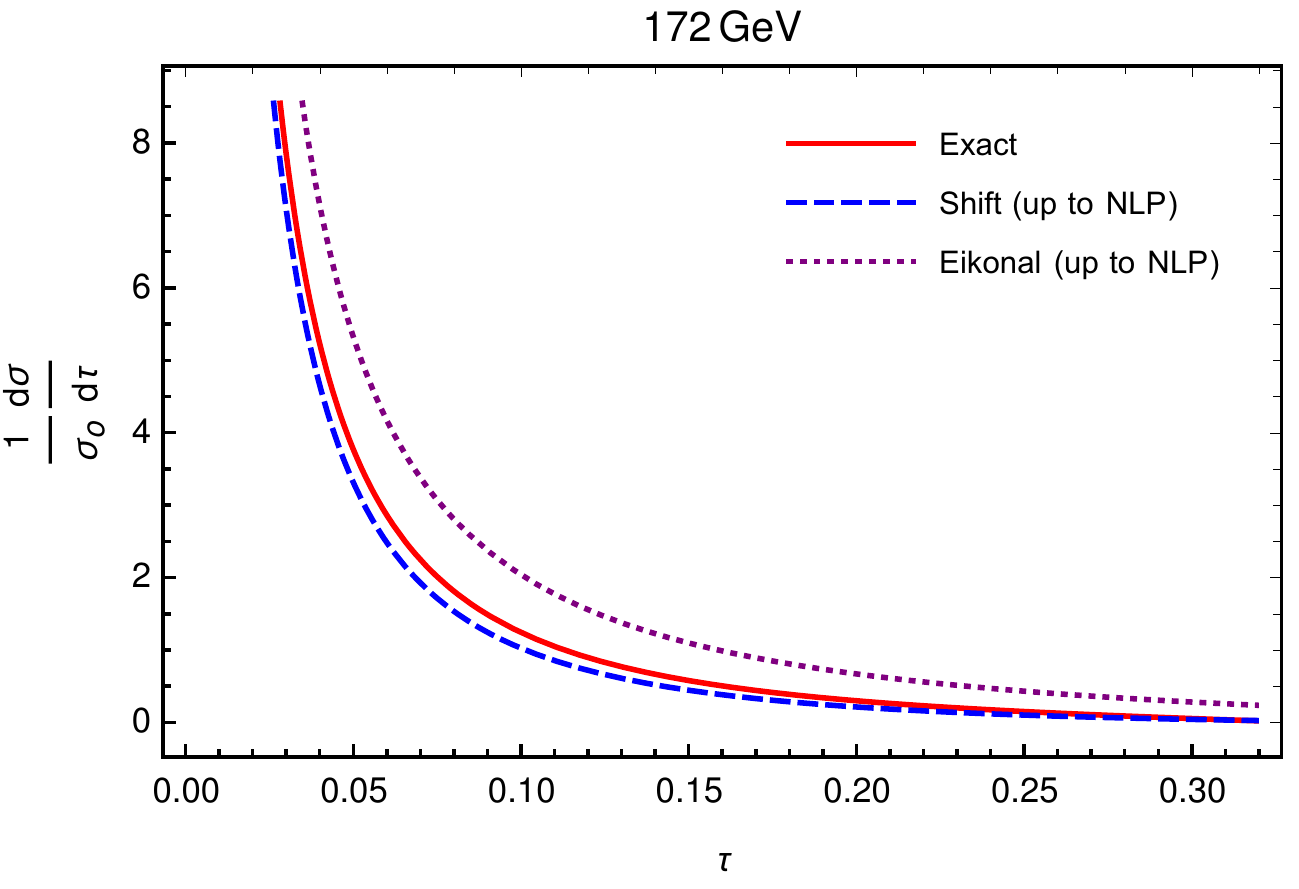}}
	\qquad 
	\subfloat[][]{\includegraphics[scale=0.6]{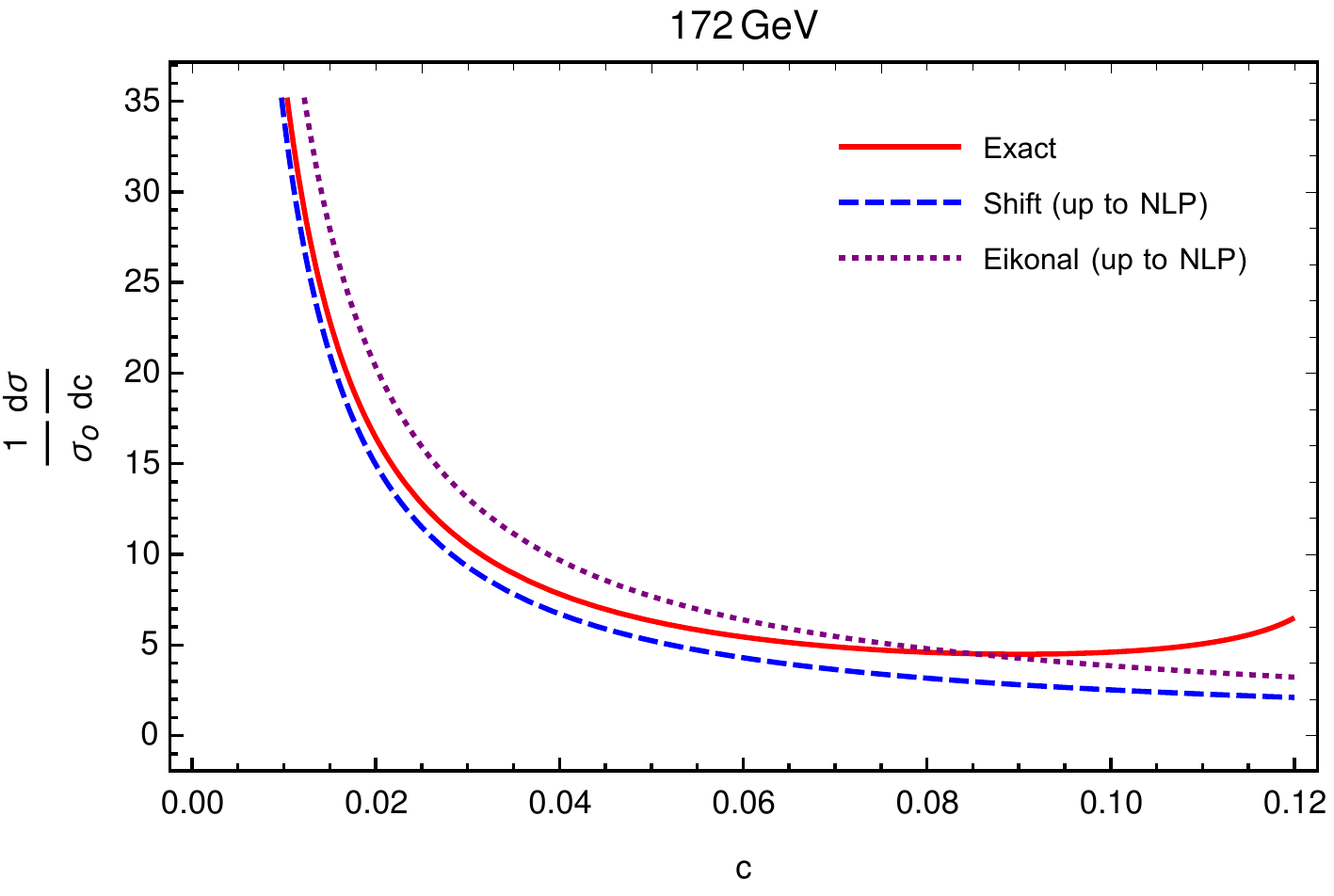}}    
	\caption{ In (a) and (b), we plot the thrust and $ c $-parameter distributions, respectively. We show the exact result (solid red curves) given in eqs.~\eqref{taudstr} and~~\eqref{eq:elliptForm}, the shifted approximation result (dashed blue curves) up to NLP terms given in eqs.~\eqref{nlpfull} and~\eqref{shiftedCPdstr} for $ \tau  $ and  $ c $ respectively. We also plot the LP term (dotted purple curves) given in eqs.~({\ref{eq:thrusteikdstr}}) and~({\ref{eq:cparaeikdstr}}) }
	\label{plot:eikonal_Combined_PLot}
\end{figure*}
\subsection{Eikonal approximation to thrust and $c$-parameter distribution}

In sections~\ref{sec:numer-assess-eikon-thrust}
and~\ref{sec:numer-assess-eikon-cp}, we compared the shifted
approximation with the LP expression of the exact distribution in 
figs.~(\ref{plot:Thrust_Combined_PLot}\textcolor{prd_blue}{a}) and~(\ref{plot:CPARA_Combined_PLot}\textcolor{prd_blue}{a}), for
the thrust and $c $-parameter distribution, respectively. Here
we add the results from the simpler eikonal approximation to the comparison.
The eikonal case follows from the approximated matrix element squared
\begin{align}
\overline{\sum} |\mathcal{M}_{\text{eik}}(x_1,x_2)|^2 \,= \, & 8(e^2e_q)^2  g_s^2 C_F N_c \frac{1}{3Q^2} \nn
\\ & \times \left( \frac{2}{(1-x_1)(1-x_2)}\right)\,.
	\label{meikonal}
\end{align}

Using the above expression and the exact definitions of thrust and
$c$-parameter in eqs.~(\ref{thrustDef}) and~(\ref{rana1}), one finds
for their respective distributions 
\begin{align}
	\frac{1}{\sigma_{0}(s)} \frac{d\sigma}{d\tau} \Bigg\vert_{\text{NLO}}&  \,=\,\frac{2 \alpha_{s}}{3 \pi}\left(\frac{-4\log{\tau}}{\tau}-8-4\log{\tau} + \mathcal{O}(\tau)  \right), \label{eq:thrusteikdstr} \\
	\frac{1}{\sigma_{0}(s)} \frac{d\sigma}{dc} \Bigg\vert_{\text{NLO}} & \,=\,\frac{2 \alpha_{s}}{3 \pi}\left(\frac{-4\log{c}}{c}-8-40\log{c} + \mathcal{O}(c)  \right) \label{eq:cparaeikdstr}. 
\end{align}

The eikonal matrix element squared generates LL terms at LP correctly,
together with some LL and NLL at NLP. It does not
capture any NLL term at LP since it lacks contributions from hard-collinear gluon emission. In
figs.~(\ref{plot:eikonal_Combined_PLot}\textcolor{prd_blue}{a}) and~(\ref{plot:eikonal_Combined_PLot}\textcolor{prd_blue}{b}), we plot
these eikonal results together with 
the thrust and c-parameter distributions computed from the exact
approach as given in eq.~\eqref{taudstr} and eq.~(\ref{eq:elliptForm})
along with the shifted approximation results up to NLP from
eqs.~\eqref{nlpfull} and~(\ref{shiftedCPdstr}). 
Clearly for both event shapes the shifted kinematics methods provides
a significantly better approximation than the eikonal approximation.


\subsection{Table of results}

In table \ref{distributionTable}, we summarize our results for thrust and the
$c$-parameter using the different approximations of the matrix elements squared:
eikonal, exact, shifted kinematics, as well as the remainder/soft
quark part.

\bibliographystyle{apsrev4-2}

\bibliography{aapmsamp}

\providecommand{\noopsort}[1]{}\providecommand{\singleletter}[1]{#1}%
\begin{thebibliography}{112}%
\makeatletter
\providecommand \@ifxundefined [1]{%
 \@ifx{#1\undefined}
}%
\providecommand \@ifnum [1]{%
 \ifnum #1\expandafter \@firstoftwo
 \else \expandafter \@secondoftwo
 \fi
}%
\providecommand \@ifx [1]{%
 \ifx #1\expandafter \@firstoftwo
 \else \expandafter \@secondoftwo
 \fi
}%
\providecommand \natexlab [1]{#1}%
\providecommand \enquote  [1]{``#1''}%
\providecommand \bibnamefont  [1]{#1}%
\providecommand \bibfnamefont [1]{#1}%
\providecommand \citenamefont [1]{#1}%
\providecommand \href@noop [0]{\@secondoftwo}%
\providecommand \href [0]{\begingroup \@sanitize@url \@href}%
\providecommand \@href[1]{\@@startlink{#1}\@@href}%
\providecommand \@@href[1]{\endgroup#1\@@endlink}%
\providecommand \@sanitize@url [0]{\catcode `\\12\catcode `\$12\catcode
  `\&12\catcode `\#12\catcode `\^12\catcode `\_12\catcode `\%12\relax}%
\providecommand \@@startlink[1]{}%
\providecommand \@@endlink[0]{}%
\providecommand \url  [0]{\begingroup\@sanitize@url \@url }%
\providecommand \@url [1]{\endgroup\@href {#1}{\urlprefix }}%
\providecommand \urlprefix  [0]{URL }%
\providecommand \Eprint [0]{\href }%
\providecommand \doibase [0]{https://doi.org/}%
\providecommand \selectlanguage [0]{\@gobble}%
\providecommand \bibinfo  [0]{\@secondoftwo}%
\providecommand \bibfield  [0]{\@secondoftwo}%
\providecommand \translation [1]{[#1]}%
\providecommand \BibitemOpen [0]{}%
\providecommand \bibitemStop [0]{}%
\providecommand \bibitemNoStop [0]{.\EOS\space}%
\providecommand \EOS [0]{\spacefactor3000\relax}%
\providecommand \BibitemShut  [1]{\csname bibitem#1\endcsname}%
\let\auto@bib@innerbib\@empty
\bibitem [{\citenamefont {Kinoshita}\ and\ \citenamefont
  {Ukawa}(1975)}]{Kinoshita:1975bt}%
  \BibitemOpen
  \bibfield  {author} {\bibinfo {author} {\bibfnamefont {T.}~\bibnamefont
  {Kinoshita}}\ and\ \bibinfo {author} {\bibfnamefont {A.}~\bibnamefont
  {Ukawa}},\ }\href {https://doi.org/10.1007/BFb0013300} {\bibfield  {journal}
  {\bibinfo  {journal} {Lect. Notes Phys.}\ }\textbf {\bibinfo {volume} {39}},\
  \bibinfo {pages} {55} (\bibinfo {year} {1975})}\BibitemShut {NoStop}%
\bibitem [{\citenamefont {Lee}\ and\ \citenamefont
  {Nauenberg}(1964)}]{PhysRev.133.B1549}%
  \BibitemOpen
  \bibfield  {author} {\bibinfo {author} {\bibfnamefont {T.~D.}\ \bibnamefont
  {Lee}}\ and\ \bibinfo {author} {\bibfnamefont {M.}~\bibnamefont
  {Nauenberg}},\ }\href {https://doi.org/10.1103/PhysRev.133.B1549} {\bibfield
  {journal} {\bibinfo  {journal} {Phys. Rev.}\ }\textbf {\bibinfo {volume}
  {133}},\ \bibinfo {pages} {B1549} (\bibinfo {year} {1964})}\BibitemShut
  {NoStop}%
\bibitem [{\citenamefont {Parisi}(1980)}]{Parisi:1979xd}%
  \BibitemOpen
  \bibfield  {author} {\bibinfo {author} {\bibfnamefont {G.}~\bibnamefont
  {Parisi}},\ }\href {https://doi.org/10.1016/0370-2693(80)90746-7} {\bibfield
  {journal} {\bibinfo  {journal} {Phys. Lett. B}\ }\textbf {\bibinfo {volume}
  {90}},\ \bibinfo {pages} {295} (\bibinfo {year} {1980})}\BibitemShut
  {NoStop}%
\bibitem [{\citenamefont {Curci}\ and\ \citenamefont
  {Greco}(1980)}]{Curci:1979am}%
  \BibitemOpen
  \bibfield  {author} {\bibinfo {author} {\bibfnamefont {G.}~\bibnamefont
  {Curci}}\ and\ \bibinfo {author} {\bibfnamefont {M.}~\bibnamefont {Greco}},\
  }\href {https://doi.org/10.1016/0370-2693(80)90331-7} {\bibfield  {journal}
  {\bibinfo  {journal} {Phys. Lett. B}\ }\textbf {\bibinfo {volume} {92}},\
  \bibinfo {pages} {175} (\bibinfo {year} {1980})}\BibitemShut {NoStop}%
\bibitem [{\citenamefont {Sterman}(1987)}]{STERMAN1987310}%
  \BibitemOpen
  \bibfield  {author} {\bibinfo {author} {\bibfnamefont {G.}~\bibnamefont
  {Sterman}},\ }\href
  {https://doi.org/https://doi.org/10.1016/0550-3213(87)90258-6} {\bibfield
  {journal} {\bibinfo  {journal} {Nuclear Physics B}\ }\textbf {\bibinfo
  {volume} {281}},\ \bibinfo {pages} {310} (\bibinfo {year}
  {1987})}\BibitemShut {NoStop}%
\bibitem [{\citenamefont {Catani}\ and\ \citenamefont
  {Trentadue}(1989)}]{Catani:1989ne}%
  \BibitemOpen
  \bibfield  {author} {\bibinfo {author} {\bibfnamefont {S.}~\bibnamefont
  {Catani}}\ and\ \bibinfo {author} {\bibfnamefont {L.}~\bibnamefont
  {Trentadue}},\ }\href {https://doi.org/10.1016/0550-3213(89)90273-3}
  {\bibfield  {journal} {\bibinfo  {journal} {Nucl. Phys. B}\ }\textbf
  {\bibinfo {volume} {327}},\ \bibinfo {pages} {323} (\bibinfo {year}
  {1989})}\BibitemShut {NoStop}%
\bibitem [{\citenamefont {Catani}\ and\ \citenamefont
  {Trentadue}(1991)}]{Catani:1990rp}%
  \BibitemOpen
  \bibfield  {author} {\bibinfo {author} {\bibfnamefont {S.}~\bibnamefont
  {Catani}}\ and\ \bibinfo {author} {\bibfnamefont {L.}~\bibnamefont
  {Trentadue}},\ }\href {https://doi.org/10.1016/0550-3213(91)90506-S}
  {\bibfield  {journal} {\bibinfo  {journal} {Nucl. Phys. B}\ }\textbf
  {\bibinfo {volume} {353}},\ \bibinfo {pages} {183} (\bibinfo {year}
  {1991})}\BibitemShut {NoStop}%
\bibitem [{\citenamefont {Gatheral}(1983)}]{Gatheral:1983cz}%
  \BibitemOpen
  \bibfield  {author} {\bibinfo {author} {\bibfnamefont {J.~G.~M.}\
  \bibnamefont {Gatheral}},\ }\href
  {https://doi.org/10.1016/0370-2693(83)90112-0} {\bibfield  {journal}
  {\bibinfo  {journal} {Phys. Lett. B}\ }\textbf {\bibinfo {volume} {133}},\
  \bibinfo {pages} {90} (\bibinfo {year} {1983})}\BibitemShut {NoStop}%
\bibitem [{\citenamefont {Frenkel}\ and\ \citenamefont
  {Taylor}(1984)}]{Frenkel:1984pz}%
  \BibitemOpen
  \bibfield  {author} {\bibinfo {author} {\bibfnamefont {J.}~\bibnamefont
  {Frenkel}}\ and\ \bibinfo {author} {\bibfnamefont {J.~C.}\ \bibnamefont
  {Taylor}},\ }\href {https://doi.org/10.1016/0550-3213(84)90294-3} {\bibfield
  {journal} {\bibinfo  {journal} {Nucl. Phys. B}\ }\textbf {\bibinfo {volume}
  {246}},\ \bibinfo {pages} {231} (\bibinfo {year} {1984})}\BibitemShut
  {NoStop}%
\bibitem [{\citenamefont {Sterman}(1981)}]{Sterman:1981jc}%
  \BibitemOpen
  \bibfield  {author} {\bibinfo {author} {\bibfnamefont {G.~F.}\ \bibnamefont
  {Sterman}},\ }\href {https://doi.org/10.1063/1.33099} {\bibfield  {journal}
  {\bibinfo  {journal} {AIP Conf. Proc.}\ }\textbf {\bibinfo {volume} {74}},\
  \bibinfo {pages} {22} (\bibinfo {year} {1981})}\BibitemShut {NoStop}%
\bibitem [{\citenamefont {Korchemsky}\ and\ \citenamefont
  {Marchesini}(1993{\natexlab{a}})}]{Korchemsky:1992xv}%
  \BibitemOpen
  \bibfield  {author} {\bibinfo {author} {\bibfnamefont {G.~P.}\ \bibnamefont
  {Korchemsky}}\ and\ \bibinfo {author} {\bibfnamefont {G.}~\bibnamefont
  {Marchesini}},\ }\href {https://doi.org/10.1016/0550-3213(93)90167-N}
  {\bibfield  {journal} {\bibinfo  {journal} {Nucl. Phys. B}\ }\textbf
  {\bibinfo {volume} {406}},\ \bibinfo {pages} {225} (\bibinfo {year}
  {1993}{\natexlab{a}})},\ \Eprint {https://arxiv.org/abs/hep-ph/9210281}
  {arXiv:hep-ph/9210281} \BibitemShut {NoStop}%
\bibitem [{\citenamefont {Korchemsky}\ and\ \citenamefont
  {Marchesini}(1993{\natexlab{b}})}]{Korchemsky:1993uz}%
  \BibitemOpen
  \bibfield  {author} {\bibinfo {author} {\bibfnamefont {G.~P.}\ \bibnamefont
  {Korchemsky}}\ and\ \bibinfo {author} {\bibfnamefont {G.}~\bibnamefont
  {Marchesini}},\ }\href {https://doi.org/10.1016/0370-2693(93)90015-A}
  {\bibfield  {journal} {\bibinfo  {journal} {Phys. Lett. B}\ }\textbf
  {\bibinfo {volume} {313}},\ \bibinfo {pages} {433} (\bibinfo {year}
  {1993}{\natexlab{b}})}\BibitemShut {NoStop}%
\bibitem [{\citenamefont {Forte}\ and\ \citenamefont
  {Ridolfi}(2003)}]{Forte:2002ni}%
  \BibitemOpen
  \bibfield  {author} {\bibinfo {author} {\bibfnamefont {S.}~\bibnamefont
  {Forte}}\ and\ \bibinfo {author} {\bibfnamefont {G.}~\bibnamefont
  {Ridolfi}},\ }\href {https://doi.org/10.1016/S0550-3213(02)01034-9}
  {\bibfield  {journal} {\bibinfo  {journal} {Nucl. Phys. B}\ }\textbf
  {\bibinfo {volume} {650}},\ \bibinfo {pages} {229} (\bibinfo {year}
  {2003})},\ \Eprint {https://arxiv.org/abs/hep-ph/0209154}
  {arXiv:hep-ph/0209154} \BibitemShut {NoStop}%
\bibitem [{\citenamefont {Contopanagos}\ \emph {et~al.}(1997)\citenamefont
  {Contopanagos}, \citenamefont {Laenen},\ and\ \citenamefont
  {Sterman}}]{Contopanagos:1996nh}%
  \BibitemOpen
  \bibfield  {author} {\bibinfo {author} {\bibfnamefont {H.}~\bibnamefont
  {Contopanagos}}, \bibinfo {author} {\bibfnamefont {E.}~\bibnamefont
  {Laenen}},\ and\ \bibinfo {author} {\bibfnamefont {G.~F.}\ \bibnamefont
  {Sterman}},\ }\href {https://doi.org/10.1016/S0550-3213(96)00567-6}
  {\bibfield  {journal} {\bibinfo  {journal} {Nucl. Phys. B}\ }\textbf
  {\bibinfo {volume} {484}},\ \bibinfo {pages} {303} (\bibinfo {year}
  {1997})},\ \Eprint {https://arxiv.org/abs/hep-ph/9604313}
  {arXiv:hep-ph/9604313} \BibitemShut {NoStop}%
\bibitem [{\citenamefont {Becher}\ and\ \citenamefont
  {Neubert}(2006)}]{Becher:2006nr}%
  \BibitemOpen
  \bibfield  {author} {\bibinfo {author} {\bibfnamefont {T.}~\bibnamefont
  {Becher}}\ and\ \bibinfo {author} {\bibfnamefont {M.}~\bibnamefont
  {Neubert}},\ }\href {https://doi.org/10.1103/PhysRevLett.97.082001}
  {\bibfield  {journal} {\bibinfo  {journal} {Phys. Rev. Lett.}\ }\textbf
  {\bibinfo {volume} {97}},\ \bibinfo {pages} {082001} (\bibinfo {year}
  {2006})},\ \Eprint {https://arxiv.org/abs/hep-ph/0605050}
  {arXiv:hep-ph/0605050} \BibitemShut {NoStop}%
\bibitem [{\citenamefont {Schwartz}(2008)}]{Schwartz:2007ib}%
  \BibitemOpen
  \bibfield  {author} {\bibinfo {author} {\bibfnamefont {M.~D.}\ \bibnamefont
  {Schwartz}},\ }\href {https://doi.org/10.1103/PhysRevD.77.014026} {\bibfield
  {journal} {\bibinfo  {journal} {Phys.Rev.}\ }\textbf {\bibinfo {volume}
  {D77}},\ \bibinfo {pages} {014026} (\bibinfo {year} {2008})}\BibitemShut
  {NoStop}%
\bibitem [{\citenamefont {Bauer}\ \emph {et~al.}(2008)\citenamefont {Bauer},
  \citenamefont {Fleming}, \citenamefont {Lee},\ and\ \citenamefont
  {Sterman}}]{PhysRevD.78.034027}%
  \BibitemOpen
  \bibfield  {author} {\bibinfo {author} {\bibfnamefont {C.~W.}\ \bibnamefont
  {Bauer}}, \bibinfo {author} {\bibfnamefont {S.}~\bibnamefont {Fleming}},
  \bibinfo {author} {\bibfnamefont {C.}~\bibnamefont {Lee}},\ and\ \bibinfo
  {author} {\bibfnamefont {G.}~\bibnamefont {Sterman}},\ }\href
  {https://doi.org/10.1103/PhysRevD.78.034027} {\bibfield  {journal} {\bibinfo
  {journal} {Phys. Rev. D}\ }\textbf {\bibinfo {volume} {78}},\ \bibinfo
  {pages} {034027} (\bibinfo {year} {2008})}\BibitemShut {NoStop}%
\bibitem [{\citenamefont {Chiu}\ \emph {et~al.}(2009)\citenamefont {Chiu},
  \citenamefont {Fuhrer}, \citenamefont {Kelley},\ and\ \citenamefont
  {Manohar}}]{PhysRevD.80.094013}%
  \BibitemOpen
  \bibfield  {author} {\bibinfo {author} {\bibfnamefont {J.-y.}\ \bibnamefont
  {Chiu}}, \bibinfo {author} {\bibfnamefont {A.}~\bibnamefont {Fuhrer}},
  \bibinfo {author} {\bibfnamefont {R.}~\bibnamefont {Kelley}},\ and\ \bibinfo
  {author} {\bibfnamefont {A.~V.}\ \bibnamefont {Manohar}},\ }\href
  {https://doi.org/10.1103/PhysRevD.80.094013} {\bibfield  {journal} {\bibinfo
  {journal} {Phys. Rev. D}\ }\textbf {\bibinfo {volume} {80}},\ \bibinfo
  {pages} {094013} (\bibinfo {year} {2009})}\BibitemShut {NoStop}%
\bibitem [{\citenamefont {Gardi}(2002)}]{Gardi:2001di}%
  \BibitemOpen
  \bibfield  {author} {\bibinfo {author} {\bibfnamefont {E.}~\bibnamefont
  {Gardi}},\ }\href {https://doi.org/10.1016/S0550-3213(01)00594-6} {\bibfield
  {journal} {\bibinfo  {journal} {Nucl. Phys. B}\ }\textbf {\bibinfo {volume}
  {622}},\ \bibinfo {pages} {365} (\bibinfo {year} {2002})},\ \Eprint
  {https://arxiv.org/abs/hep-ph/0108222} {arXiv:hep-ph/0108222} \BibitemShut
  {NoStop}%
\bibitem [{\citenamefont {Agarwal}\ \emph {et~al.}(2021)\citenamefont
  {Agarwal}, \citenamefont {Mukhopadhyay}, \citenamefont {Pal},\ and\
  \citenamefont {Tripathi}}]{Agarwal:2020uxi}%
  \BibitemOpen
  \bibfield  {author} {\bibinfo {author} {\bibfnamefont {N.}~\bibnamefont
  {Agarwal}}, \bibinfo {author} {\bibfnamefont {A.}~\bibnamefont
  {Mukhopadhyay}}, \bibinfo {author} {\bibfnamefont {S.}~\bibnamefont {Pal}},\
  and\ \bibinfo {author} {\bibfnamefont {A.}~\bibnamefont {Tripathi}},\ }\href
  {https://doi.org/10.1007/JHEP03(2021)155} {\bibfield  {journal} {\bibinfo
  {journal} {JHEP}\ }\textbf {\bibinfo {volume} {03}},\ \bibinfo {pages}
  {155}},\ \Eprint {https://arxiv.org/abs/2012.06842} {arXiv:2012.06842
  [hep-ph]} \BibitemShut {NoStop}%
\bibitem [{\citenamefont {Laenen}(2004)}]{Laenen:2004pm}%
  \BibitemOpen
  \bibfield  {author} {\bibinfo {author} {\bibfnamefont {E.}~\bibnamefont
  {Laenen}},\ }\href {https://doi.org/10.1007/BF02704892} {\bibfield  {journal}
  {\bibinfo  {journal} {Pramana}\ }\textbf {\bibinfo {volume} {63}},\ \bibinfo
  {pages} {1225} (\bibinfo {year} {2004})}\BibitemShut {NoStop}%
\bibitem [{\citenamefont {Luisoni}\ and\ \citenamefont
  {Marzani}(2015)}]{Luisoni_2015}%
  \BibitemOpen
  \bibfield  {author} {\bibinfo {author} {\bibfnamefont {G.}~\bibnamefont
  {Luisoni}}\ and\ \bibinfo {author} {\bibfnamefont {S.}~\bibnamefont
  {Marzani}},\ }\href {https://doi.org/10.1088/0954-3899/42/10/103101}
  {\bibfield  {journal} {\bibinfo  {journal} {Journal of Physics G: Nuclear and
  Particle Physics}\ }\textbf {\bibinfo {volume} {42}},\ \bibinfo {pages}
  {103101} (\bibinfo {year} {2015})}\BibitemShut {NoStop}%
\bibitem [{\citenamefont {Becher}\ \emph {et~al.}(2015)\citenamefont {Becher},
  \citenamefont {Broggio},\ and\ \citenamefont {Ferroglia}}]{Becher:2014oda}%
  \BibitemOpen
  \bibfield  {author} {\bibinfo {author} {\bibfnamefont {T.}~\bibnamefont
  {Becher}}, \bibinfo {author} {\bibfnamefont {A.}~\bibnamefont {Broggio}},\
  and\ \bibinfo {author} {\bibfnamefont {A.}~\bibnamefont {Ferroglia}},\ }\href
  {https://doi.org/10.1007/978-3-319-14848-9} {\emph {\bibinfo {title}
  {{Introduction to Soft-Collinear Effective Theory}}}},\ Vol.\ \bibinfo
  {volume} {896}\ (\bibinfo  {publisher} {Springer},\ \bibinfo {year} {2015})\
  \Eprint {https://arxiv.org/abs/1410.1892} {arXiv:1410.1892 [hep-ph]}
  \BibitemShut {NoStop}%
\bibitem [{\citenamefont {Campbell}\ \emph {et~al.}(2017)\citenamefont
  {Campbell}, \citenamefont {Huston},\ and\ \citenamefont
  {Krauss}}]{Campbell:2017hsr}%
  \BibitemOpen
  \bibfield  {author} {\bibinfo {author} {\bibfnamefont {J.}~\bibnamefont
  {Campbell}}, \bibinfo {author} {\bibfnamefont {J.}~\bibnamefont {Huston}},\
  and\ \bibinfo {author} {\bibfnamefont {F.}~\bibnamefont {Krauss}},\
  }\href@noop {} {\emph {\bibinfo {title} {{The Black Book of Quantum
  Chromodynamics}: {A Primer for the LHC Era}}}}\ (\bibinfo  {publisher}
  {Oxford University Press},\ \bibinfo {year} {2017})\BibitemShut {NoStop}%
\bibitem [{\citenamefont {Agarwal}\ \emph {et~al.}(2023)\citenamefont
  {Agarwal}, \citenamefont {Magnea}, \citenamefont {Signorile-Signorile},\ and\
  \citenamefont {Tripathi}}]{Agarwal:2021ais}%
  \BibitemOpen
  \bibfield  {author} {\bibinfo {author} {\bibfnamefont {N.}~\bibnamefont
  {Agarwal}}, \bibinfo {author} {\bibfnamefont {L.}~\bibnamefont {Magnea}},
  \bibinfo {author} {\bibfnamefont {C.}~\bibnamefont {Signorile-Signorile}},\
  and\ \bibinfo {author} {\bibfnamefont {A.}~\bibnamefont {Tripathi}},\ }\href
  {https://doi.org/10.1016/j.physrep.2022.10.001} {\bibfield  {journal}
  {\bibinfo  {journal} {Phys. Rept.}\ }\textbf {\bibinfo {volume} {994}},\
  \bibinfo {pages} {1} (\bibinfo {year} {2023})},\ \Eprint
  {https://arxiv.org/abs/2112.07099} {arXiv:2112.07099 [hep-ph]} \BibitemShut
  {NoStop}%
\bibitem [{\citenamefont {Bauer}\ \emph {et~al.}(2000)\citenamefont {Bauer},
  \citenamefont {Fleming},\ and\ \citenamefont {Luke}}]{Bauer:2000ew}%
  \BibitemOpen
  \bibfield  {author} {\bibinfo {author} {\bibfnamefont {C.~W.}\ \bibnamefont
  {Bauer}}, \bibinfo {author} {\bibfnamefont {S.}~\bibnamefont {Fleming}},\
  and\ \bibinfo {author} {\bibfnamefont {M.~E.}\ \bibnamefont {Luke}},\ }\href
  {https://doi.org/10.1103/PhysRevD.63.014006} {\bibfield  {journal} {\bibinfo
  {journal} {Phys. Rev. D}\ }\textbf {\bibinfo {volume} {63}},\ \bibinfo
  {pages} {014006} (\bibinfo {year} {2000})},\ \Eprint
  {https://arxiv.org/abs/hep-ph/0005275} {arXiv:hep-ph/0005275} \BibitemShut
  {NoStop}%
\bibitem [{\citenamefont {Bauer}\ and\ \citenamefont
  {Stewart}(2001)}]{Bauer:2001ct}%
  \BibitemOpen
  \bibfield  {author} {\bibinfo {author} {\bibfnamefont {C.~W.}\ \bibnamefont
  {Bauer}}\ and\ \bibinfo {author} {\bibfnamefont {I.~W.}\ \bibnamefont
  {Stewart}},\ }\href {https://doi.org/10.1016/S0370-2693(01)00902-9}
  {\bibfield  {journal} {\bibinfo  {journal} {Phys. Lett. B}\ }\textbf
  {\bibinfo {volume} {516}},\ \bibinfo {pages} {134} (\bibinfo {year}
  {2001})},\ \Eprint {https://arxiv.org/abs/hep-ph/0107001}
  {arXiv:hep-ph/0107001} \BibitemShut {NoStop}%
\bibitem [{\citenamefont {Bauer}\ \emph {et~al.}(2001)\citenamefont {Bauer},
  \citenamefont {Fleming}, \citenamefont {Pirjol},\ and\ \citenamefont
  {Stewart}}]{Bauer:2000yr}%
  \BibitemOpen
  \bibfield  {author} {\bibinfo {author} {\bibfnamefont {C.~W.}\ \bibnamefont
  {Bauer}}, \bibinfo {author} {\bibfnamefont {S.}~\bibnamefont {Fleming}},
  \bibinfo {author} {\bibfnamefont {D.}~\bibnamefont {Pirjol}},\ and\ \bibinfo
  {author} {\bibfnamefont {I.~W.}\ \bibnamefont {Stewart}},\ }\href
  {https://doi.org/10.1103/PhysRevD.63.114020} {\bibfield  {journal} {\bibinfo
  {journal} {Phys. Rev. D}\ }\textbf {\bibinfo {volume} {63}},\ \bibinfo
  {pages} {114020} (\bibinfo {year} {2001})},\ \Eprint
  {https://arxiv.org/abs/hep-ph/0011336} {arXiv:hep-ph/0011336} \BibitemShut
  {NoStop}%
\bibitem [{\citenamefont {Bauer}\ \emph {et~al.}(2002)\citenamefont {Bauer},
  \citenamefont {Pirjol},\ and\ \citenamefont {Stewart}}]{Bauer:2001yt}%
  \BibitemOpen
  \bibfield  {author} {\bibinfo {author} {\bibfnamefont {C.~W.}\ \bibnamefont
  {Bauer}}, \bibinfo {author} {\bibfnamefont {D.}~\bibnamefont {Pirjol}},\ and\
  \bibinfo {author} {\bibfnamefont {I.~W.}\ \bibnamefont {Stewart}},\ }\href
  {https://doi.org/10.1103/PhysRevD.65.054022} {\bibfield  {journal} {\bibinfo
  {journal} {Phys. Rev. D}\ }\textbf {\bibinfo {volume} {65}},\ \bibinfo
  {pages} {054022} (\bibinfo {year} {2002})},\ \Eprint
  {https://arxiv.org/abs/hep-ph/0109045} {arXiv:hep-ph/0109045} \BibitemShut
  {NoStop}%
\bibitem [{\citenamefont {Bonocore}\ \emph {et~al.}(2015)\citenamefont
  {Bonocore}, \citenamefont {Laenen}, \citenamefont {Magnea}, \citenamefont
  {Melville}, \citenamefont {Vernazza},\ and\ \citenamefont
  {White}}]{Bonocore2015}%
  \BibitemOpen
  \bibfield  {author} {\bibinfo {author} {\bibfnamefont {D.}~\bibnamefont
  {Bonocore}}, \bibinfo {author} {\bibfnamefont {E.}~\bibnamefont {Laenen}},
  \bibinfo {author} {\bibfnamefont {L.}~\bibnamefont {Magnea}}, \bibinfo
  {author} {\bibfnamefont {S.}~\bibnamefont {Melville}}, \bibinfo {author}
  {\bibfnamefont {L.}~\bibnamefont {Vernazza}},\ and\ \bibinfo {author}
  {\bibfnamefont {C.~D.}\ \bibnamefont {White}},\ }\href
  {https://doi.org/10.1007/JHEP06(2015)008} {\bibfield  {journal} {\bibinfo
  {journal} {Journal of High Energy Physics}\ }\textbf {\bibinfo {volume}
  {2015}},\ \bibinfo {pages} {8} (\bibinfo {year} {2015})}\BibitemShut
  {NoStop}%
\bibitem [{\citenamefont {Bonocore}\ \emph {et~al.}(2016)\citenamefont
  {Bonocore}, \citenamefont {Laenen}, \citenamefont {Magnea}, \citenamefont
  {Vernazza},\ and\ \citenamefont {White}}]{Bonocore2016}%
  \BibitemOpen
  \bibfield  {author} {\bibinfo {author} {\bibfnamefont {D.}~\bibnamefont
  {Bonocore}}, \bibinfo {author} {\bibfnamefont {E.}~\bibnamefont {Laenen}},
  \bibinfo {author} {\bibfnamefont {L.}~\bibnamefont {Magnea}}, \bibinfo
  {author} {\bibfnamefont {L.}~\bibnamefont {Vernazza}},\ and\ \bibinfo
  {author} {\bibfnamefont {C.~D.}\ \bibnamefont {White}},\ }\href
  {https://doi.org/10.1007/JHEP12(2016)121} {\bibfield  {journal} {\bibinfo
  {journal} {Journal of High Energy Physics}\ }\textbf {\bibinfo {volume}
  {2016}},\ \bibinfo {pages} {121} (\bibinfo {year} {2016})}\BibitemShut
  {NoStop}%
\bibitem [{\citenamefont {Moult}\ \emph {et~al.}(2018)\citenamefont {Moult},
  \citenamefont {Stewart}, \citenamefont {Vita},\ and\ \citenamefont
  {Zhu}}]{Moult:2018jjd}%
  \BibitemOpen
  \bibfield  {author} {\bibinfo {author} {\bibfnamefont {I.}~\bibnamefont
  {Moult}}, \bibinfo {author} {\bibfnamefont {I.~W.}\ \bibnamefont {Stewart}},
  \bibinfo {author} {\bibfnamefont {G.}~\bibnamefont {Vita}},\ and\ \bibinfo
  {author} {\bibfnamefont {H.~X.}\ \bibnamefont {Zhu}},\ }\href
  {https://doi.org/10.1007/JHEP08(2018)013} {\bibfield  {journal} {\bibinfo
  {journal} {JHEP}\ }\textbf {\bibinfo {volume} {08}},\ \bibinfo {pages}
  {013}},\ \Eprint {https://arxiv.org/abs/1804.04665} {arXiv:1804.04665
  [hep-ph]} \BibitemShut {NoStop}%
\bibitem [{\citenamefont {Beneke}\ \emph
  {et~al.}(2020{\natexlab{a}})\citenamefont {Beneke}, \citenamefont {Broggio},
  \citenamefont {Jaskiewicz},\ and\ \citenamefont {Vernazza}}]{Beneke:2019oqx}%
  \BibitemOpen
  \bibfield  {author} {\bibinfo {author} {\bibfnamefont {M.}~\bibnamefont
  {Beneke}}, \bibinfo {author} {\bibfnamefont {A.}~\bibnamefont {Broggio}},
  \bibinfo {author} {\bibfnamefont {S.}~\bibnamefont {Jaskiewicz}},\ and\
  \bibinfo {author} {\bibfnamefont {L.}~\bibnamefont {Vernazza}},\ }\href
  {https://doi.org/10.1007/JHEP07(2020)078} {\bibfield  {journal} {\bibinfo
  {journal} {JHEP}\ }\textbf {\bibinfo {volume} {07}},\ \bibinfo {pages}
  {078}},\ \Eprint {https://arxiv.org/abs/1912.01585} {arXiv:1912.01585
  [hep-ph]} \BibitemShut {NoStop}%
\bibitem [{\citenamefont {Moult}\ \emph {et~al.}(2019)\citenamefont {Moult},
  \citenamefont {Stewart},\ and\ \citenamefont {Vita}}]{Moult:2019mog}%
  \BibitemOpen
  \bibfield  {author} {\bibinfo {author} {\bibfnamefont {I.}~\bibnamefont
  {Moult}}, \bibinfo {author} {\bibfnamefont {I.~W.}\ \bibnamefont {Stewart}},\
  and\ \bibinfo {author} {\bibfnamefont {G.}~\bibnamefont {Vita}},\ }\href
  {https://doi.org/10.1007/JHEP11(2019)153} {\bibfield  {journal} {\bibinfo
  {journal} {JHEP}\ }\textbf {\bibinfo {volume} {11}},\ \bibinfo {pages}
  {153}},\ \Eprint {https://arxiv.org/abs/1905.07411} {arXiv:1905.07411
  [hep-ph]} \BibitemShut {NoStop}%
\bibitem [{\citenamefont {Bahjat-Abbas}\ \emph {et~al.}(2019)\citenamefont
  {Bahjat-Abbas}, \citenamefont {Bonocore}, \citenamefont {Sinninghe~Damst\'e},
  \citenamefont {Laenen}, \citenamefont {Magnea}, \citenamefont {Vernazza},\
  and\ \citenamefont {White}}]{Bahjat-Abbas:2019fqa}%
  \BibitemOpen
  \bibfield  {author} {\bibinfo {author} {\bibfnamefont {N.}~\bibnamefont
  {Bahjat-Abbas}}, \bibinfo {author} {\bibfnamefont {D.}~\bibnamefont
  {Bonocore}}, \bibinfo {author} {\bibfnamefont {J.}~\bibnamefont
  {Sinninghe~Damst\'e}}, \bibinfo {author} {\bibfnamefont {E.}~\bibnamefont
  {Laenen}}, \bibinfo {author} {\bibfnamefont {L.}~\bibnamefont {Magnea}},
  \bibinfo {author} {\bibfnamefont {L.}~\bibnamefont {Vernazza}},\ and\
  \bibinfo {author} {\bibfnamefont {C.~D.}\ \bibnamefont {White}},\ }\href
  {https://doi.org/10.1007/JHEP11(2019)002} {\bibfield  {journal} {\bibinfo
  {journal} {JHEP}\ }\textbf {\bibinfo {volume} {11}},\ \bibinfo {pages}
  {002}},\ \Eprint {https://arxiv.org/abs/1905.13710} {arXiv:1905.13710
  [hep-ph]} \BibitemShut {NoStop}%
\bibitem [{\citenamefont {Beneke}\ \emph
  {et~al.}(2020{\natexlab{b}})\citenamefont {Beneke}, \citenamefont {Garny},
  \citenamefont {Jaskiewicz}, \citenamefont {Szafron}, \citenamefont
  {Vernazza},\ and\ \citenamefont {Wang}}]{Beneke:2019mua}%
  \BibitemOpen
  \bibfield  {author} {\bibinfo {author} {\bibfnamefont {M.}~\bibnamefont
  {Beneke}}, \bibinfo {author} {\bibfnamefont {M.}~\bibnamefont {Garny}},
  \bibinfo {author} {\bibfnamefont {S.}~\bibnamefont {Jaskiewicz}}, \bibinfo
  {author} {\bibfnamefont {R.}~\bibnamefont {Szafron}}, \bibinfo {author}
  {\bibfnamefont {L.}~\bibnamefont {Vernazza}},\ and\ \bibinfo {author}
  {\bibfnamefont {J.}~\bibnamefont {Wang}},\ }\href
  {https://doi.org/10.1007/JHEP01(2020)094} {\bibfield  {journal} {\bibinfo
  {journal} {JHEP}\ }\textbf {\bibinfo {volume} {01}},\ \bibinfo {pages}
  {094}},\ \Eprint {https://arxiv.org/abs/1910.12685} {arXiv:1910.12685
  [hep-ph]} \BibitemShut {NoStop}%
\bibitem [{\citenamefont {Moult}\ \emph
  {et~al.}(2020{\natexlab{a}})\citenamefont {Moult}, \citenamefont {Vita},\
  and\ \citenamefont {Yan}}]{Moult:2019vou}%
  \BibitemOpen
  \bibfield  {author} {\bibinfo {author} {\bibfnamefont {I.}~\bibnamefont
  {Moult}}, \bibinfo {author} {\bibfnamefont {G.}~\bibnamefont {Vita}},\ and\
  \bibinfo {author} {\bibfnamefont {K.}~\bibnamefont {Yan}},\ }\href
  {https://doi.org/10.1007/JHEP07(2020)005} {\bibfield  {journal} {\bibinfo
  {journal} {JHEP}\ }\textbf {\bibinfo {volume} {07}},\ \bibinfo {pages}
  {005}},\ \Eprint {https://arxiv.org/abs/1912.02188} {arXiv:1912.02188
  [hep-ph]} \BibitemShut {NoStop}%
\bibitem [{\citenamefont {Ajjath}\ \emph {et~al.}(2021)\citenamefont {Ajjath},
  \citenamefont {Mukherjee}, \citenamefont {Ravindran}, \citenamefont
  {Sankar},\ and\ \citenamefont {Tiwari}}]{Ajjath:2020sjk}%
  \BibitemOpen
  \bibfield  {author} {\bibinfo {author} {\bibfnamefont {A.~H.}\ \bibnamefont
  {Ajjath}}, \bibinfo {author} {\bibfnamefont {P.}~\bibnamefont {Mukherjee}},
  \bibinfo {author} {\bibfnamefont {V.}~\bibnamefont {Ravindran}}, \bibinfo
  {author} {\bibfnamefont {A.}~\bibnamefont {Sankar}},\ and\ \bibinfo {author}
  {\bibfnamefont {S.}~\bibnamefont {Tiwari}},\ }\href
  {https://doi.org/10.1007/JHEP04(2021)131} {\bibfield  {journal} {\bibinfo
  {journal} {JHEP}\ }\textbf {\bibinfo {volume} {04}},\ \bibinfo {pages}
  {131}},\ \Eprint {https://arxiv.org/abs/2007.12214} {arXiv:2007.12214
  [hep-ph]} \BibitemShut {NoStop}%
\bibitem [{\citenamefont {Beneke}\ \emph
  {et~al.}(2020{\natexlab{c}})\citenamefont {Beneke}, \citenamefont {Garny},
  \citenamefont {Jaskiewicz}, \citenamefont {Szafron}, \citenamefont
  {Vernazza},\ and\ \citenamefont {Wang}}]{Beneke:2020ibj}%
  \BibitemOpen
  \bibfield  {author} {\bibinfo {author} {\bibfnamefont {M.}~\bibnamefont
  {Beneke}}, \bibinfo {author} {\bibfnamefont {M.}~\bibnamefont {Garny}},
  \bibinfo {author} {\bibfnamefont {S.}~\bibnamefont {Jaskiewicz}}, \bibinfo
  {author} {\bibfnamefont {R.}~\bibnamefont {Szafron}}, \bibinfo {author}
  {\bibfnamefont {L.}~\bibnamefont {Vernazza}},\ and\ \bibinfo {author}
  {\bibfnamefont {J.}~\bibnamefont {Wang}},\ }\href
  {https://doi.org/10.1007/JHEP10(2020)196} {\bibfield  {journal} {\bibinfo
  {journal} {JHEP}\ }\textbf {\bibinfo {volume} {10}},\ \bibinfo {pages}
  {196}},\ \Eprint {https://arxiv.org/abs/2008.04943} {arXiv:2008.04943
  [hep-ph]} \BibitemShut {NoStop}%
\bibitem [{\citenamefont {Ajjath}\ \emph
  {et~al.}(2022{\natexlab{a}})\citenamefont {Ajjath}, \citenamefont
  {Mukherjee},\ and\ \citenamefont {Ravindran}}]{Ajjath:2020ulr}%
  \BibitemOpen
  \bibfield  {author} {\bibinfo {author} {\bibfnamefont {A.~H.}\ \bibnamefont
  {Ajjath}}, \bibinfo {author} {\bibfnamefont {P.}~\bibnamefont {Mukherjee}},\
  and\ \bibinfo {author} {\bibfnamefont {V.}~\bibnamefont {Ravindran}},\ }\href
  {https://doi.org/10.1103/PhysRevD.105.094035} {\bibfield  {journal} {\bibinfo
   {journal} {Phys. Rev. D}\ }\textbf {\bibinfo {volume} {105}},\ \bibinfo
  {pages} {094035} (\bibinfo {year} {2022}{\natexlab{a}})},\ \Eprint
  {https://arxiv.org/abs/2006.06726} {arXiv:2006.06726 [hep-ph]} \BibitemShut
  {NoStop}%
\bibitem [{\citenamefont {van Beekveld}\ \emph
  {et~al.}(2021{\natexlab{a}})\citenamefont {van Beekveld}, \citenamefont
  {Vernazza},\ and\ \citenamefont {White}}]{vanBeekveld:2021mxn}%
  \BibitemOpen
  \bibfield  {author} {\bibinfo {author} {\bibfnamefont {M.}~\bibnamefont {van
  Beekveld}}, \bibinfo {author} {\bibfnamefont {L.}~\bibnamefont {Vernazza}},\
  and\ \bibinfo {author} {\bibfnamefont {C.~D.}\ \bibnamefont {White}},\ }\href
  {https://doi.org/10.1007/JHEP12(2021)087} {\bibfield  {journal} {\bibinfo
  {journal} {JHEP}\ }\textbf {\bibinfo {volume} {12}},\ \bibinfo {pages}
  {087}},\ \Eprint {https://arxiv.org/abs/2109.09752} {arXiv:2109.09752
  [hep-ph]} \BibitemShut {NoStop}%
\bibitem [{\citenamefont {Liu}\ and\ \citenamefont
  {Neubert}(2020{\natexlab{a}})}]{Liu:2019oav}%
  \BibitemOpen
  \bibfield  {author} {\bibinfo {author} {\bibfnamefont {Z.~L.}\ \bibnamefont
  {Liu}}\ and\ \bibinfo {author} {\bibfnamefont {M.}~\bibnamefont {Neubert}},\
  }\href {https://doi.org/10.1007/JHEP04(2020)033} {\bibfield  {journal}
  {\bibinfo  {journal} {JHEP}\ }\textbf {\bibinfo {volume} {04}},\ \bibinfo
  {pages} {033}},\ \Eprint {https://arxiv.org/abs/1912.08818} {arXiv:1912.08818
  [hep-ph]} \BibitemShut {NoStop}%
\bibitem [{\citenamefont {Krämer}\ \emph {et~al.}(1998)\citenamefont
  {Krämer}, \citenamefont {Laenen},\ and\ \citenamefont
  {Spira}}]{KRAMER1998523}%
  \BibitemOpen
  \bibfield  {author} {\bibinfo {author} {\bibfnamefont {M.}~\bibnamefont
  {Krämer}}, \bibinfo {author} {\bibfnamefont {E.}~\bibnamefont {Laenen}},\
  and\ \bibinfo {author} {\bibfnamefont {M.}~\bibnamefont {Spira}},\ }\href
  {https://doi.org/https://doi.org/10.1016/S0550-3213(97)00679-2} {\bibfield
  {journal} {\bibinfo  {journal} {Nuclear Physics B}\ }\textbf {\bibinfo
  {volume} {511}},\ \bibinfo {pages} {523} (\bibinfo {year}
  {1998})}\BibitemShut {NoStop}%
\bibitem [{\citenamefont {Ball}\ \emph {et~al.}(2013)\citenamefont {Ball},
  \citenamefont {Bonvini}, \citenamefont {Forte}, \citenamefont {Marzani},\
  and\ \citenamefont {Ridolfi}}]{BALL2013746}%
  \BibitemOpen
  \bibfield  {author} {\bibinfo {author} {\bibfnamefont {R.~D.}\ \bibnamefont
  {Ball}}, \bibinfo {author} {\bibfnamefont {M.}~\bibnamefont {Bonvini}},
  \bibinfo {author} {\bibfnamefont {S.}~\bibnamefont {Forte}}, \bibinfo
  {author} {\bibfnamefont {S.}~\bibnamefont {Marzani}},\ and\ \bibinfo {author}
  {\bibfnamefont {G.}~\bibnamefont {Ridolfi}},\ }\href
  {https://doi.org/https://doi.org/10.1016/j.nuclphysb.2013.06.012} {\bibfield
  {journal} {\bibinfo  {journal} {Nuclear Physics B}\ }\textbf {\bibinfo
  {volume} {874}},\ \bibinfo {pages} {746} (\bibinfo {year}
  {2013})}\BibitemShut {NoStop}%
\bibitem [{\citenamefont {Anastasiou}\ \emph {et~al.}(2015)\citenamefont
  {Anastasiou}, \citenamefont {Duhr}, \citenamefont {Dulat}, \citenamefont
  {Herzog},\ and\ \citenamefont {Mistlberger}}]{PhysRevLett.114.212001}%
  \BibitemOpen
  \bibfield  {author} {\bibinfo {author} {\bibfnamefont {C.}~\bibnamefont
  {Anastasiou}}, \bibinfo {author} {\bibfnamefont {C.}~\bibnamefont {Duhr}},
  \bibinfo {author} {\bibfnamefont {F.}~\bibnamefont {Dulat}}, \bibinfo
  {author} {\bibfnamefont {F.}~\bibnamefont {Herzog}},\ and\ \bibinfo {author}
  {\bibfnamefont {B.}~\bibnamefont {Mistlberger}},\ }\href
  {https://doi.org/10.1103/PhysRevLett.114.212001} {\bibfield  {journal}
  {\bibinfo  {journal} {Phys. Rev. Lett.}\ }\textbf {\bibinfo {volume} {114}},\
  \bibinfo {pages} {212001} (\bibinfo {year} {2015})}\BibitemShut {NoStop}%
\bibitem [{\citenamefont {van Beekveld}\ \emph {et~al.}(2020)\citenamefont {van
  Beekveld}, \citenamefont {Beenakker}, \citenamefont {Laenen},\ and\
  \citenamefont {White}}]{vanBeekveld:2019prq}%
  \BibitemOpen
  \bibfield  {author} {\bibinfo {author} {\bibfnamefont {M.}~\bibnamefont {van
  Beekveld}}, \bibinfo {author} {\bibfnamefont {W.}~\bibnamefont {Beenakker}},
  \bibinfo {author} {\bibfnamefont {E.}~\bibnamefont {Laenen}},\ and\ \bibinfo
  {author} {\bibfnamefont {C.~D.}\ \bibnamefont {White}},\ }\href
  {https://doi.org/10.1007/JHEP03(2020)106} {\bibfield  {journal} {\bibinfo
  {journal} {JHEP}\ }\textbf {\bibinfo {volume} {03}},\ \bibinfo {pages}
  {106}},\ \Eprint {https://arxiv.org/abs/1905.08741} {arXiv:1905.08741
  [hep-ph]} \BibitemShut {NoStop}%
\bibitem [{\citenamefont {van Beekveld}\ \emph
  {et~al.}(2021{\natexlab{b}})\citenamefont {van Beekveld}, \citenamefont
  {Laenen}, \citenamefont {Sinninghe~Damst\'e},\ and\ \citenamefont
  {Vernazza}}]{vanBeekveld:2021hhv}%
  \BibitemOpen
  \bibfield  {author} {\bibinfo {author} {\bibfnamefont {M.}~\bibnamefont {van
  Beekveld}}, \bibinfo {author} {\bibfnamefont {E.}~\bibnamefont {Laenen}},
  \bibinfo {author} {\bibfnamefont {J.}~\bibnamefont {Sinninghe~Damst\'e}},\
  and\ \bibinfo {author} {\bibfnamefont {L.}~\bibnamefont {Vernazza}},\ }\href
  {https://doi.org/10.1007/JHEP05(2021)114} {\bibfield  {journal} {\bibinfo
  {journal} {JHEP}\ }\textbf {\bibinfo {volume} {05}},\ \bibinfo {pages}
  {114}},\ \Eprint {https://arxiv.org/abs/2101.07270} {arXiv:2101.07270
  [hep-ph]} \BibitemShut {NoStop}%
\bibitem [{\citenamefont {Ajjath}\ \emph
  {et~al.}(2022{\natexlab{b}})\citenamefont {Ajjath}, \citenamefont
  {Mukherjee}, \citenamefont {Ravindran}, \citenamefont {Sankar},\ and\
  \citenamefont {Tiwari}}]{Ajjath2022}%
  \BibitemOpen
  \bibfield  {author} {\bibinfo {author} {\bibfnamefont {A.~H.}\ \bibnamefont
  {Ajjath}}, \bibinfo {author} {\bibfnamefont {P.}~\bibnamefont {Mukherjee}},
  \bibinfo {author} {\bibfnamefont {V.}~\bibnamefont {Ravindran}}, \bibinfo
  {author} {\bibfnamefont {A.}~\bibnamefont {Sankar}},\ and\ \bibinfo {author}
  {\bibfnamefont {S.}~\bibnamefont {Tiwari}},\ }\href
  {https://doi.org/10.1140/epjc/s10052-022-10174-7} {\bibfield  {journal}
  {\bibinfo  {journal} {European Physical Journal C}\ }\textbf {\bibinfo
  {volume} {82}},\ \bibinfo {pages} {234} (\bibinfo {year}
  {2022}{\natexlab{b}})}\BibitemShut {NoStop}%
\bibitem [{\citenamefont {Laenen}\ \emph {et~al.}(2009)\citenamefont {Laenen},
  \citenamefont {Stavenga},\ and\ \citenamefont {White}}]{Laenen_2009}%
  \BibitemOpen
  \bibfield  {author} {\bibinfo {author} {\bibfnamefont {E.}~\bibnamefont
  {Laenen}}, \bibinfo {author} {\bibfnamefont {G.}~\bibnamefont {Stavenga}},\
  and\ \bibinfo {author} {\bibfnamefont {C.~D.}\ \bibnamefont {White}},\ }\href
  {https://doi.org/10.1088/1126-6708/2009/03/054} {\bibfield  {journal}
  {\bibinfo  {journal} {Journal of High Energy Physics}\ }\textbf {\bibinfo
  {volume} {2009}},\ \bibinfo {pages} {054} (\bibinfo {year}
  {2009})}\BibitemShut {NoStop}%
\bibitem [{\citenamefont {Laenen}\ \emph {et~al.}(2011)\citenamefont {Laenen},
  \citenamefont {Magnea}, \citenamefont {Stavenga},\ and\ \citenamefont
  {White}}]{Laenen2011}%
  \BibitemOpen
  \bibfield  {author} {\bibinfo {author} {\bibfnamefont {E.}~\bibnamefont
  {Laenen}}, \bibinfo {author} {\bibfnamefont {L.}~\bibnamefont {Magnea}},
  \bibinfo {author} {\bibfnamefont {G.}~\bibnamefont {Stavenga}},\ and\
  \bibinfo {author} {\bibfnamefont {C.~D.}\ \bibnamefont {White}},\ }\href
  {https://doi.org/10.1007/JHEP01(2011)141} {\bibfield  {journal} {\bibinfo
  {journal} {Journal of High Energy Physics}\ }\textbf {\bibinfo {volume}
  {2011}},\ \bibinfo {pages} {141} (\bibinfo {year} {2011})}\BibitemShut
  {NoStop}%
\bibitem [{\citenamefont {Soar}\ \emph {et~al.}(2010)\citenamefont {Soar},
  \citenamefont {Vogt}, \citenamefont {Moch},\ and\ \citenamefont
  {Vermaseren}}]{Soar2010OnHD}%
  \BibitemOpen
  \bibfield  {author} {\bibinfo {author} {\bibfnamefont {G.}~\bibnamefont
  {Soar}}, \bibinfo {author} {\bibfnamefont {A.}~\bibnamefont {Vogt}}, \bibinfo
  {author} {\bibfnamefont {S.}~\bibnamefont {Moch}},\ and\ \bibinfo {author}
  {\bibfnamefont {J.~A.~M.}\ \bibnamefont {Vermaseren}},\ }\href@noop {}
  {\bibfield  {journal} {\bibinfo  {journal} {Nuclear Physics}\ }\textbf
  {\bibinfo {volume} {832}},\ \bibinfo {pages} {152} (\bibinfo {year}
  {2010})}\BibitemShut {NoStop}%
\bibitem [{\citenamefont {de~Florian}\ \emph {et~al.}(2014)\citenamefont
  {de~Florian}, \citenamefont {Mazzitelli}, \citenamefont {Moch},\ and\
  \citenamefont {Vogt}}]{Florian2014ApproximateNH}%
  \BibitemOpen
  \bibfield  {author} {\bibinfo {author} {\bibfnamefont {D.}~\bibnamefont
  {de~Florian}}, \bibinfo {author} {\bibfnamefont {J.}~\bibnamefont
  {Mazzitelli}}, \bibinfo {author} {\bibfnamefont {S.}~\bibnamefont {Moch}},\
  and\ \bibinfo {author} {\bibfnamefont {A.}~\bibnamefont {Vogt}},\ }\href@noop
  {} {\bibfield  {journal} {\bibinfo  {journal} {Journal of High Energy
  Physics}\ }\textbf {\bibinfo {volume} {2014}},\ \bibinfo {pages} {1}
  (\bibinfo {year} {2014})}\BibitemShut {NoStop}%
\bibitem [{\citenamefont {Presti}\ \emph {et~al.}(2014)\citenamefont {Presti},
  \citenamefont {Almasy},\ and\ \citenamefont {Vogt}}]{Presti2014LeadingLL}%
  \BibitemOpen
  \bibfield  {author} {\bibinfo {author} {\bibfnamefont {N.~A.~L.}\
  \bibnamefont {Presti}}, \bibinfo {author} {\bibfnamefont {A.~A.}\
  \bibnamefont {Almasy}},\ and\ \bibinfo {author} {\bibfnamefont
  {A.}~\bibnamefont {Vogt}},\ }\href@noop {} {\bibfield  {journal} {\bibinfo
  {journal} {Physics Letters B}\ }\textbf {\bibinfo {volume} {737}},\ \bibinfo
  {pages} {120} (\bibinfo {year} {2014})}\BibitemShut {NoStop}%
\bibitem [{\citenamefont {Bonocore}(2020)}]{Bonocore2020AsymptoticDO}%
  \BibitemOpen
  \bibfield  {author} {\bibinfo {author} {\bibfnamefont {D.}~\bibnamefont
  {Bonocore}},\ }\href@noop {} {\bibfield  {journal} {\bibinfo  {journal}
  {arXiv: High Energy Physics - Theory}\ } (\bibinfo {year}
  {2020})}\BibitemShut {NoStop}%
\bibitem [{\citenamefont {Gervais}(2017{\natexlab{a}})}]{PhysRevD.95.125009}%
  \BibitemOpen
  \bibfield  {author} {\bibinfo {author} {\bibfnamefont {H.}~\bibnamefont
  {Gervais}},\ }\href {https://doi.org/10.1103/PhysRevD.95.125009} {\bibfield
  {journal} {\bibinfo  {journal} {Phys. Rev. D}\ }\textbf {\bibinfo {volume}
  {95}},\ \bibinfo {pages} {125009} (\bibinfo {year}
  {2017}{\natexlab{a}})}\BibitemShut {NoStop}%
\bibitem [{\citenamefont {Gervais}(2017{\natexlab{b}})}]{PhysRevD.96.065007}%
  \BibitemOpen
  \bibfield  {author} {\bibinfo {author} {\bibfnamefont {H.}~\bibnamefont
  {Gervais}},\ }\href {https://doi.org/10.1103/PhysRevD.96.065007} {\bibfield
  {journal} {\bibinfo  {journal} {Phys. Rev. D}\ }\textbf {\bibinfo {volume}
  {96}},\ \bibinfo {pages} {065007} (\bibinfo {year}
  {2017}{\natexlab{b}})}\BibitemShut {NoStop}%
\bibitem [{\citenamefont {Gervais}(2017{\natexlab{c}})}]{Gervais2017SoftRT}%
  \BibitemOpen
  \bibfield  {author} {\bibinfo {author} {\bibfnamefont {H.~P.}\ \bibnamefont
  {Gervais}}\ }(\bibinfo {year} {2017})\BibitemShut {NoStop}%
\bibitem [{\citenamefont {Laenen}\ \emph {et~al.}(2021)\citenamefont {Laenen},
  \citenamefont {Sinninghe~Damst\'e}, \citenamefont {Vernazza}, \citenamefont
  {Waalewijn},\ and\ \citenamefont {Zoppi}}]{PhysRevD.103.034022}%
  \BibitemOpen
  \bibfield  {author} {\bibinfo {author} {\bibfnamefont {E.}~\bibnamefont
  {Laenen}}, \bibinfo {author} {\bibfnamefont {J.}~\bibnamefont
  {Sinninghe~Damst\'e}}, \bibinfo {author} {\bibfnamefont {L.}~\bibnamefont
  {Vernazza}}, \bibinfo {author} {\bibfnamefont {W.}~\bibnamefont
  {Waalewijn}},\ and\ \bibinfo {author} {\bibfnamefont {L.}~\bibnamefont
  {Zoppi}},\ }\href {https://doi.org/10.1103/PhysRevD.103.034022} {\bibfield
  {journal} {\bibinfo  {journal} {Phys. Rev. D}\ }\textbf {\bibinfo {volume}
  {103}},\ \bibinfo {pages} {034022} (\bibinfo {year} {2021})}\BibitemShut
  {NoStop}%
\bibitem [{\citenamefont {Beneke}\ \emph {et~al.}(2022)\citenamefont {Beneke},
  \citenamefont {Garny}, \citenamefont {Jaskiewicz}, \citenamefont {Strohm},
  \citenamefont {Szafron}, \citenamefont {Vernazza},\ and\ \citenamefont
  {Wang}}]{Beneke:2022obx}%
  \BibitemOpen
  \bibfield  {author} {\bibinfo {author} {\bibfnamefont {M.}~\bibnamefont
  {Beneke}}, \bibinfo {author} {\bibfnamefont {M.}~\bibnamefont {Garny}},
  \bibinfo {author} {\bibfnamefont {S.}~\bibnamefont {Jaskiewicz}}, \bibinfo
  {author} {\bibfnamefont {J.}~\bibnamefont {Strohm}}, \bibinfo {author}
  {\bibfnamefont {R.}~\bibnamefont {Szafron}}, \bibinfo {author} {\bibfnamefont
  {L.}~\bibnamefont {Vernazza}},\ and\ \bibinfo {author} {\bibfnamefont
  {J.}~\bibnamefont {Wang}},\ }\href {https://doi.org/10.1007/JHEP07(2022)144}
  {\bibfield  {journal} {\bibinfo  {journal} {JHEP}\ }\textbf {\bibinfo
  {volume} {07}},\ \bibinfo {pages} {144}},\ \Eprint
  {https://arxiv.org/abs/2205.04479} {arXiv:2205.04479 [hep-ph]} \BibitemShut
  {NoStop}%
\bibitem [{\citenamefont {Kolodrubetz}\ \emph {et~al.}(2016)\citenamefont
  {Kolodrubetz}, \citenamefont {Moult},\ and\ \citenamefont
  {Stewart}}]{Kolodrubetz:2016uim}%
  \BibitemOpen
  \bibfield  {author} {\bibinfo {author} {\bibfnamefont {D.~W.}\ \bibnamefont
  {Kolodrubetz}}, \bibinfo {author} {\bibfnamefont {I.}~\bibnamefont {Moult}},\
  and\ \bibinfo {author} {\bibfnamefont {I.~W.}\ \bibnamefont {Stewart}},\
  }\href {https://doi.org/10.1007/JHEP05(2016)139} {\bibfield  {journal}
  {\bibinfo  {journal} {JHEP}\ }\textbf {\bibinfo {volume} {05}},\ \bibinfo
  {pages} {139}},\ \Eprint {https://arxiv.org/abs/1601.02607} {arXiv:1601.02607
  [hep-ph]} \BibitemShut {NoStop}%
\bibitem [{\citenamefont {Moult}\ \emph
  {et~al.}(2017{\natexlab{a}})\citenamefont {Moult}, \citenamefont {Rothen},
  \citenamefont {Stewart}, \citenamefont {Tackmann},\ and\ \citenamefont
  {Zhu}}]{Moult:2016fqy}%
  \BibitemOpen
  \bibfield  {author} {\bibinfo {author} {\bibfnamefont {I.}~\bibnamefont
  {Moult}}, \bibinfo {author} {\bibfnamefont {L.}~\bibnamefont {Rothen}},
  \bibinfo {author} {\bibfnamefont {I.~W.}\ \bibnamefont {Stewart}}, \bibinfo
  {author} {\bibfnamefont {F.~J.}\ \bibnamefont {Tackmann}},\ and\ \bibinfo
  {author} {\bibfnamefont {H.~X.}\ \bibnamefont {Zhu}},\ }\href
  {https://doi.org/10.1103/PhysRevD.95.074023} {\bibfield  {journal} {\bibinfo
  {journal} {Phys. Rev. D}\ }\textbf {\bibinfo {volume} {95}},\ \bibinfo
  {pages} {074023} (\bibinfo {year} {2017}{\natexlab{a}})},\ \Eprint
  {https://arxiv.org/abs/1612.00450} {arXiv:1612.00450 [hep-ph]} \BibitemShut
  {NoStop}%
\bibitem [{\citenamefont {Feige}\ \emph {et~al.}(2017)\citenamefont {Feige},
  \citenamefont {Kolodrubetz}, \citenamefont {Moult},\ and\ \citenamefont
  {Stewart}}]{Feige:2017zci}%
  \BibitemOpen
  \bibfield  {author} {\bibinfo {author} {\bibfnamefont {I.}~\bibnamefont
  {Feige}}, \bibinfo {author} {\bibfnamefont {D.~W.}\ \bibnamefont
  {Kolodrubetz}}, \bibinfo {author} {\bibfnamefont {I.}~\bibnamefont {Moult}},\
  and\ \bibinfo {author} {\bibfnamefont {I.~W.}\ \bibnamefont {Stewart}},\
  }\href {https://doi.org/10.1007/JHEP11(2017)142} {\bibfield  {journal}
  {\bibinfo  {journal} {JHEP}\ }\textbf {\bibinfo {volume} {11}},\ \bibinfo
  {pages} {142}},\ \Eprint {https://arxiv.org/abs/1703.03411} {arXiv:1703.03411
  [hep-ph]} \BibitemShut {NoStop}%
\bibitem [{\citenamefont {Beneke}\ \emph
  {et~al.}(2018{\natexlab{a}})\citenamefont {Beneke}, \citenamefont {Garny},
  \citenamefont {Szafron},\ and\ \citenamefont {Wang}}]{Beneke:2017ztn}%
  \BibitemOpen
  \bibfield  {author} {\bibinfo {author} {\bibfnamefont {M.}~\bibnamefont
  {Beneke}}, \bibinfo {author} {\bibfnamefont {M.}~\bibnamefont {Garny}},
  \bibinfo {author} {\bibfnamefont {R.}~\bibnamefont {Szafron}},\ and\ \bibinfo
  {author} {\bibfnamefont {J.}~\bibnamefont {Wang}},\ }\href
  {https://doi.org/10.1007/JHEP03(2018)001} {\bibfield  {journal} {\bibinfo
  {journal} {JHEP}\ }\textbf {\bibinfo {volume} {03}},\ \bibinfo {pages}
  {001}},\ \Eprint {https://arxiv.org/abs/1712.04416} {arXiv:1712.04416
  [hep-ph]} \BibitemShut {NoStop}%
\bibitem [{\citenamefont {Beneke}\ \emph
  {et~al.}(2018{\natexlab{b}})\citenamefont {Beneke}, \citenamefont {Garny},
  \citenamefont {Szafron},\ and\ \citenamefont {Wang}}]{Beneke:2018rbh}%
  \BibitemOpen
  \bibfield  {author} {\bibinfo {author} {\bibfnamefont {M.}~\bibnamefont
  {Beneke}}, \bibinfo {author} {\bibfnamefont {M.}~\bibnamefont {Garny}},
  \bibinfo {author} {\bibfnamefont {R.}~\bibnamefont {Szafron}},\ and\ \bibinfo
  {author} {\bibfnamefont {J.}~\bibnamefont {Wang}},\ }\href
  {https://doi.org/10.1007/JHEP11(2018)112} {\bibfield  {journal} {\bibinfo
  {journal} {JHEP}\ }\textbf {\bibinfo {volume} {11}},\ \bibinfo {pages}
  {112}},\ \Eprint {https://arxiv.org/abs/1808.04742} {arXiv:1808.04742
  [hep-ph]} \BibitemShut {NoStop}%
\bibitem [{\citenamefont {Bhattacharya}\ \emph {et~al.}(2019)\citenamefont
  {Bhattacharya}, \citenamefont {Moult}, \citenamefont {Stewart},\ and\
  \citenamefont {Vita}}]{Bhattacharya:2018vph}%
  \BibitemOpen
  \bibfield  {author} {\bibinfo {author} {\bibfnamefont {A.}~\bibnamefont
  {Bhattacharya}}, \bibinfo {author} {\bibfnamefont {I.}~\bibnamefont {Moult}},
  \bibinfo {author} {\bibfnamefont {I.~W.}\ \bibnamefont {Stewart}},\ and\
  \bibinfo {author} {\bibfnamefont {G.}~\bibnamefont {Vita}},\ }\href
  {https://doi.org/10.1007/JHEP05(2019)192} {\bibfield  {journal} {\bibinfo
  {journal} {JHEP}\ }\textbf {\bibinfo {volume} {05}},\ \bibinfo {pages}
  {192}},\ \Eprint {https://arxiv.org/abs/1812.06950} {arXiv:1812.06950
  [hep-ph]} \BibitemShut {NoStop}%
\bibitem [{\citenamefont {Beneke}\ \emph
  {et~al.}(2019{\natexlab{a}})\citenamefont {Beneke}, \citenamefont {Garny},
  \citenamefont {Szafron},\ and\ \citenamefont {Wang}}]{Beneke:2019kgv}%
  \BibitemOpen
  \bibfield  {author} {\bibinfo {author} {\bibfnamefont {M.}~\bibnamefont
  {Beneke}}, \bibinfo {author} {\bibfnamefont {M.}~\bibnamefont {Garny}},
  \bibinfo {author} {\bibfnamefont {R.}~\bibnamefont {Szafron}},\ and\ \bibinfo
  {author} {\bibfnamefont {J.}~\bibnamefont {Wang}},\ }\href
  {https://doi.org/10.1007/JHEP09(2019)101} {\bibfield  {journal} {\bibinfo
  {journal} {JHEP}\ }\textbf {\bibinfo {volume} {09}},\ \bibinfo {pages}
  {101}},\ \Eprint {https://arxiv.org/abs/1907.05463} {arXiv:1907.05463
  [hep-ph]} \BibitemShut {NoStop}%
\bibitem [{\citenamefont {Bodwin}\ \emph {et~al.}(2021)\citenamefont {Bodwin},
  \citenamefont {Ee}, \citenamefont {Lee},\ and\ \citenamefont
  {Wang}}]{Bodwin:2021epw}%
  \BibitemOpen
  \bibfield  {author} {\bibinfo {author} {\bibfnamefont {G.~T.}\ \bibnamefont
  {Bodwin}}, \bibinfo {author} {\bibfnamefont {J.-H.}\ \bibnamefont {Ee}},
  \bibinfo {author} {\bibfnamefont {J.}~\bibnamefont {Lee}},\ and\ \bibinfo
  {author} {\bibfnamefont {X.-P.}\ \bibnamefont {Wang}},\ }\href
  {https://doi.org/10.1103/PhysRevD.104.116025} {\bibfield  {journal} {\bibinfo
   {journal} {Phys. Rev. D}\ }\textbf {\bibinfo {volume} {104}},\ \bibinfo
  {pages} {116025} (\bibinfo {year} {2021})},\ \Eprint
  {https://arxiv.org/abs/2107.07941} {arXiv:2107.07941 [hep-ph]} \BibitemShut
  {NoStop}%
\bibitem [{\citenamefont {Liu}\ \emph {et~al.}(2021)\citenamefont {Liu},
  \citenamefont {Mecaj}, \citenamefont {Neubert},\ and\ \citenamefont
  {Wang}}]{Liu:2020tzd}%
  \BibitemOpen
  \bibfield  {author} {\bibinfo {author} {\bibfnamefont {Z.~L.}\ \bibnamefont
  {Liu}}, \bibinfo {author} {\bibfnamefont {B.}~\bibnamefont {Mecaj}}, \bibinfo
  {author} {\bibfnamefont {M.}~\bibnamefont {Neubert}},\ and\ \bibinfo {author}
  {\bibfnamefont {X.}~\bibnamefont {Wang}},\ }\href
  {https://doi.org/10.1103/PhysRevD.104.014004} {\bibfield  {journal} {\bibinfo
   {journal} {Phys. Rev. D}\ }\textbf {\bibinfo {volume} {104}},\ \bibinfo
  {pages} {014004} (\bibinfo {year} {2021})},\ \Eprint
  {https://arxiv.org/abs/2009.04456} {arXiv:2009.04456 [hep-ph]} \BibitemShut
  {NoStop}%
\bibitem [{\citenamefont {Boughezal}\ \emph {et~al.}(2017)\citenamefont
  {Boughezal}, \citenamefont {Liu},\ and\ \citenamefont
  {Petriello}}]{Boughezal:2016zws}%
  \BibitemOpen
  \bibfield  {author} {\bibinfo {author} {\bibfnamefont {R.}~\bibnamefont
  {Boughezal}}, \bibinfo {author} {\bibfnamefont {X.}~\bibnamefont {Liu}},\
  and\ \bibinfo {author} {\bibfnamefont {F.}~\bibnamefont {Petriello}},\ }\href
  {https://doi.org/10.1007/JHEP03(2017)160} {\bibfield  {journal} {\bibinfo
  {journal} {JHEP}\ }\textbf {\bibinfo {volume} {03}},\ \bibinfo {pages}
  {160}},\ \Eprint {https://arxiv.org/abs/1612.02911} {arXiv:1612.02911
  [hep-ph]} \BibitemShut {NoStop}%
\bibitem [{\citenamefont {Moult}\ \emph
  {et~al.}(2017{\natexlab{b}})\citenamefont {Moult}, \citenamefont {Stewart},\
  and\ \citenamefont {Vita}}]{Moult:2017rpl}%
  \BibitemOpen
  \bibfield  {author} {\bibinfo {author} {\bibfnamefont {I.}~\bibnamefont
  {Moult}}, \bibinfo {author} {\bibfnamefont {I.~W.}\ \bibnamefont {Stewart}},\
  and\ \bibinfo {author} {\bibfnamefont {G.}~\bibnamefont {Vita}},\ }\href
  {https://doi.org/10.1007/JHEP07(2017)067} {\bibfield  {journal} {\bibinfo
  {journal} {JHEP}\ }\textbf {\bibinfo {volume} {07}},\ \bibinfo {pages}
  {067}},\ \Eprint {https://arxiv.org/abs/1703.03408} {arXiv:1703.03408
  [hep-ph]} \BibitemShut {NoStop}%
\bibitem [{\citenamefont {Chang}\ \emph {et~al.}(2018)\citenamefont {Chang},
  \citenamefont {Stewart},\ and\ \citenamefont {Vita}}]{Chang:2017atu}%
  \BibitemOpen
  \bibfield  {author} {\bibinfo {author} {\bibfnamefont {C.-H.}\ \bibnamefont
  {Chang}}, \bibinfo {author} {\bibfnamefont {I.~W.}\ \bibnamefont {Stewart}},\
  and\ \bibinfo {author} {\bibfnamefont {G.}~\bibnamefont {Vita}},\ }\href
  {https://doi.org/10.1007/JHEP04(2018)041} {\bibfield  {journal} {\bibinfo
  {journal} {JHEP}\ }\textbf {\bibinfo {volume} {04}},\ \bibinfo {pages}
  {041}},\ \Eprint {https://arxiv.org/abs/1712.04343} {arXiv:1712.04343
  [hep-ph]} \BibitemShut {NoStop}%
\bibitem [{\citenamefont {Beneke}\ \emph
  {et~al.}(2019{\natexlab{b}})\citenamefont {Beneke}, \citenamefont {Broggio},
  \citenamefont {Garny}, \citenamefont {Jaskiewicz}, \citenamefont {Szafron},
  \citenamefont {Vernazza},\ and\ \citenamefont {Wang}}]{Beneke:2018gvs}%
  \BibitemOpen
  \bibfield  {author} {\bibinfo {author} {\bibfnamefont {M.}~\bibnamefont
  {Beneke}}, \bibinfo {author} {\bibfnamefont {A.}~\bibnamefont {Broggio}},
  \bibinfo {author} {\bibfnamefont {M.}~\bibnamefont {Garny}}, \bibinfo
  {author} {\bibfnamefont {S.}~\bibnamefont {Jaskiewicz}}, \bibinfo {author}
  {\bibfnamefont {R.}~\bibnamefont {Szafron}}, \bibinfo {author} {\bibfnamefont
  {L.}~\bibnamefont {Vernazza}},\ and\ \bibinfo {author} {\bibfnamefont
  {J.}~\bibnamefont {Wang}},\ }\href {https://doi.org/10.1007/JHEP03(2019)043}
  {\bibfield  {journal} {\bibinfo  {journal} {JHEP}\ }\textbf {\bibinfo
  {volume} {03}},\ \bibinfo {pages} {043}},\ \Eprint
  {https://arxiv.org/abs/1809.10631} {arXiv:1809.10631 [hep-ph]} \BibitemShut
  {NoStop}%
\bibitem [{\citenamefont {Ebert}\ \emph {et~al.}(2019)\citenamefont {Ebert},
  \citenamefont {Moult}, \citenamefont {Stewart}, \citenamefont {Tackmann},
  \citenamefont {Vita},\ and\ \citenamefont {Zhu}}]{Ebert:2018gsn}%
  \BibitemOpen
  \bibfield  {author} {\bibinfo {author} {\bibfnamefont {M.~A.}\ \bibnamefont
  {Ebert}}, \bibinfo {author} {\bibfnamefont {I.}~\bibnamefont {Moult}},
  \bibinfo {author} {\bibfnamefont {I.~W.}\ \bibnamefont {Stewart}}, \bibinfo
  {author} {\bibfnamefont {F.~J.}\ \bibnamefont {Tackmann}}, \bibinfo {author}
  {\bibfnamefont {G.}~\bibnamefont {Vita}},\ and\ \bibinfo {author}
  {\bibfnamefont {H.~X.}\ \bibnamefont {Zhu}},\ }\href
  {https://doi.org/10.1007/JHEP04(2019)123} {\bibfield  {journal} {\bibinfo
  {journal} {JHEP}\ }\textbf {\bibinfo {volume} {04}},\ \bibinfo {pages}
  {123}},\ \Eprint {https://arxiv.org/abs/1812.08189} {arXiv:1812.08189
  [hep-ph]} \BibitemShut {NoStop}%
\bibitem [{\citenamefont {Moult}\ \emph
  {et~al.}(2020{\natexlab{b}})\citenamefont {Moult}, \citenamefont {Stewart},
  \citenamefont {Vita},\ and\ \citenamefont {Zhu}}]{Moult:2019uhz}%
  \BibitemOpen
  \bibfield  {author} {\bibinfo {author} {\bibfnamefont {I.}~\bibnamefont
  {Moult}}, \bibinfo {author} {\bibfnamefont {I.~W.}\ \bibnamefont {Stewart}},
  \bibinfo {author} {\bibfnamefont {G.}~\bibnamefont {Vita}},\ and\ \bibinfo
  {author} {\bibfnamefont {H.~X.}\ \bibnamefont {Zhu}},\ }\href
  {https://doi.org/10.1007/JHEP05(2020)089} {\bibfield  {journal} {\bibinfo
  {journal} {JHEP}\ }\textbf {\bibinfo {volume} {05}},\ \bibinfo {pages}
  {089}},\ \Eprint {https://arxiv.org/abs/1910.14038} {arXiv:1910.14038
  [hep-ph]} \BibitemShut {NoStop}%
\bibitem [{\citenamefont {Liu}\ and\ \citenamefont
  {Neubert}(2020{\natexlab{b}})}]{Liu:2020ydl}%
  \BibitemOpen
  \bibfield  {author} {\bibinfo {author} {\bibfnamefont {Z.~L.}\ \bibnamefont
  {Liu}}\ and\ \bibinfo {author} {\bibfnamefont {M.}~\bibnamefont {Neubert}},\
  }\href {https://doi.org/10.1007/JHEP06(2020)060} {\bibfield  {journal}
  {\bibinfo  {journal} {JHEP}\ }\textbf {\bibinfo {volume} {06}},\ \bibinfo
  {pages} {060}},\ \Eprint {https://arxiv.org/abs/2003.03393} {arXiv:2003.03393
  [hep-ph]} \BibitemShut {NoStop}%
\bibitem [{\citenamefont {Liu}\ \emph {et~al.}(2020)\citenamefont {Liu},
  \citenamefont {Mecaj}, \citenamefont {Neubert}, \citenamefont {Wang},\ and\
  \citenamefont {Fleming}}]{Liu:2020eqe}%
  \BibitemOpen
  \bibfield  {author} {\bibinfo {author} {\bibfnamefont {Z.~L.}\ \bibnamefont
  {Liu}}, \bibinfo {author} {\bibfnamefont {B.}~\bibnamefont {Mecaj}}, \bibinfo
  {author} {\bibfnamefont {M.}~\bibnamefont {Neubert}}, \bibinfo {author}
  {\bibfnamefont {X.}~\bibnamefont {Wang}},\ and\ \bibinfo {author}
  {\bibfnamefont {S.}~\bibnamefont {Fleming}},\ }\href
  {https://doi.org/10.1007/JHEP07(2020)104} {\bibfield  {journal} {\bibinfo
  {journal} {JHEP}\ }\textbf {\bibinfo {volume} {07}},\ \bibinfo {pages}
  {104}},\ \Eprint {https://arxiv.org/abs/2005.03013} {arXiv:2005.03013
  [hep-ph]} \BibitemShut {NoStop}%
\bibitem [{\citenamefont {Wang}(2019)}]{Wang:2019mym}%
  \BibitemOpen
  \bibfield  {author} {\bibinfo {author} {\bibfnamefont {J.}~\bibnamefont
  {Wang}},\ }\href@noop {} {\  (\bibinfo {year} {2019})},\ \Eprint
  {https://arxiv.org/abs/1912.09920} {arXiv:1912.09920 [hep-ph]} \BibitemShut
  {NoStop}%
\bibitem [{\citenamefont {Banfi}\ \emph {et~al.}(2002)\citenamefont {Banfi},
  \citenamefont {Salam},\ and\ \citenamefont {Zanderighi}}]{Banfi:2001bz}%
  \BibitemOpen
  \bibfield  {author} {\bibinfo {author} {\bibfnamefont {A.}~\bibnamefont
  {Banfi}}, \bibinfo {author} {\bibfnamefont {G.}~\bibnamefont {Salam}},\ and\
  \bibinfo {author} {\bibfnamefont {G.}~\bibnamefont {Zanderighi}},\
  }\href@noop {} {\bibfield  {journal} {\bibinfo  {journal} {JHEP}\ }\textbf
  {\bibinfo {volume} {0201}},\ \bibinfo {pages} {018}},\ \Eprint
  {https://arxiv.org/abs/hep-ph/0112156} {arXiv:hep-ph/0112156 [hep-ph]}
  \BibitemShut {NoStop}%
\bibitem [{\citenamefont {Catani}\ \emph {et~al.}(1992)\citenamefont {Catani},
  \citenamefont {Turnock},\ and\ \citenamefont {Webber}}]{Catani:1992jc}%
  \BibitemOpen
  \bibfield  {author} {\bibinfo {author} {\bibfnamefont {S.}~\bibnamefont
  {Catani}}, \bibinfo {author} {\bibfnamefont {G.}~\bibnamefont {Turnock}},\
  and\ \bibinfo {author} {\bibfnamefont {B.}~\bibnamefont {Webber}},\ }\href
  {https://doi.org/10.1016/0370-2693(92)91565-Q} {\bibfield  {journal}
  {\bibinfo  {journal} {Phys. Lett. B}\ }\textbf {\bibinfo {volume} {295}},\
  \bibinfo {pages} {269} (\bibinfo {year} {1992})}\BibitemShut {NoStop}%
\bibitem [{\citenamefont {Catani}\ \emph {et~al.}(1993)\citenamefont {Catani},
  \citenamefont {Trentadue}, \citenamefont {Turnock},\ and\ \citenamefont
  {Webber}}]{Catani:1992ua}%
  \BibitemOpen
  \bibfield  {author} {\bibinfo {author} {\bibfnamefont {S.}~\bibnamefont
  {Catani}}, \bibinfo {author} {\bibfnamefont {L.}~\bibnamefont {Trentadue}},
  \bibinfo {author} {\bibfnamefont {G.}~\bibnamefont {Turnock}},\ and\ \bibinfo
  {author} {\bibfnamefont {B.}~\bibnamefont {Webber}},\ }\href
  {https://doi.org/10.1016/0550-3213(93)90271-P} {\bibfield  {journal}
  {\bibinfo  {journal} {Nucl.Phys.}\ }\textbf {\bibinfo {volume} {B407}},\
  \bibinfo {pages} {3} (\bibinfo {year} {1993})}\BibitemShut {NoStop}%
\bibitem [{\citenamefont {Ellis}\ \emph {et~al.}(2011)\citenamefont {Ellis},
  \citenamefont {Stirling},\ and\ \citenamefont {Webber}}]{Ellis:1991qj}%
  \BibitemOpen
  \bibfield  {author} {\bibinfo {author} {\bibfnamefont {R.}~\bibnamefont
  {Ellis}}, \bibinfo {author} {\bibfnamefont {W.}~\bibnamefont {Stirling}},\
  and\ \bibinfo {author} {\bibfnamefont {B.}~\bibnamefont {Webber}},\
  }\href@noop {} {\emph {\bibinfo {title} {{QCD and collider physics}}}},\
  Vol.~\bibinfo {volume} {8}\ (\bibinfo  {publisher} {Cambridge University
  Press},\ \bibinfo {year} {2011})\BibitemShut {NoStop}%
\bibitem [{\citenamefont {Dokshitzer}\ \emph {et~al.}(1998)\citenamefont
  {Dokshitzer}, \citenamefont {Lucenti}, \citenamefont {Marchesini},\ and\
  \citenamefont {Salam}}]{Dokshitzer_1998}%
  \BibitemOpen
  \bibfield  {author} {\bibinfo {author} {\bibfnamefont {Y.~L.}\ \bibnamefont
  {Dokshitzer}}, \bibinfo {author} {\bibfnamefont {A.}~\bibnamefont {Lucenti}},
  \bibinfo {author} {\bibfnamefont {G.}~\bibnamefont {Marchesini}},\ and\
  \bibinfo {author} {\bibfnamefont {G.~P.}\ \bibnamefont {Salam}},\ }\href
  {https://doi.org/10.1088/1126-6708/1998/01/011} {\bibfield  {journal}
  {\bibinfo  {journal} {Journal of High Energy Physics}\ }\textbf {\bibinfo
  {volume} {1998}},\ \bibinfo {pages} {011} (\bibinfo {year}
  {1998})}\BibitemShut {NoStop}%
\bibitem [{\citenamefont {Dasgupta}\ and\ \citenamefont
  {Salam}(2002)}]{Dasgupta_2002}%
  \BibitemOpen
  \bibfield  {author} {\bibinfo {author} {\bibfnamefont {M.}~\bibnamefont
  {Dasgupta}}\ and\ \bibinfo {author} {\bibfnamefont {G.~P.}\ \bibnamefont
  {Salam}},\ }\href {https://doi.org/10.1088/1126-6708/2002/03/017} {\bibfield
  {journal} {\bibinfo  {journal} {Journal of High Energy Physics}\ }\textbf
  {\bibinfo {volume} {2002}},\ \bibinfo {pages} {017} (\bibinfo {year}
  {2002})}\BibitemShut {NoStop}%
\bibitem [{\citenamefont {Antonelli}\ \emph {et~al.}(2000)\citenamefont
  {Antonelli}, \citenamefont {Dasgupta},\ and\ \citenamefont
  {Salam}}]{Antonelli_2000}%
  \BibitemOpen
  \bibfield  {author} {\bibinfo {author} {\bibfnamefont {V.}~\bibnamefont
  {Antonelli}}, \bibinfo {author} {\bibfnamefont {M.}~\bibnamefont
  {Dasgupta}},\ and\ \bibinfo {author} {\bibfnamefont {G.~P.}\ \bibnamefont
  {Salam}},\ }\href {https://doi.org/10.1088/1126-6708/2000/02/001} {\bibfield
  {journal} {\bibinfo  {journal} {Journal of High Energy Physics}\ }\textbf
  {\bibinfo {volume} {2000}},\ \bibinfo {pages} {001} (\bibinfo {year}
  {2000})}\BibitemShut {NoStop}%
\bibitem [{\citenamefont {Ridder}\ \emph {et~al.}(2007)\citenamefont {Ridder},
  \citenamefont {Gehrmann}, \citenamefont {Glover},\ and\ \citenamefont
  {Heinrich}}]{Ridder_2007}%
  \BibitemOpen
  \bibfield  {author} {\bibinfo {author} {\bibfnamefont {A.~G.-D.}\
  \bibnamefont {Ridder}}, \bibinfo {author} {\bibfnamefont {T.}~\bibnamefont
  {Gehrmann}}, \bibinfo {author} {\bibfnamefont {E.}~\bibnamefont {Glover}},\
  and\ \bibinfo {author} {\bibfnamefont {G.}~\bibnamefont {Heinrich}},\ }\href
  {https://doi.org/10.1088/1126-6708/2007/12/094} {\bibfield  {journal}
  {\bibinfo  {journal} {Journal of High Energy Physics}\ }\textbf {\bibinfo
  {volume} {2007}},\ \bibinfo {pages} {094} (\bibinfo {year}
  {2007})}\BibitemShut {NoStop}%
\bibitem [{\citenamefont {Becher}\ and\ \citenamefont
  {Schwartz}(2008{\natexlab{a}})}]{Becher_2008}%
  \BibitemOpen
  \bibfield  {author} {\bibinfo {author} {\bibfnamefont {T.}~\bibnamefont
  {Becher}}\ and\ \bibinfo {author} {\bibfnamefont {M.~D.}\ \bibnamefont
  {Schwartz}},\ }\href {https://doi.org/10.1088/1126-6708/2008/07/034}
  {\bibfield  {journal} {\bibinfo  {journal} {Journal of High Energy Physics}\
  }\textbf {\bibinfo {volume} {2008}},\ \bibinfo {pages} {034} (\bibinfo {year}
  {2008}{\natexlab{a}})}\BibitemShut {NoStop}%
\bibitem [{\citenamefont {Weinzierl}(2008)}]{PhysRevLett.101.162001}%
  \BibitemOpen
  \bibfield  {author} {\bibinfo {author} {\bibfnamefont {S.}~\bibnamefont
  {Weinzierl}},\ }\href {https://doi.org/10.1103/PhysRevLett.101.162001}
  {\bibfield  {journal} {\bibinfo  {journal} {Phys. Rev. Lett.}\ }\textbf
  {\bibinfo {volume} {101}},\ \bibinfo {pages} {162001} (\bibinfo {year}
  {2008})}\BibitemShut {NoStop}%
\bibitem [{\citenamefont {Gehrmann-De~Ridder}\ \emph
  {et~al.}(2007)\citenamefont {Gehrmann-De~Ridder}, \citenamefont {Gehrmann},
  \citenamefont {Glover},\ and\ \citenamefont
  {Heinrich}}]{GehrmannDeRidder:2007bj}%
  \BibitemOpen
  \bibfield  {author} {\bibinfo {author} {\bibfnamefont {A.}~\bibnamefont
  {Gehrmann-De~Ridder}}, \bibinfo {author} {\bibfnamefont {T.}~\bibnamefont
  {Gehrmann}}, \bibinfo {author} {\bibfnamefont {E.}~\bibnamefont {Glover}},\
  and\ \bibinfo {author} {\bibfnamefont {G.}~\bibnamefont {Heinrich}},\ }\href
  {https://doi.org/10.1103/PhysRevLett.99.132002} {\bibfield  {journal}
  {\bibinfo  {journal} {Phys.Rev.Lett.}\ }\textbf {\bibinfo {volume} {99}},\
  \bibinfo {pages} {132002} (\bibinfo {year} {2007})},\ \Eprint
  {https://arxiv.org/abs/0707.1285} {arXiv:0707.1285 [hep-ph]} \BibitemShut
  {NoStop}%
\bibitem [{\citenamefont {Becher}\ and\ \citenamefont
  {Schwartz}(2008{\natexlab{b}})}]{Becher:2008cf}%
  \BibitemOpen
  \bibfield  {author} {\bibinfo {author} {\bibfnamefont {T.}~\bibnamefont
  {Becher}}\ and\ \bibinfo {author} {\bibfnamefont {M.~D.}\ \bibnamefont
  {Schwartz}},\ }\href {https://doi.org/10.1088/1126-6708/2008/07/034}
  {\bibfield  {journal} {\bibinfo  {journal} {JHEP}\ }\textbf {\bibinfo
  {volume} {0807}},\ \bibinfo {pages} {034}},\ \Eprint
  {https://arxiv.org/abs/0803.0342} {arXiv:0803.0342 [hep-ph]} \BibitemShut
  {NoStop}%
\bibitem [{\citenamefont {Dasgupta}\ and\ \citenamefont
  {Salam}(2004)}]{Dasgupta:2003iq}%
  \BibitemOpen
  \bibfield  {author} {\bibinfo {author} {\bibfnamefont {M.}~\bibnamefont
  {Dasgupta}}\ and\ \bibinfo {author} {\bibfnamefont {G.~P.}\ \bibnamefont
  {Salam}},\ }\href {https://doi.org/10.1088/0954-3899/30/5/R01} {\bibfield
  {journal} {\bibinfo  {journal} {J.Phys.}\ }\textbf {\bibinfo {volume}
  {G30}},\ \bibinfo {pages} {R143} (\bibinfo {year} {2004})},\ \Eprint
  {https://arxiv.org/abs/hep-ph/0312283} {arXiv:hep-ph/0312283 [hep-ph]}
  \BibitemShut {NoStop}%
\bibitem [{\citenamefont {Catani}\ \emph {et~al.}(1991)\citenamefont {Catani},
  \citenamefont {Turnock}, \citenamefont {Webber},\ and\ \citenamefont
  {Trentadue}}]{Catani:1991kz}%
  \BibitemOpen
  \bibfield  {author} {\bibinfo {author} {\bibfnamefont {S.}~\bibnamefont
  {Catani}}, \bibinfo {author} {\bibfnamefont {G.}~\bibnamefont {Turnock}},
  \bibinfo {author} {\bibfnamefont {B.~R.}\ \bibnamefont {Webber}},\ and\
  \bibinfo {author} {\bibfnamefont {L.}~\bibnamefont {Trentadue}},\ }\href
  {https://doi.org/10.1016/0370-2693(91)90494-B} {\bibfield  {journal}
  {\bibinfo  {journal} {Phys. Lett. B}\ }\textbf {\bibinfo {volume} {263}},\
  \bibinfo {pages} {491} (\bibinfo {year} {1991})}\BibitemShut {NoStop}%
\bibitem [{\citenamefont {Del~Duca}\ \emph {et~al.}(2017)\citenamefont
  {Del~Duca}, \citenamefont {Laenen}, \citenamefont {Magnea}, \citenamefont
  {Vernazza},\ and\ \citenamefont {White}}]{DelDuca:2017twk}%
  \BibitemOpen
  \bibfield  {author} {\bibinfo {author} {\bibfnamefont {V.}~\bibnamefont
  {Del~Duca}}, \bibinfo {author} {\bibfnamefont {E.}~\bibnamefont {Laenen}},
  \bibinfo {author} {\bibfnamefont {L.}~\bibnamefont {Magnea}}, \bibinfo
  {author} {\bibfnamefont {L.}~\bibnamefont {Vernazza}},\ and\ \bibinfo
  {author} {\bibfnamefont {C.}~\bibnamefont {White}},\ }\href
  {https://doi.org/10.1007/JHEP11(2017)057} {\bibfield  {journal} {\bibinfo
  {journal} {JHEP}\ }\textbf {\bibinfo {volume} {11}},\ \bibinfo {pages}
  {057}},\ \Eprint {https://arxiv.org/abs/1706.04018} {arXiv:1706.04018
  [hep-ph]} \BibitemShut {NoStop}%
\bibitem [{\citenamefont {Low}(1958)}]{Low:1958sn}%
  \BibitemOpen
  \bibfield  {author} {\bibinfo {author} {\bibfnamefont {F.~E.}\ \bibnamefont
  {Low}},\ }\href {https://doi.org/10.1103/PhysRev.110.974} {\bibfield
  {journal} {\bibinfo  {journal} {Phys. Rev.}\ }\textbf {\bibinfo {volume}
  {110}},\ \bibinfo {pages} {974} (\bibinfo {year} {1958})}\BibitemShut
  {NoStop}%
\bibitem [{\citenamefont {Burnett}\ and\ \citenamefont
  {Kroll}(1968)}]{Burnett:1967km}%
  \BibitemOpen
  \bibfield  {author} {\bibinfo {author} {\bibfnamefont {T.~H.}\ \bibnamefont
  {Burnett}}\ and\ \bibinfo {author} {\bibfnamefont {N.~M.}\ \bibnamefont
  {Kroll}},\ }\href {https://doi.org/10.1103/PhysRevLett.20.86} {\bibfield
  {journal} {\bibinfo  {journal} {Phys. Rev. Lett.}\ }\textbf {\bibinfo
  {volume} {20}},\ \bibinfo {pages} {86} (\bibinfo {year} {1968})}\BibitemShut
  {NoStop}%
\bibitem [{\citenamefont {Del~Duca}(1990)}]{DelDuca:1990gz}%
  \BibitemOpen
  \bibfield  {author} {\bibinfo {author} {\bibfnamefont {V.}~\bibnamefont
  {Del~Duca}},\ }\href {https://doi.org/10.1016/0550-3213(90)90392-Q}
  {\bibfield  {journal} {\bibinfo  {journal} {Nucl. Phys. B}\ }\textbf
  {\bibinfo {volume} {345}},\ \bibinfo {pages} {369} (\bibinfo {year}
  {1990})}\BibitemShut {NoStop}%
\bibitem [{\citenamefont {Farhi}(1977)}]{Farhi:1977sg}%
  \BibitemOpen
  \bibfield  {author} {\bibinfo {author} {\bibfnamefont {E.}~\bibnamefont
  {Farhi}},\ }\href {https://doi.org/10.1103/PhysRevLett.39.1587} {\bibfield
  {journal} {\bibinfo  {journal} {Phys. Rev. Lett.}\ }\textbf {\bibinfo
  {volume} {39}},\ \bibinfo {pages} {1587} (\bibinfo {year}
  {1977})}\BibitemShut {NoStop}%
\bibitem [{\citenamefont {Heister}\ \emph {et~al.}(2004)\citenamefont {Heister}
  \emph {et~al.}}]{ALEPH:2003obs}%
  \BibitemOpen
  \bibfield  {author} {\bibinfo {author} {\bibfnamefont {A.}~\bibnamefont
  {Heister}} \emph {et~al.} (\bibinfo {collaboration} {ALEPH}),\ }\href
  {https://doi.org/10.1140/epjc/s2004-01891-4} {\bibfield  {journal} {\bibinfo
  {journal} {Eur. Phys. J. C}\ }\textbf {\bibinfo {volume} {35}},\ \bibinfo
  {pages} {457} (\bibinfo {year} {2004})}\BibitemShut {NoStop}%
\bibitem [{ALE()}]{ALEPHsite}%
  \BibitemOpen
  \href@noop {} {\bibinfo  {journal}
  {https://aleph.web.cern.ch/aleph/aleph/newpub/physics.html}\ }\BibitemShut
  {NoStop}%
\bibitem [{\citenamefont {Catani}\ and\ \citenamefont
  {Webber}(1998)}]{Catani:1998sf}%
  \BibitemOpen
\bibfield  {journal} {  }\bibfield  {author} {\bibinfo {author} {\bibfnamefont
  {S.}~\bibnamefont {Catani}}\ and\ \bibinfo {author} {\bibfnamefont
  {B.}~\bibnamefont {Webber}},\ }\href
  {https://doi.org/10.1016/S0370-2693(98)00359-1} {\bibfield  {journal}
  {\bibinfo  {journal} {Phys. Lett. B}\ }\textbf {\bibinfo {volume} {427}},\
  \bibinfo {pages} {377} (\bibinfo {year} {1998})},\ \Eprint
  {https://arxiv.org/abs/hep-ph/9801350} {arXiv:hep-ph/9801350} \BibitemShut
  {NoStop}%
\bibitem [{\citenamefont {Gardi}\ and\ \citenamefont
  {Magnea}(2003)}]{Gardi:2003iv}%
  \BibitemOpen
  \bibfield  {author} {\bibinfo {author} {\bibfnamefont {E.}~\bibnamefont
  {Gardi}}\ and\ \bibinfo {author} {\bibfnamefont {L.}~\bibnamefont {Magnea}},\
  }\href@noop {} {\bibfield  {journal} {\bibinfo  {journal} {JHEP}\ }\textbf
  {\bibinfo {volume} {0308}},\ \bibinfo {pages} {030}},\ \Eprint
  {https://arxiv.org/abs/hep-ph/0306094} {arXiv:hep-ph/0306094 [hep-ph]}
  \BibitemShut {NoStop}%
\bibitem [{\citenamefont {Donoghue}\ \emph {et~al.}(1979)\citenamefont
  {Donoghue}, \citenamefont {Low},\ and\ \citenamefont {Pi}}]{Donoghue:1979vi}%
  \BibitemOpen
  \bibfield  {author} {\bibinfo {author} {\bibfnamefont {J.~F.}\ \bibnamefont
  {Donoghue}}, \bibinfo {author} {\bibfnamefont {F.}~\bibnamefont {Low}},\ and\
  \bibinfo {author} {\bibfnamefont {S.-Y.}\ \bibnamefont {Pi}},\ }\href
  {https://doi.org/10.1103/PhysRevD.20.2759} {\bibfield  {journal} {\bibinfo
  {journal} {Phys. Rev. D}\ }\textbf {\bibinfo {volume} {20}},\ \bibinfo
  {pages} {2759} (\bibinfo {year} {1979})}\BibitemShut {NoStop}%
\bibitem [{\citenamefont {Fox}\ and\ \citenamefont
  {Wolfram}(1978)}]{Fox:1978vu}%
  \BibitemOpen
  \bibfield  {author} {\bibinfo {author} {\bibfnamefont {G.~C.}\ \bibnamefont
  {Fox}}\ and\ \bibinfo {author} {\bibfnamefont {S.}~\bibnamefont {Wolfram}},\
  }\href {https://doi.org/10.1103/PhysRevLett.41.1581} {\bibfield  {journal}
  {\bibinfo  {journal} {Phys. Rev. Lett.}\ }\textbf {\bibinfo {volume} {41}},\
  \bibinfo {pages} {1581} (\bibinfo {year} {1978})}\BibitemShut {NoStop}%
\bibitem [{\citenamefont {Parisi}(1974)}]{Parisi:1974sq}%
  \BibitemOpen
  \bibfield  {author} {\bibinfo {author} {\bibfnamefont {G.}~\bibnamefont
  {Parisi}},\ }\href {https://doi.org/10.1016/0370-2693(74)90692-3} {\bibfield
  {journal} {\bibinfo  {journal} {Phys. Lett. B}\ }\textbf {\bibinfo {volume}
  {50}},\ \bibinfo {pages} {367} (\bibinfo {year} {1974})}\BibitemShut
  {NoStop}%
\bibitem [{\citenamefont {Ellis}\ \emph {et~al.}(1981)\citenamefont {Ellis},
  \citenamefont {Ross},\ and\ \citenamefont {Terrano}}]{Ellis:1980wv}%
  \BibitemOpen
  \bibfield  {author} {\bibinfo {author} {\bibfnamefont {R.~K.}\ \bibnamefont
  {Ellis}}, \bibinfo {author} {\bibfnamefont {D.~A.}\ \bibnamefont {Ross}},\
  and\ \bibinfo {author} {\bibfnamefont {A.~E.}\ \bibnamefont {Terrano}},\
  }\href {https://doi.org/10.1016/0550-3213(81)90165-6} {\bibfield  {journal}
  {\bibinfo  {journal} {Nucl. Phys. B}\ }\textbf {\bibinfo {volume} {178}},\
  \bibinfo {pages} {421} (\bibinfo {year} {1981})}\BibitemShut {NoStop}%
\bibitem [{\citenamefont {Catani}\ and\ \citenamefont
  {Webber}(1997)}]{Catani:1997xc}%
  \BibitemOpen
  \bibfield  {author} {\bibinfo {author} {\bibfnamefont {S.}~\bibnamefont
  {Catani}}\ and\ \bibinfo {author} {\bibfnamefont {B.~R.}\ \bibnamefont
  {Webber}},\ }\href {https://doi.org/10.1088/1126-6708/1997/10/005} {\bibfield
   {journal} {\bibinfo  {journal} {JHEP}\ }\textbf {\bibinfo {volume} {10}},\
  \bibinfo {pages} {005}},\ \Eprint {https://arxiv.org/abs/hep-ph/9710333}
  {arXiv:hep-ph/9710333} \BibitemShut {NoStop}%
\bibitem [{\citenamefont {Erdelyi}(1953)}]{ElliptBook}%
  \BibitemOpen
  \bibfield  {author} {\bibinfo {author} {\bibfnamefont {A.}~\bibnamefont
  {Erdelyi}},\ }\href@noop {} {\bibfield  {journal} {\bibinfo  {journal}
  {McGraw-Hill}\ }\textbf {\bibinfo {volume} {II}} (\bibinfo {year}
  {1953})}\BibitemShut {NoStop}%
\bibitem [{\citenamefont {Hoang}\ \emph {et~al.}(2015)\citenamefont {Hoang},
  \citenamefont {Kolodrubetz}, \citenamefont {Mateu},\ and\ \citenamefont
  {Stewart}}]{PhysRevD.91.094017}%
  \BibitemOpen
  \bibfield  {author} {\bibinfo {author} {\bibfnamefont {A.~H.}\ \bibnamefont
  {Hoang}}, \bibinfo {author} {\bibfnamefont {D.~W.}\ \bibnamefont
  {Kolodrubetz}}, \bibinfo {author} {\bibfnamefont {V.}~\bibnamefont {Mateu}},\
  and\ \bibinfo {author} {\bibfnamefont {I.~W.}\ \bibnamefont {Stewart}},\
  }\href {https://doi.org/10.1103/PhysRevD.91.094017} {\bibfield  {journal}
  {\bibinfo  {journal} {Phys. Rev. D}\ }\textbf {\bibinfo {volume} {91}},\
  \bibinfo {pages} {094017} (\bibinfo {year} {2015})}\BibitemShut {NoStop}%
\bibitem [{\citenamefont {Dokshitzer}\ and\ \citenamefont
  {Webber}(1995)}]{Dokshitzer:1995zt}%
  \BibitemOpen
  \bibfield  {author} {\bibinfo {author} {\bibfnamefont {Y.~L.}\ \bibnamefont
  {Dokshitzer}}\ and\ \bibinfo {author} {\bibfnamefont {B.~R.}\ \bibnamefont
  {Webber}},\ }\href {https://doi.org/10.1016/0370-2693(95)00548-Y} {\bibfield
  {journal} {\bibinfo  {journal} {Phys. Lett. B}\ }\textbf {\bibinfo {volume}
  {352}},\ \bibinfo {pages} {451} (\bibinfo {year} {1995})},\ \Eprint
  {https://arxiv.org/abs/hep-ph/9504219} {arXiv:hep-ph/9504219} \BibitemShut
  {NoStop}%
\bibitem [{\citenamefont {Akhoury}\ and\ \citenamefont
  {Zakharov}(1995)}]{Akhoury:1995sp}%
  \BibitemOpen
  \bibfield  {author} {\bibinfo {author} {\bibfnamefont {R.}~\bibnamefont
  {Akhoury}}\ and\ \bibinfo {author} {\bibfnamefont {V.~I.}\ \bibnamefont
  {Zakharov}},\ }\href {https://doi.org/10.1016/0370-2693(95)00866-J}
  {\bibfield  {journal} {\bibinfo  {journal} {Phys. Lett. B}\ }\textbf
  {\bibinfo {volume} {357}},\ \bibinfo {pages} {646} (\bibinfo {year}
  {1995})},\ \Eprint {https://arxiv.org/abs/hep-ph/9504248}
  {arXiv:hep-ph/9504248} \BibitemShut {NoStop}%
\bibitem [{\citenamefont {Korchemsky}\ and\ \citenamefont
  {Sterman}(1995)}]{Korchemsky:1994is}%
  \BibitemOpen
  \bibfield  {author} {\bibinfo {author} {\bibfnamefont {G.~P.}\ \bibnamefont
  {Korchemsky}}\ and\ \bibinfo {author} {\bibfnamefont {G.~F.}\ \bibnamefont
  {Sterman}},\ }\href {https://doi.org/10.1016/0550-3213(94)00006-Z} {\bibfield
   {journal} {\bibinfo  {journal} {Nucl. Phys. B}\ }\textbf {\bibinfo {volume}
  {437}},\ \bibinfo {pages} {415} (\bibinfo {year} {1995})},\ \Eprint
  {https://arxiv.org/abs/hep-ph/9411211} {arXiv:hep-ph/9411211} \BibitemShut
  {NoStop}%
\bibitem [{\citenamefont {Lee}\ and\ \citenamefont
  {Sterman}(2007)}]{PhysRevD.75.014022}%
  \BibitemOpen
  \bibfield  {author} {\bibinfo {author} {\bibfnamefont {C.}~\bibnamefont
  {Lee}}\ and\ \bibinfo {author} {\bibfnamefont {G.}~\bibnamefont {Sterman}},\
  }\href {https://doi.org/10.1103/PhysRevD.75.014022} {\bibfield  {journal}
  {\bibinfo  {journal} {Phys. Rev. D}\ }\textbf {\bibinfo {volume} {75}},\
  \bibinfo {pages} {014022} (\bibinfo {year} {2007})}\BibitemShut {NoStop}%
\bibitem [{\citenamefont {van Beekveld}\ \emph {et~al.}(2019)\citenamefont {van
  Beekveld}, \citenamefont {Beenakker}, \citenamefont {Laenen}, \citenamefont
  {Misra},\ and\ \citenamefont {White}}]{vanBeekveld:2019lwy}%
  \BibitemOpen
  \bibfield  {author} {\bibinfo {author} {\bibfnamefont {M.}~\bibnamefont {van
  Beekveld}}, \bibinfo {author} {\bibfnamefont {W.}~\bibnamefont {Beenakker}},
  \bibinfo {author} {\bibfnamefont {E.}~\bibnamefont {Laenen}}, \bibinfo
  {author} {\bibfnamefont {A.}~\bibnamefont {Misra}},\ and\ \bibinfo {author}
  {\bibfnamefont {C.~D.}\ \bibnamefont {White}},\ }\href
  {https://doi.org/10.22323/1.375.0053} {\bibfield  {journal} {\bibinfo
  {journal} {PoS}\ }\textbf {\bibinfo {volume} {RADCOR2019}},\ \bibinfo {pages}
  {053} (\bibinfo {year} {2019})}\BibitemShut {NoStop}%
\end{thebibliography}%

\end{document}